\documentclass[twocolumn]{aastex6}

\shorttitle{Observable Emission Features of GRMHD Jets}

\begin{document}
\title{Observable Emission Features of Black Hole GRMHD Jets on Event Horizon Scales}
\author{Hung-Yi Pu$^{1,2 \star}$, Kinwah Wu$^{3,4 \star}$, Ziri Younsi$^{5 \star}$, Keiichi Asada$^{2 \star}$, Yosuke Mizuno$^{5 \star}$, Masanori Nakamura$^{2 \star}$}

\affil{\altaffilmark{1}
Perimeter Institute for Theoretical Physics, 31 Caroline Street North, Waterloo, ON, N2L 2Y5, Canada
}	
\affil{\altaffilmark{2}
	Institute of Astronomy \& Astrophysics, Academia Sinica, 11F of Astronomy-Mathematics Building, AS/NTU No. 1, Taipei 10617, Taiwan}

\affil{\altaffilmark{3}
	Mullard Space Science Laboratory, University College London, Holmbury St. Mary, Dorking, Surrey, RH5 6NT, UK}	

\affil{\altaffilmark{4}
	School of Physics, University of Sydney, Sydney, NSW 2006, Australia}		
	
\affil{\altaffilmark{5}
	Institut f\"ur Theoretische Physik, Max-von-Laue-Stra{\ss}e 1, D-60438 Frankfurt am Main, Germany}

\email{$^{\star}$hpu@perimeterinstitute.ca; kinwah.wu@ucl.ac.uk;}
\email{younsi@th.physik.uni-frankfurt.de;asada@asiaa.sinica.edu.tw;}
\email{mizuno@th.physik.uni-frankfurt.de; nakamura@asiaa.sinica.edu.tw }

\begin{abstract}
The general-relativistic magnetohydrodynamical (GRMHD) formulation for black hole-powered jets naturally gives rise to a stagnation surface, wherefrom inflows and outflows along magnetic field lines that thread the black hole event horizon originate. 
We derive a conservative formulation for the transport of energetic electrons which are initially injected at the stagnation surface and subsequently transported along flow streamlines.  
With this formulation the energy spectra evolution of the electrons along the flow in the presence of radiative and adiabatic cooling is determined.
For flows regulated by synchrotron radiative losses and adiabatic cooling,
  the effective radio emission region is found to be finite, and geometrically it is more extended along the jet central axis.  
Moreover, the emission from regions adjacent to the stagnation surface is expected to be the most luminous 
  as this is where the freshly injected energetic electrons concentrate.   
An observable stagnation surface is thus a strong prediction of the GRMHD jet model with the prescribed non-thermal electron injection. 
Future millimeter/sub-millimeter (mm/sub-mm) very-long-baseline interferometric (VLBI) observations of supermassive black hole candidates, 
  such as the one at the center of M87, can verify this GRMHD jet model and its associated non-thermal electron injection mechanism.
\end{abstract}


\section{Introduction}
\label{sec:intro}

It is a pressing question as to how the radiation that is observed in relativistic jets in active galactic nuclei (AGN) is generated \citep[e.g.,][]{bla79,mar80,zen97,lai02,hon10,lev11,mos11,ito13,mas13,pot13,wan14,sco14,shi14,hov14,tur15,asa16,hir16,koa16,kha16,pri16}.  
Although there is a common consensus that the emitters are energetic particles, 
how these particles are accelerated to such high energies, 
how they dissipate their energy, 
and how they are transported with the jets themselves are still the subject of feverish investigation \citep[e.g.,][]{ bla87}.  
It is argued that relativistic outflows from black holes are associated with accretion flows \citep[][]{bla76, fen04,mei05,fer06,tru11,pu12,wu13,sba14,ish14}.  
In the case of collimated relativistic jets, 
  magnetic fields must play an important role \citep[][]{cam86a,cam86b,cam87,fen01,vla04,kom07,lyu09,nak13,hom15}, 
  and it is argued that jets are powered
  at the expense of the black hole, 
  wherein energy is extracted from a reservoir of rotational energy from the black hole itself, 
  either by electromagnetic means \citep[][]{bla77,kom04,kom05,tom14} or through magnetohydrodynamical processes \citep[][]{phi83,tak90,koi02,mac04,haw06}. 
Models have been proposed for both of these cases, 
  and they both in principle possess certain testable predictions. 
In particular, for the latter, 
  numerical GRMHD simulations \citep[e.g.\ ][]{M06} and analytical GRMHD studies \citep[e.g.\ ][]{tak90,pu16} 
  consistently show the presence of a stagnation or separation surface (a separatrix). 
This surface separates the (inner) inflow region from the (outer) outflow region, 
  which both follow the same global, black hole-threading magnetic field lines. 
 The relatively slow radial velocities near the stagnation surface imply a high concentration of fluid particles.    
If energetic particles are injected in the vicinity of the stagnation surface or near the black hole event horizon, 
   they must accumulate in high concentrations near the stagnation surface, 
  provided that the cooling time scale is not significantly shorter than the dynamical time scale of the jet fluid flow.  
  This surface is a unique feature of relativistic GRMHD jets, 
  and in contrast to an ideal force-free magnetic jet \citep[e.g.][]{mck07,tch08,bro09}.  

In a GRMHD jet model \citep[e.g.,][]{cam86b,tak90}, the fluid mass is comprised mainly of baryons, 
  although this does not rule out the possibility that there could also be a substantial number of positively-charged leptons.
For such a baryonic jet, 
  one would expect that the mm/sub-mm radio emission is predominantly synchrotron radiation from energetic leptons
  which are likely non-thermal in origin. 
In a simple approximation, the radiative power of leptonic synchrotron emission
  is related to the magnetic energy density as  
  $P\propto\bar{\gamma}^{2}\bar{\beta}^{2} B^{2}$, 
  where $\bar{\gamma}$ 
  is the effective Lorentz factor, $\bar{\beta}$ is the effective velocity of the leptons 
  (normalised to the speed of light), 
  and $B$ is the magnetic field strength. 
While the non-thermal synchrotron model is in general consistent with polarimetric and spectral observations  
  of the radio emission from AGN jets \citep[e.g.][]{bro10,cla11}, 
  the exact origin of non-thermal particles (e.g.~electrons in the context of baryonic jets) 
  is still an unresolved issue. 

In this study, we aim to determine the observational signatures of the stagnation surface presented in model GRMHD jets 
  in the context of non-thermal synchrotron radiation from baryon-dominated relativistic flows.  
In spite of the many uncertainties in the detailed microphysics, 
  many of these are unimportant in determining the qualitative aspects of the observable features.  
Due to the absence of physical models for the injection of non-thermal electrons into jets,   
  certain assumptions have been adopted in the pioneering works in the field,  
  such as phenomenological mass loading models \citep{bro09} 
  or a fixed ratio of the internal energy of non-thermal electrons to the magnetic energy \citep{dex12}. 
These assumptions have restricted the ``thermal properties" (such as the the energy cutoff and the spectral index) of the non-thermal synchrotron-emitting particles to be rigidly associated with the jet flow. 
We relax such restrictions by allowing the energy content of the particles to evolve as they are transported in the flow, 
  through appropriate {\it loss} processes, such as synchrotron and adiabatic processes.    
This improvement enables us to provide more reliable observational signatures for the GRMHD jet.  
With forthcoming mm/sub-mm VLBI observations of AGN jets, in particular the one in M87 
  by the Event Horizon Telescope (EHT) \citep[][]{doe12,lu14,kin15,ric15} and the Greenland Telescope (GLT) \citep[]{nak13a,ino14,raf16},  
  definitive observable emission signatures like those presented in this work 
  will prove essential in testing theoretical models. 
  
The paper is organised as follows. 
The modelling of the spatial and energy population of non-thermal electrons along a background GRMHD flow is presented in \S\,\ref{sec:model_nth}. 
A model to qualitatively describe a GRMHD flow is described in \S\,\ref{sec:model}. The results and implications of the spatial variation of non-thermal electrons are given in \S\,\ref{sec:result}.  Finally, the summary and conclusions are presented in \S\,\ref{sec:implication}.

\section{Spatial and energy population of non-thermal electrons}\label{sec:model_nth}

In this section we formulate the population of non-thermal energetic particles and their evolution in a stationary jet flow. 
As the flow within the jet follows streamlines, 
  the particle number density along each flow streamline evolves from an initial value\footnote{
In a one-zone formulation  
   the energy spectral evolution of a population of energetic particles is governed by the transport equation 
\[
  \frac{\partial n(\gamma)}{\partial t} + \frac{\partial }{\partial \gamma} {{\dot \gamma} n(\gamma)}  
     =  {\dot Q}_{\rm inj}(\gamma) - {\dot Q}_{\rm loss}(\gamma)  \,,
      \] 
\citep[see e.g.][]{gin64},  
   where $n(\gamma)$ is the particle number density, and
   ${\dot Q}_{\rm inj}$ and ${\dot Q}_{\rm loss}$ are the particle injection and particle loss rates, respectively. 
This equation has assumed that the angular distribution of the particles in the momentum space is preserved, 
  which is justified in a magnetically confined region if radiative losses are insignificant.  
In the MHD jet model considered here,  
  the energetic (non-thermal) particles are carried along the flow streamlines (which are parallel to the magnetic field lines) by the fluid.    
This formulation is therefore not applicable. 
A covariant formulation which takes into account the spatial transport is required 
  when constructing the transport equation for the energy spectral evolution of the particle population \citep[see, e.g., ][]{cha17}. 
}.
This enables the construction of a simple analytic spectral evolutionary model for the non-thermal particles 
  along the flow within the entire jet,  provided that:
  (i) the non-thermal particles do not thermalise on time scales shorter than the dynamical time scale of the flow,  
  (ii) there is no further injection and no leakage of particles along the flow 
      and the diffusion of non-thermal particles across flow streamlines is negligible, 
  (iii) collisions between non-thermal particles, 
      and the scattering and drift along magnetic field lines (and flow streamlines) are unimportant. 
In a stationary state 
  the evolution of the non-thermal particle spectral density along a specific flow streamline 
  can therefore be expressed as 
 \begin{eqnarray} \label{eq:g1}
   n_{\rm nth} \left(\gamma , s\right) = \lambda_0\;\! n_{\rm th}(s) \! \ \mathcal{G}(\gamma)\bigg\vert_s \ ,  
 \end{eqnarray}    
  where $s$ is the poloidal location on the flow streamline, 
 $\lambda_0$ is a parameter specifying the relative fraction of the non-thermal particles, 
   and $\mathcal{G}(\gamma)$ is the normalized energy spectral distribution function of the non-thermal particles,   
   evaluated in the co-moving frame of the bulk flow.   
The absence of particle injection, leakage and thermalisation 
  implies that $\lambda_0$ will remain constant along each flow streamline.  
The number of thermal particles is given by the density of the fluid in the flow, 
  which is the solution of the GRMHD equations of the jet.
  
\subsection{Transport of the particle distribution function}
\label{sec:trans_F}

Note that Liouville's theorem for the conservation of phase space volume along geodesics  
  is not applicable to determine $\mathcal{G}(\gamma)$, 
  as it is not a free-falling system locally or globally.  
More specifically,    
  the flow of the non-thermal radiating particles (presumably electrons) 
  in the configuration space is constrained by the bulk motion of MHD fluid in the jet  
  and the flow in the momentum space is determined 
  by the radiative loss and the thermo-hydrodynamics of the jet fluid. 
We may however formulate the transport equation to determine the profile of $\mathcal{G}(\gamma)|_s$ along $s$ 
  for the non-thermal particles in terms of the particles' cumulative energy spectral distribution function 
  by taking advantage of the probability invariance and the particle number conservation along the flow.  

We start by defining the cumulative energy spectral distribution function of the non-thermal particles, evaluated at location $s$, as 
\begin{eqnarray} \label{eq:g}
  \mathcal{F}(\gamma,s) & = & \int_1^\gamma {\rm d}\gamma' \, \mathcal{G}(\gamma')\bigg\vert_s \ , 
\end{eqnarray} 
   whose normalization is 
\begin{equation} \label{eq:g_nor}
  \int^{\gamma_{\rm max}}_{1}  {\rm d}\gamma \  \mathcal{G}(\gamma) \ \bigg\vert_s =1 \ .
\end{equation}
This cumulative energy distribution function $\mathcal{F}(\gamma, s)$ 
   is essentially the area under the curve $\mathcal{G}(\gamma')|_s$, bounded by $\gamma'=1$ and  $\gamma$.  

\begin{figure} 
\begin{centering}
\includegraphics[width=0.95\columnwidth]{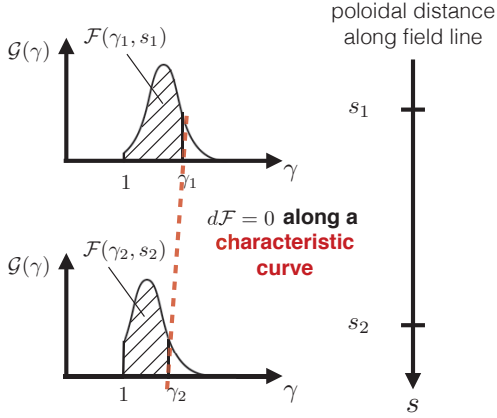}
\caption{Illustration of tracing the spatial variation of non-thermal electron energy, $\gamma$,  by using $\mathcal{G}(\gamma)|_{s}$ 
(see \S~\ref{sec:trans_F})  and a moving boundary such that $d\mathcal{F}=0$ (red dashed line).
$\mathcal{F}(\gamma,s)$ is the cumulative  energy spectral distribution function (equation (\ref{eq:g})), represented by the shaded region.
The moving boundary $d\mathcal{F}=0$ can be described by a characteristic curve (equation (\ref{eq:dg_ds})).
See \S~\ref{sec:trans_F} for details.}\label{fig:df}
\end{centering}
\end{figure}

In the absence of particle scattering,   
   the condition $\gamma_{\rm b} > \gamma_{\rm a}$ is invariant along $s$ for any arbitrary pair of particles ``a'' and ``b''. 
This ensures that $\mathcal{F}(\gamma_{2},s_{2})=\mathcal{F}(\gamma_{1},s_{1})$ 
  with only the bounds to determine $\mathcal{F}(\gamma, s)$ being changed from $\gamma_{1}$ to $\gamma_{2}$,  
  when a population of particles located at $s_{1}$ is transported along the characteristic to a new location $s_{2}$  (see illustration in Figure~\ref{fig:df}).   
The invariance of $\mathcal{F}(\gamma,s)$ along $s$ implies that 
   the cumulative distribution function satisfies the advective transport equation 
\begin{equation}
  \left[  \frac{\partial  }{\partial s} + \frac{{\rm d} \gamma}{{\rm d}s} \frac{\partial   }{\partial \gamma} \right] 
  \mathcal{F}(\gamma, s) =0 \; ,  
\end{equation}  
   and $\mathcal{G}(\gamma)$ is traced by the curves ${\rm d}\mathcal{F}=0$, 
   which are the characteristic curves.   
Hence, the transport equation can be solved 
   using the method of characteristics once the energy gain/loss rate along $s$, ${\rm d}\gamma/{\rm d}s$, is specified.  
   
In the reference frame co-moving with the bulk motion of the transported jet fluid,   
\begin{eqnarray} \label{eq:dg_ds}
  \frac{{\rm d} \gamma}{{\rm d}s} & = & \frac{\dot  \gamma}{v} \; \! \frac{{\rm d} \tau}{{\rm d} t}\;\! \bigg\vert_s \; , 
\end{eqnarray}  
  where ${\rm d} \tau$ is the proper time interval, and $\dot \gamma\;\! (\equiv {\rm d} \gamma /  {\rm d} \tau)$ 
  is the energy gain/loss per unit time. 
Since $s$ is increasing from the black hole to infinity, we can freely choose where $s=0$. Here and hereafter we adopt the convention $s=s_{0}=0$ at the stagnation surface, therefore ${\rm d}s > 0$ for the outflow and ${\rm d} s < 0$ for the inflow in the jet.
The speed of the jet fluid, $v={\rm d} s/{\rm d}t$, may then be written (as measured in the Boyer-Lindquist coordinate frame) as
  \begin{equation}
  v=\sqrt{\frac{\Sigma}{\Delta}}\left[ v^{r}+v^{\theta}\left( \frac{\mathrm{d} r}{\mathrm{d} \theta} \right) \right] 
   \sqrt{1+\Delta \left(\frac{{\rm d}\theta}{{\rm d} r}\right)^{2}} \; ,
  \end{equation}  
  where $\Sigma\equiv r^{2}+a^{2}\cos^{2}\theta$, $\Delta\equiv r^{2}-2 M_{\rm BH} r+a^{2}$, $M_{\rm BH}$ is the black hole mass, $a$ is the spin parameter of the black hole, $v^{i}\equiv u^{i}/u^{t}$, and $u^{\mu}$ is the four-velocity of the fluid.
The expression for ${\rm d}r/{\rm d}\theta$ in the above equation is given by the stream function  (\S\ref{sec:model}).
  
For non-thermal particles with $\mathcal{F}(\gamma, s_0)$ injected at $s=s_{0}$ in a stationary manner, 
  the transport equation may now be expressed as 
\begin{eqnarray} \label{eq:trans_f}
 \left[  \frac{\partial  }{\partial s} + \frac{{\rm d} \gamma}{{\rm d}s} \frac{\partial   }{\partial \gamma} \right] 
   \mathcal{F}(\gamma, s) 
    & = & \mathcal{F}(\gamma, s_0)\;\! \delta(s - s_{0}) \, . 
\end{eqnarray}    
Since freshly accelerated particles in high-energy relativistic systems tend to have a power-law energy spectrum, 
  we simply adopt a power-law spectrum for $\mathcal{F}(\gamma, s_0)$ to avoid unnecessary complications. 
Hence, we have    
\begin{equation}\label{eq:inj}
  \mathcal{G}(\gamma)\, \big\vert_{s_{0}}\propto\gamma_{0}^{-\alpha}\;, 
\end{equation}
  where $\gamma_{0} \equiv \gamma |_{s_{0}}$. 
The value of the spectral index $\alpha$ depends on the particle acceleration process 
  \citep[e.g.][]{bel78,bla87,sir11,sir14,wer16}. 
The characteristic curve, which pivots about ($\gamma_{0}, s_{0}$), 
  is determined from equation (\ref{eq:dg_ds}) and is hence independent of $\alpha$,  
  despite the fact that $\mathcal{F}(\gamma_{0},s_{0})$ (equation (\ref{eq:g})) is dependent on the chosen value of $\alpha$.

\subsection{Energy Loss} 
\label{sec:energy_dot}

The energy losses of relativistic electrons in jet outflows are often attributed to cooling due to  
   synchrotron radiation, Compton scattering, and/or adiabatic volume expansion.  
The first two are radiative processes determined by the microscopic properties of the particles, 
  and the last one is associated with the hydrodynamics of the bulk outflow of the transported jet fluid. 
The change in energy of the energetic particles can be expressed as 
\begin{eqnarray} 
   {\dot \gamma} &=&   {\dot \gamma}_{\rm rad}  + {\dot \gamma}_{\rm adi}  \nonumber \\
   &=&   ({\dot \gamma}_{\rm syn}+  { \dot \gamma}_{\rm com}) +   {\dot \gamma}_{\rm adi} \, ,
\end{eqnarray} 
  in both the inflow and the outflow regions.  
In this expression 
  the radiative loss terms due to synchrotron radiation and Compton scattering 
  (${\dot \gamma}_{\rm syn}$ and ${ \dot \gamma}_{\rm com}$ respectively) 
  are always negative, 
  but the mechanical term due to adiabatic change in the volume of the fluid element ($\dot \gamma_{\rm adi}$) 
  can be positive (for compression heating) or negative (for expansion cooling).   
  
The radiative losses for relativistic electrons due to synchrotron radiation or Compton scattering (assuming the electrons have an isotropic momentum distribution and energy specified by a Lorentz factor $\gamma$), 
are given \citep[][]{tuc75} by  
\begin{eqnarray}  \label{eq:cooling_syn}
  \dot \gamma_{\rm rad} & = & -\frac{4}{3} \frac{\sigma_{\rm T}}{m_{\rm e} c} \left(\gamma^2 -1\right) U_{\rm x}  \nonumber \\ 
     &\approx & -3.2\times 10^{-8} \left(\gamma^2 -1\right) U_{\rm x} \  {\rm s}^{-1} \,.
\end{eqnarray}
Here $\sigma_{\rm T}$ is the Thomson scattering cross-section  
   and $U_{\rm x}$ is the energy density of the magnetic field ($U_{\rm x}= U_{\rm mag} = B^2/8\pi$) for synchrotron radiation, 
  which is also the energy density of the photon radiation field ($U_{\rm x}=U_{\rm ph}$) for Compton scattering.  
 
No energy or particles (which transport kinetic energy) leak from this volume element $(\delta V)$ 
    in a strictly adiabatic process. 
Hence, the energy gain/loss of the particles caused by an adiabatic increase/decrease in volume is       
\begin{eqnarray} \label{eq:adi-V} 
  {\dot \gamma}_{\rm adi} & \approx & -   \frac{\left(\gamma - 1\right)}{\delta V}  \frac{{\rm d}  ( \delta V)  }{{\rm d}\tau} \bigg\vert_{\rm nth} \  {\rm s}^{-1} \,,  
\end{eqnarray} 
   where $\delta V\vert_{\rm nth}$ is a co-moving volume element containing a specific number of energetic electrons.   
 As the diffusion drift of non-thermal particles is unimportant on the time scales of interest,  
  the particle transport is dominated by advection.  
It is therefore sufficient to set $\delta V\vert_{\rm nth} =  \delta V\vert_{\rm fluid}$.
Thus, a volume expansion in the fluid ($\Delta (\delta V\vert_{\rm fluid}) >0$) gives a negative ${\dot \gamma}_{\rm adi}$ (i.e.~cooling) 
  and a volume contraction in the fluid ($\Delta (\delta V\vert_{\rm fluid}) <0$) gives a positive ${\dot \gamma}_{\rm adi}$ (i.e.~heating). 

We may express the time evolution of the jet fluid volume  
  in terms of the time evolution of 
  the mass density of the jet fluid along the flow streamline
  (since the contribution to the total fluid mass by non-thermal particles is negligible), 
  ${\rm d} \ln\rho=-{\rm d}\ln(\delta V)$, and obtain:

\begin{eqnarray}  
\frac{1}{\delta V} \frac{{\rm d} \left(\delta V \right)}{{\rm d}\tau } \bigg\vert_{\rm fluid}
     & = & -  v \left( \frac{{\rm d}t}{{\rm d}\tau}\right)
       \left[\frac{{\rm d} }{{\rm d}s} \ln \left( \frac{\rho}{\rho_0} \right) \right] \  {\rm s}^{-1} \,, 
\end{eqnarray}   
 where $\rho_{0}$ is the initial value of $\rho$ at the stagnation surface, which is introduced
 to keep the argument of the logarithm function dimensionless.
It follows that 
  \begin{equation} \label{eq:cooling_adi}
   {\dot \gamma}_{\rm adi}
     \approx    ( \gamma -1)\;\!   v  \left( \frac{{\rm d}t}{{\rm d}\tau}\right)
      \left[\frac{{\rm d} }{{\rm d}s} \ln \left( \frac{\rho}{\rho_0} \right) \right] \  {\rm s}^{-1}  \, ,  
  \end{equation}
  an expression valid for both jet inflow and jet outflow. 
 As a result, whether the adiabatic process results in a cooling or heating process is determined by
  the profile of $\rho$ along the flow. 
  If the profile of $\rho$ is decreasing (or increasing) from the stagnation surface, adiabatic cooling (heating) will take place. We demonstrate that adiabatic cooling occurs both for GRMHD outflows and for GRMHD inflows in \S\,\ref{sec:result}.

Note that ${\dot \gamma}_{\rm rad}$ is determined by the internal thermodynamical properties of the flow,  
  and hence does not depend explicitly on the geometrical extent of the system.  
However, ${\dot \gamma}_{\rm adi}$ is determined 
  by the global thermodynamical properties of the system, 
  and is hence expected to scale with the black hole mass, $M_{\rm BH}$.   
From the ratio of equations (\ref{eq:cooling_syn}) and (\ref{eq:adi-V}) we obtain 
\begin{equation} \label{eq:cooling_ratio}
  \frac{{\dot \gamma}_{\rm rad}}{{\dot \gamma}_{\rm adi}} \propto 
    \frac{(\gamma+1)U_{\rm x} (\delta V)}{\rm d({\delta V})/{\rm d}\tau} \sim t_{\rm dyn}\bigg\vert_{\{ \gamma, \; U_{\rm x}\} }  \ , 
\end{equation} 
 where $t_{\rm dyn}$ is the dynamical time scale of the flow. 
For a relativistic jet flow near the black hole event horizon, $t_{\rm dyn} \sim r_{\rm g}/c \propto M_{\rm BH}$,
where $r_{\rm g}\equiv G M_{\rm BH}/c^{2}$ is the gravitational radius of the black hole. 
Thus, one would expect that 
  the relative importance of the the radiative cooling and adiabatic cooling would 
  vary according to the black hole mass. 
In a synchrotron cooling flow, i.e. $U_{\rm x} = B^2/8\pi$, if the spatial structure of the magnetic field $B$ is specified and fixed, 
  systems with a more massive black hole would be dominated by synchrotron cooling.  
In systems with a less massive black hole, adiabatic cooling would become more important.

\subsection{Spatial variation of $\mathcal{G}(\gamma)$ and its dependence on $\alpha$}
\label{sec:evolution}

The spatial variation of the power-law for ${\rm d}(\ln \mathcal{G})/{\rm d}(\ln \gamma)$ 
at a position $s$ can be approximated by taking $\left(\gamma^{2}-1\right) \approx \gamma^{2}$ and $\left(\gamma-1\right) \approx \gamma$ for the energy loss terms, which is justified when considering only the high energy component of the electron population. 
With these approximations, the synchrotron cooling in equation (\ref{eq:dg_ds}) has the form ${\rm d}\gamma/{\rm d}s\propto \gamma^{2}$ and may be written as
\begin{equation}
\frac{{\rm d}\gamma}{\gamma^{2}} =-\xi(s)\, {\rm d}s\,,
\end{equation}
where
\begin{equation}
\xi(s) \equiv  \frac{\sigma_{\rm T}}{ 6\pi\;\! m_{\rm e}c} 
 \left(\frac{B^{2}}{v}\frac{{\rm d}\tau}{{\rm d}t}\right)\bigg\vert_{ s}  \; .
\end{equation}
The relation 
\begin{equation}
\gamma_{0}=\frac{\gamma}{1-\gamma x} \,,
\end{equation}
is then obtained after some algebra, where
\begin{equation}
x(s) \equiv \int {\rm d}s \, \xi(s) \,> 0\;  .
\end{equation}

Due to the conservation of area on the ``$\mathcal{G}(\gamma)-\gamma$'' plane, the relation 
\begin{eqnarray*}
\mathcal{G}(\gamma)\big\vert_{s}{\rm d} \gamma &=&\mathcal{G}(\gamma)\big\vert_{s_{0}}{\rm d} \gamma_{0} \nonumber \\
&=&\mathcal{G}_{0}\;\!\gamma_{0}^{-\alpha}\, {\rm d}\gamma_{0}  \, ,
 \end{eqnarray*} 
  holds, 
  which upon inserting $\mathrm{d}\gamma_{0}/\mathrm{d}\gamma$ yields
\begin{equation}
\mathcal{G}(\gamma)\big\vert_{s} = \frac{ \mathcal{G}_{0}\;\! \gamma^{-\alpha}}{(1-\gamma x)^{-\alpha+2}}\,. \nonumber
\end{equation}
The slope at a given position $s$ is therefore 
\begin{equation}\label{eq:app_syn}
\frac{{\rm d}(\ln \mathcal{G})}{{\rm d}(\ln\gamma)}\bigg\vert_{s}=- \alpha-(\alpha-2)x \ \! \gamma_{0} \,. \nonumber
\end{equation} 
Since $\alpha>0$, from this relation it may be verified that, at the high energy end, $[{\rm d}(\ln \mathcal{G})/{\rm d}(\ln \gamma)]\vert_{s}$ is always negative if $\alpha\geqslant 2$. 
When $\alpha>2$, the highest energy end may have a positive slope, resulting in a ``raised-up" profile.
An approximation and result similar to that derived here 
has previously been applied to estimate the time-dependent electron distribution 
with a form ${\rm d}\gamma/{\rm d} t \propto \gamma^{2}$ in
\citet[][]{pac70}. 

In a similar manner, one may repeat this for the adiabatic cooling (equation (\ref{eq:cooling_adi})) 
  with a form ${\rm d}\gamma/{\rm d}s\propto \gamma$, yielding
\begin{equation}\label{eq:app_adi}
\frac{{\rm d}(\ln \mathcal{G})}{{\rm d}(\ln\gamma)}\bigg\vert_{s} 
  = - \alpha \nonumber \, .
\end{equation}
Unlike synchrotron cooling, which has the form ${\rm d}\gamma/{\rm d}s\propto \gamma^{2}$, the slope variation due to adiabatic cooling is always negative. Furthermore, the initial slope $-\alpha$ is preserved at different spatial locations.

\begin{figure}
\begin{centering}
\includegraphics[width=1.\columnwidth]{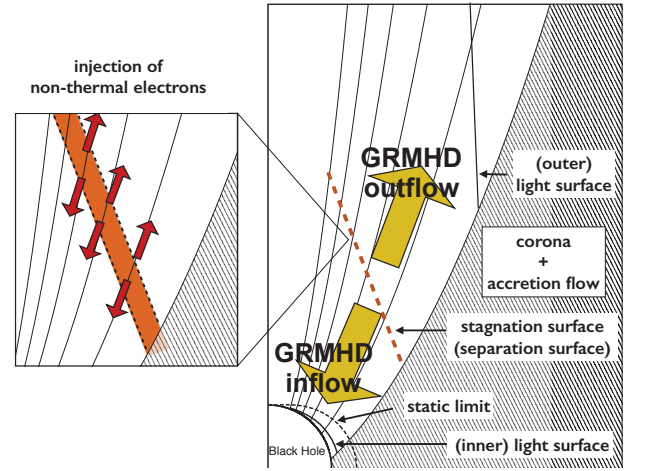}
\caption{Schematic plot of the GRMHD flow and the black hole magnetosphere. 
We assume that the non-thermal electrons (red arrows) are generated at the stagnation surface (red region), and consider their subsequent cooling when they are embedded in the background GRMHD inflow and outflow (yellow arrows).}\label{fig:model}
\end{centering}
\end{figure}

\section{Flow in an Axisymmetric and Stationary GRMHD Jet} \label{sec:model}

To investigate the energy variation of non-thermal electrons within a background GRMHD flow, we proceed to construct a ``qualitatively correct" GRMHD flow configuration as illustrated in Figure~\ref{fig:model} \citep[see also][]{M06, pu15}. 
The flow along large-scale magnetic field lines threading the central black hole is separated into inflow and outflow regions, and is surrounded by the accretion flow and its corona. 
A global magnetic field and MHD outflow solution is related to the mass loading onto the field lines  \citep[][]{bes98,bes06, glo13,pu15} and is computationally expensive to calculate. 
For a description mimicking the outflow properties, we instead employ the following working assumptions. 
We consider a class of stream functions 
\begin{equation}\label{eq:stream}
\Psi=r^{p}(1-\cos\theta)\;,\;\; (0 \leqslant p \leqslant 1)\; ,
\end{equation}  
  as described in \cite{bro09}, 
  which approximately describe the force-free magnetic field \citep{tch08}.

 \subsection{Stagnation Surface}\label{sec:ss}
Along a rigidly-rotating field line, the stagnation surface can be estimated from the location where the magnetocentrifugal force is balanced by the gravity of the black hole \cite[][]{tak90}. 
The stagnation surface can be located by searching for where $K'_{0}=0$ along the field line, where $K_{0}\equiv-\left(g_{\phi\phi}\Omega_{\rm F}^{2}+2g_{t\phi}\Omega_{\rm F}+g_{tt}\right)$, $\Omega_{\rm F}$ is the angular velocity of the field. 
Inside (outside) the stagnation surface, the flow is accelerated inward (outward) and becomes an inflow (outflow). 
For a cold flow within which the pressure may be ignored, the flow begins with zero velocity at the location where $K'_{0}=0$ (the acceleration is cancelled out). 
In a more physically-realistic scenario, the velocity of a hot flow at $K'_{0}=0$ does not begin with zero velocity. 
Here we assume that the inflow and outflow solutions (as described below) begin at $K'_{0}=0$, and that the non-thermal electrons are generated therein in a stationary manner.

 \subsection{Flow Dynamics}\label{sec:flow_dyn}
For {\em outflow} along the field lines, we also adopt the aforementioned force-free model, 
describing the fluid motion and magnetic field strength for the region where the field lines thread the black hole event horizon.
Although this semi-analytic description has the drawback 
of being a force-free jet model which cannot reflect the intrinsic particle acceleration properties of GRMHD processes (related to the mass loading), it greatly simplifies numerical calculations whilst still providing a sensible qualitative model for the flow dynamics. 
We therefore adopt this approach as a starting point in this work, leaving a more in-depth treatment to a future work.  
Noting that the drift velocity is always positive (outward), 
we artificially exclude the GRMHD inflow region inside the stagnation surface so as to match the characteristic features of a GRMHD jet. 

The dynamics of the {\em inflow} region are obtained by solving the equations of motion along a given field line, i.e.~the relativistic Bernoulli equation in the cold limit. 
As shown in \citet[][]{pu15}, a magnetically-dominated GRMHD inflow solution is insensitive to mass loading, because its Alfv\'en surface and fast surface are constrained to be close to the inner light surface and the event horizon, respectively. 
Based on such characteristic features, we have the freedom to choose a representative inflow solution for the four-velocity $u^{\mu}$ of the inflow, for a given magnetization sufficiently high to result in a negative energy inflow, from which the black hole rotational energy is extracted outward.
To avoid numerical stiffness issues caused by small values of $v$ and correspondingly large values of $\gamma$ 
  for synchrotron processes (${\rm d}\gamma/{\rm d}s\,\propto\,\dot{\gamma}_{\rm rad}/v\,\sim\,\gamma^{2}/v$, 
    see equations (\ref{eq:dg_ds}) and (\ref{eq:cooling_syn})), we consider a floor value $v_{\rm flr}$ for the cold flow solution, 
    mimicking the non-vanishing flow velocity of a hot flow (where the sound speed is not zero) near the stagnation surface. 
If the flow velocity $\left|u^{r}/u^{t}\right|$ is below the floor value $v_{\rm flr}$, both synchrotron processes and adiabatic processes are not considered.
The magnetic field structure is determined for both the inflow and outflow regions from the force-free model, and 
its normalization, $B_{0}$, is a free parameter chosen according to the particular physical scenario.

 \subsection{{\rm Fluid Mass Density}} \label{sec:rho}
In both the inflow and outflow region, as required by continuity, the fluid mass density (mass loading)  can be determined from the conserved quantities $\eta (\Psi)$ (particle flux per unit flux tube) along each field line \cite[e.g.][]{tak90}, which gives the number density of thermal electrons as
\begin{equation}
\label{eq:eta}
n_{\rm th}=n_{0}\frac{ \eta (\Psi) \Psi,_\theta }{\sqrt{-g}\ u^{r}}\,>0\;,
\end{equation}
where $g$ is the determinant of the Kerr metric, $u^{r}$ is the radial component of the four-velocity, and $n_{0}$ is a normalisation for the number of thermal electrons, $n_{\rm th}$ \citep[c.f.~][]{bro09}. 
Applying $\rho\propto n_{th}$ for equation (\ref{eq:cooling_adi}),  the value of $\eta(\Psi)$ is not relevant when considering $\rho/\rho_{0}$.

\section{Calculations and Results} \label{sec:result}
\begin{figure*} 
\begin{centering}
\includegraphics[width=0.7\columnwidth, angle=-90]{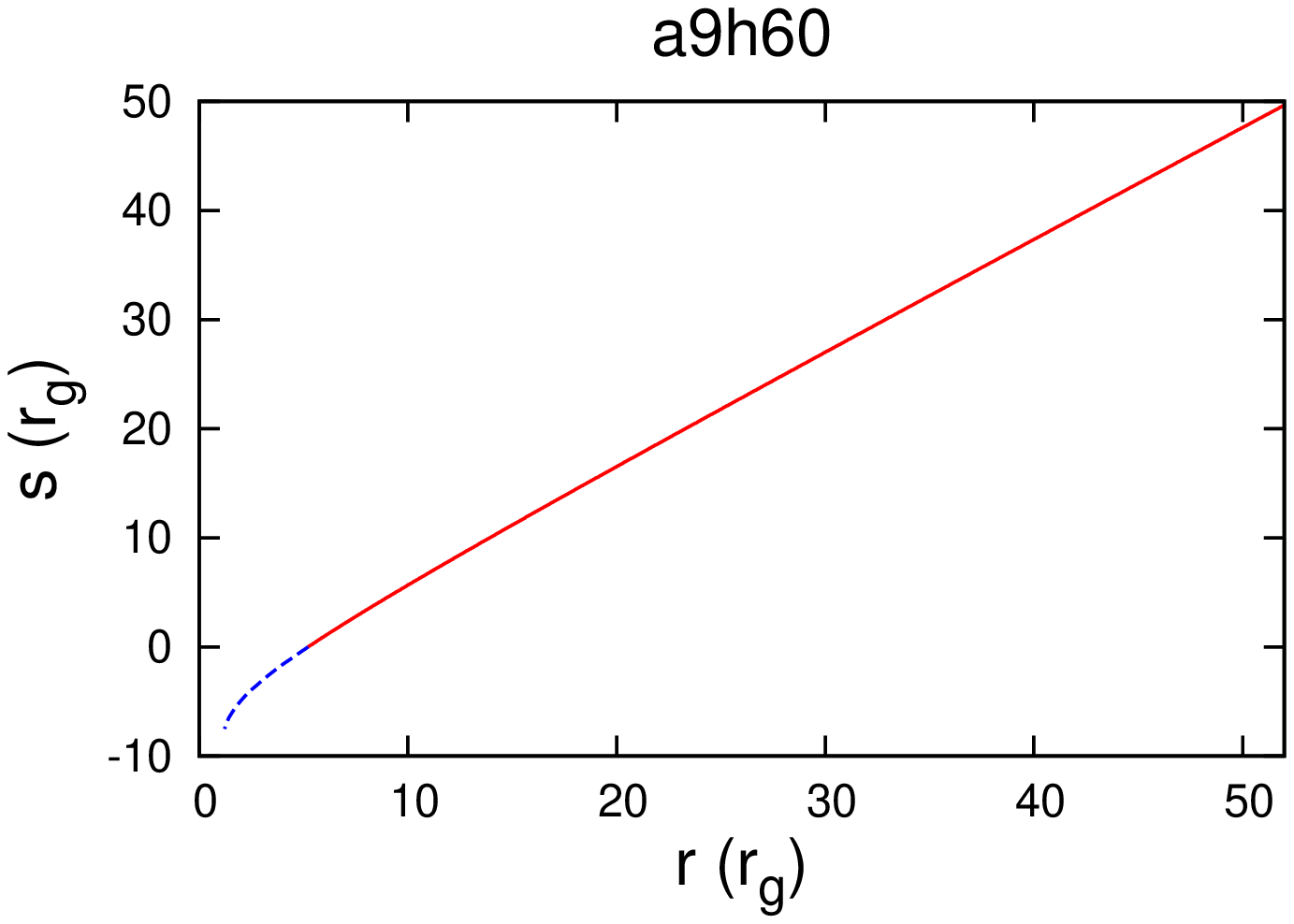}
\includegraphics[width=0.7\columnwidth, angle=-90]{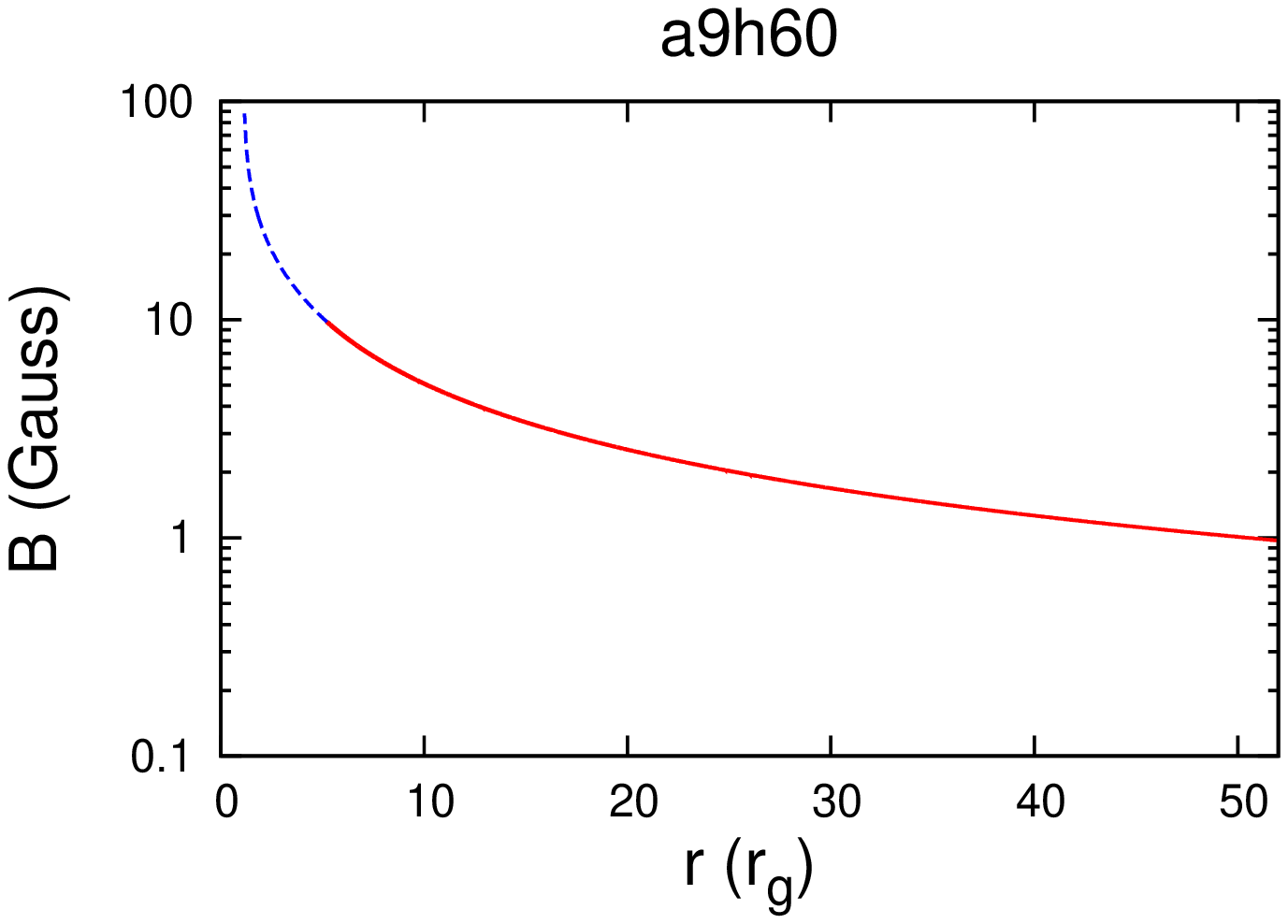}\\
\includegraphics[width=0.7\columnwidth, angle=-90]{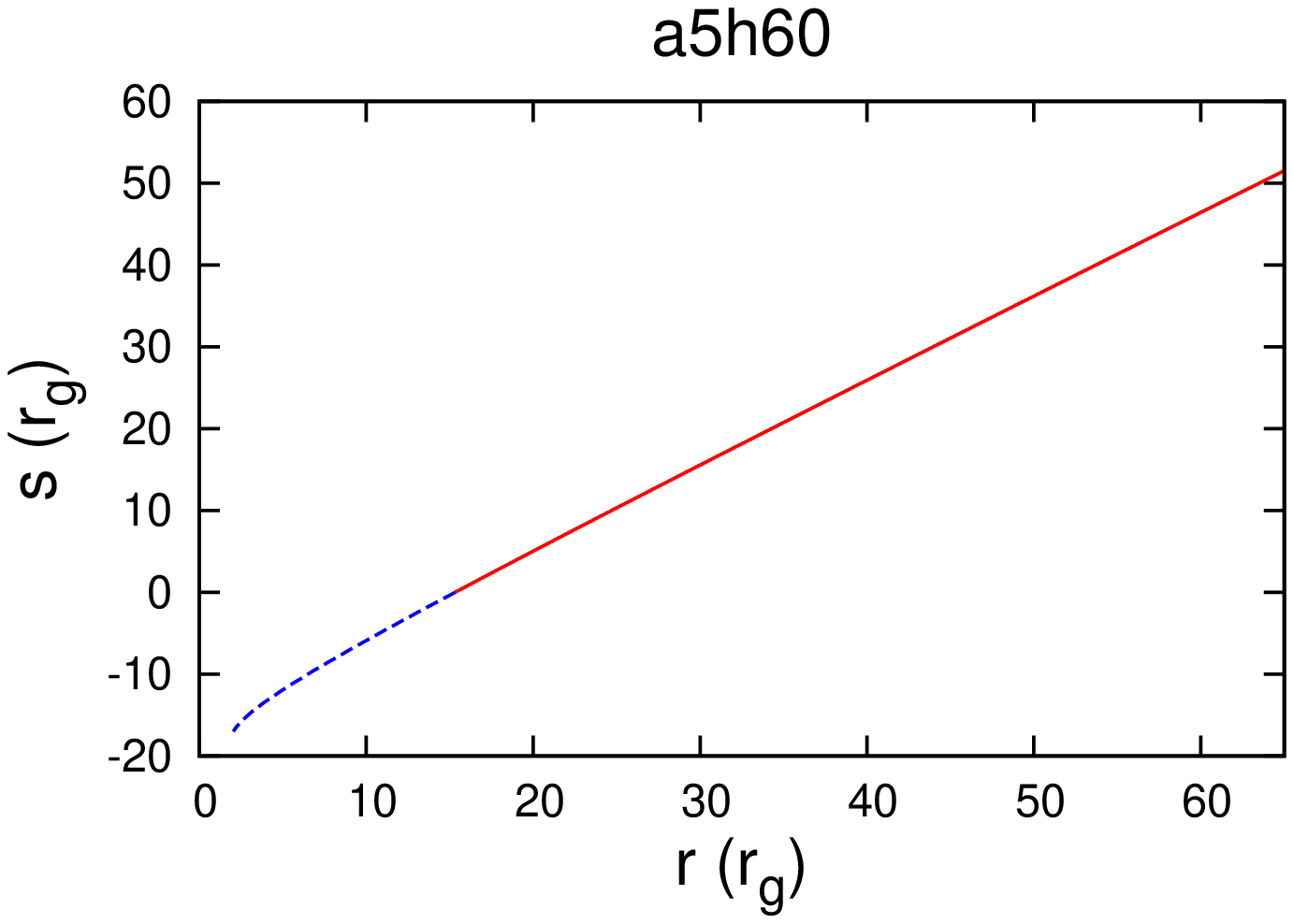}
\includegraphics[width=0.7\columnwidth, angle=-90]{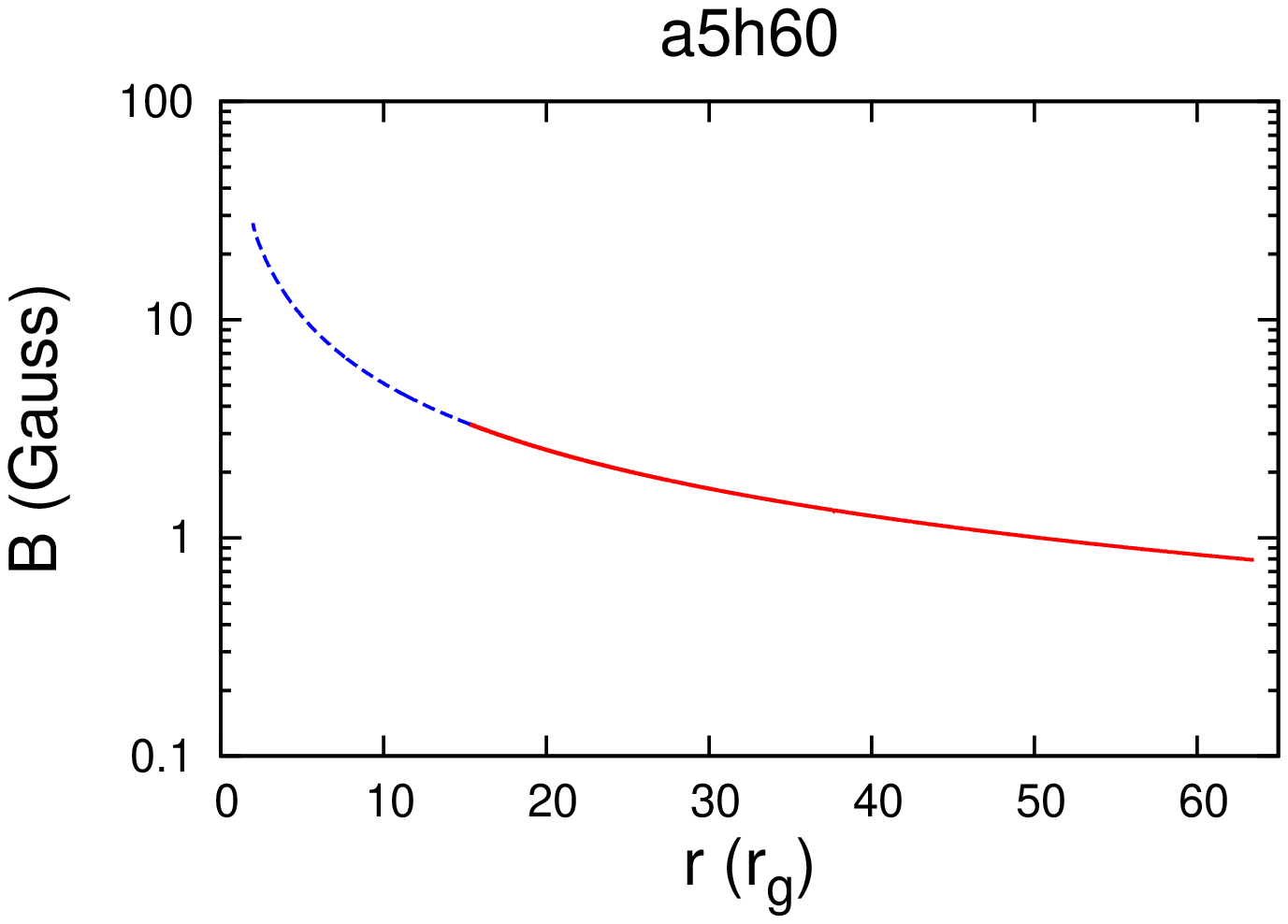}\\
\includegraphics[width=0.7\columnwidth, angle=-90]{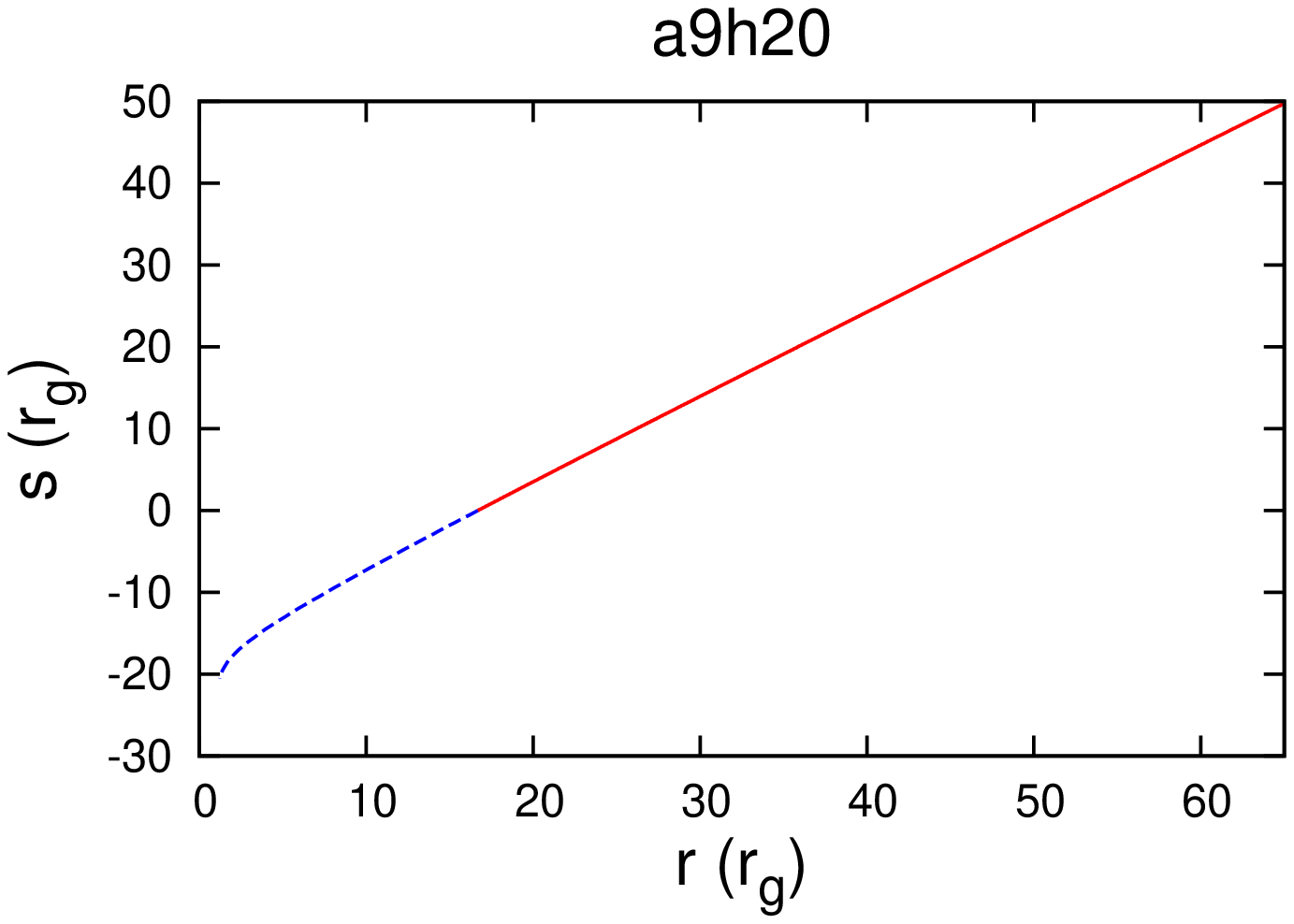}
\includegraphics[width=0.7\columnwidth, angle=-90]{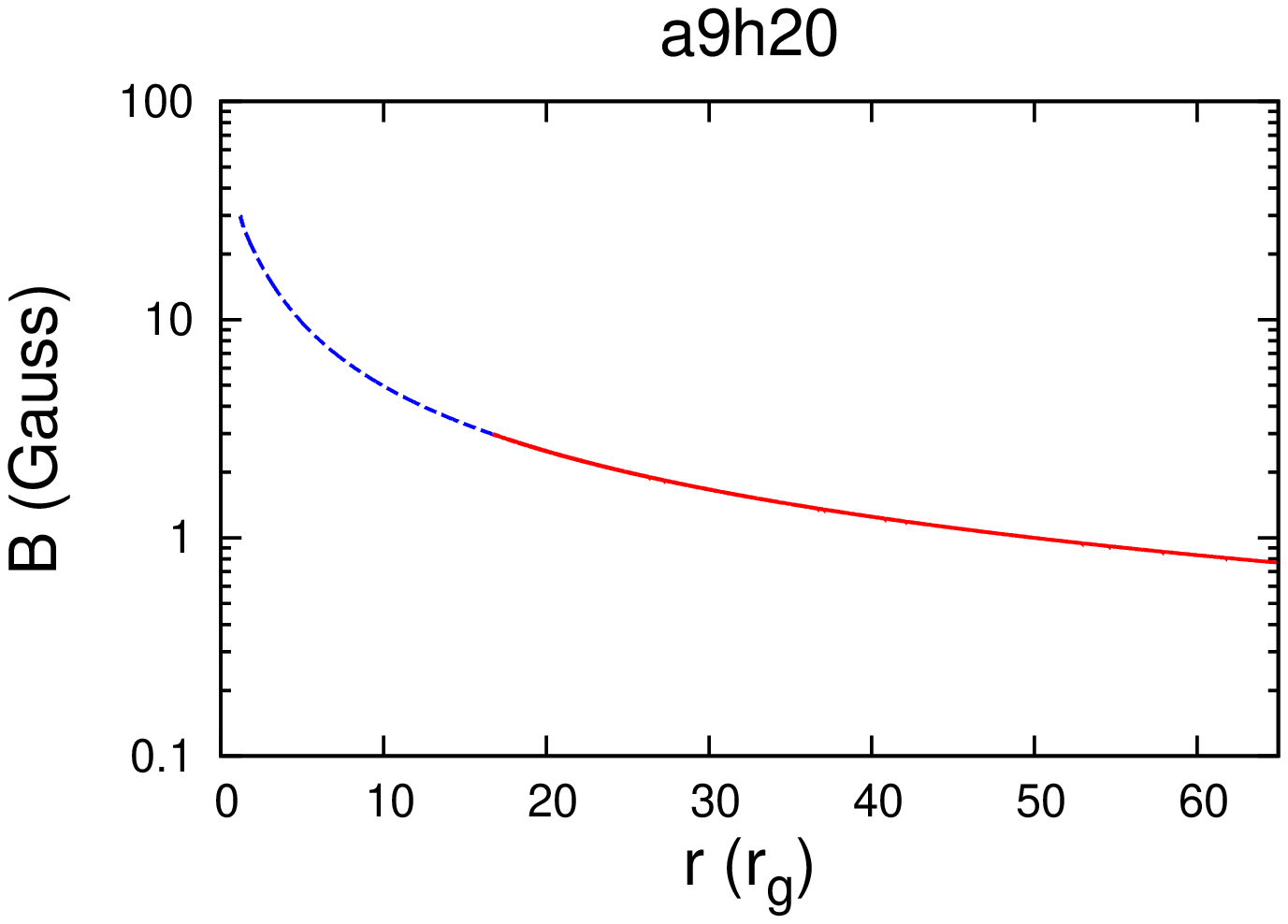}
\caption{
The poloidal distance $s$ and the magnetic field strength $B$ (in Gauss) along the field line for each model, in terms of the Boyer-Lindquist radial coordinate, $r$, in units of $r_{\rm g}$. 
Inflow and outflow regions are indicated by the blue dashed and red solid segments, respectively. The stagnation surface is located at $s=0$.}
\label{fig:srb}
\end{centering}
\end{figure*}

\begin{figure} 
\begin{centering}
\includegraphics[width=0.7\columnwidth, angle=-90]{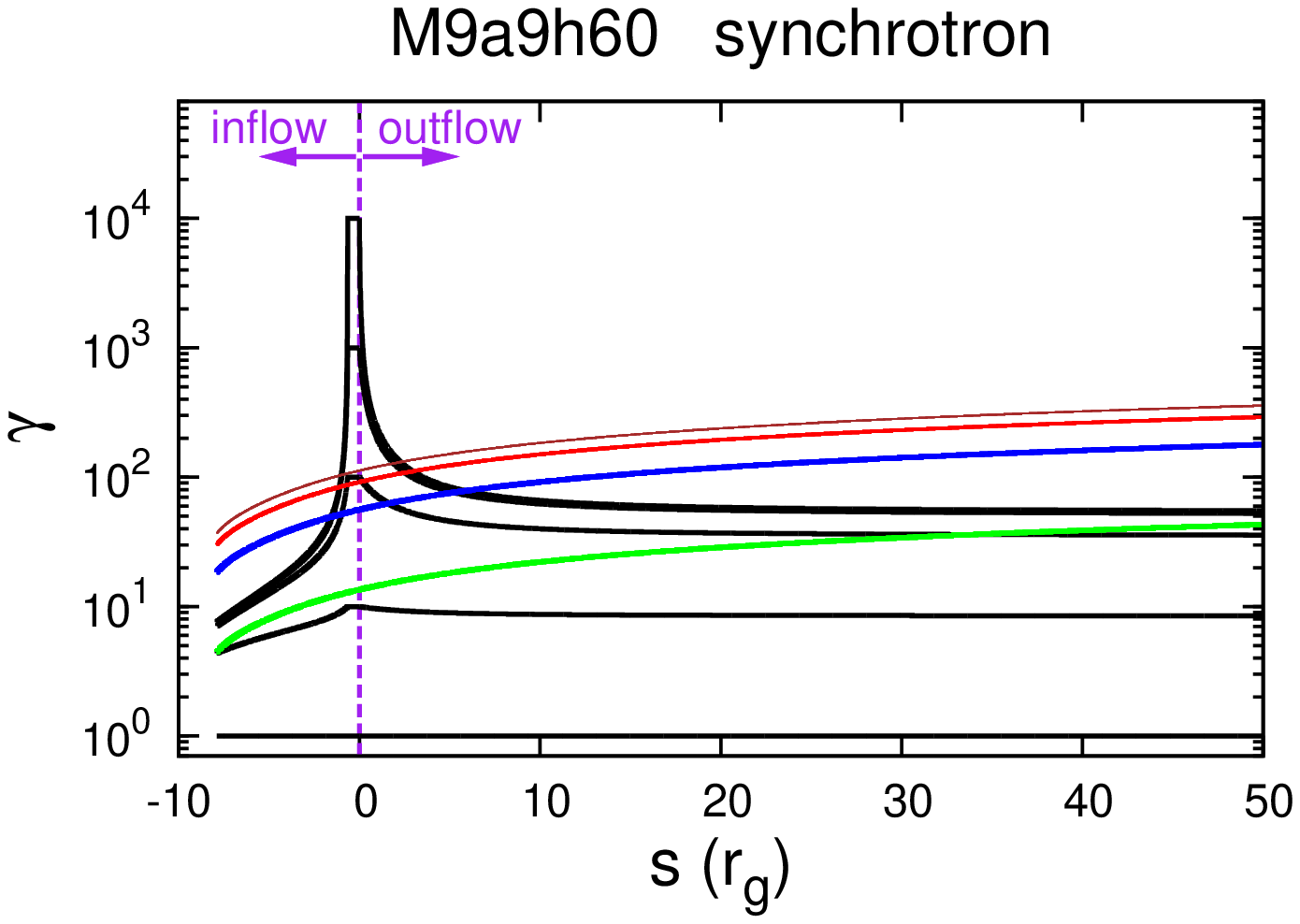}\\
\includegraphics[width=0.7\columnwidth, angle=-90]{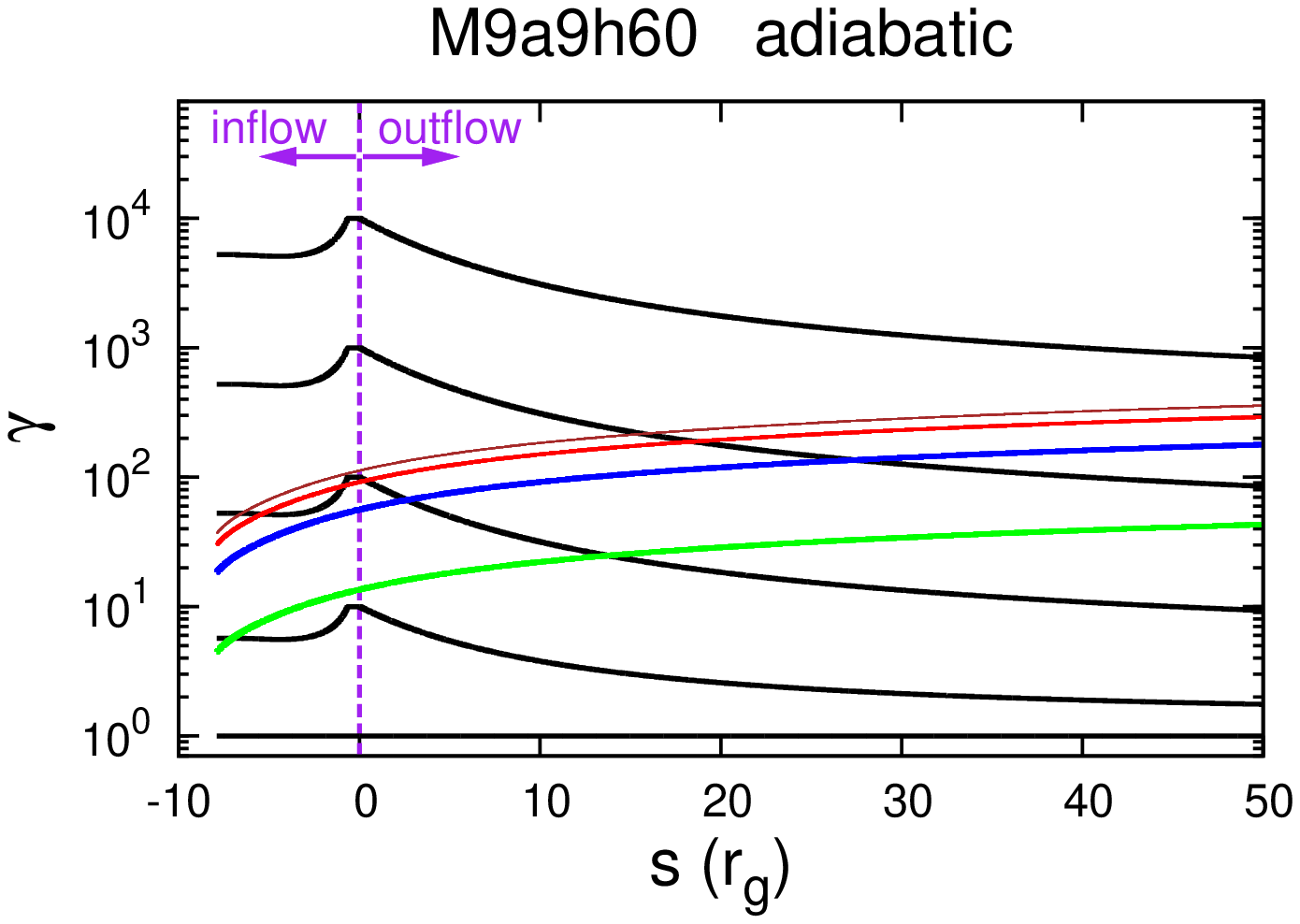}\\
\includegraphics[width=0.7\columnwidth, angle=-90]{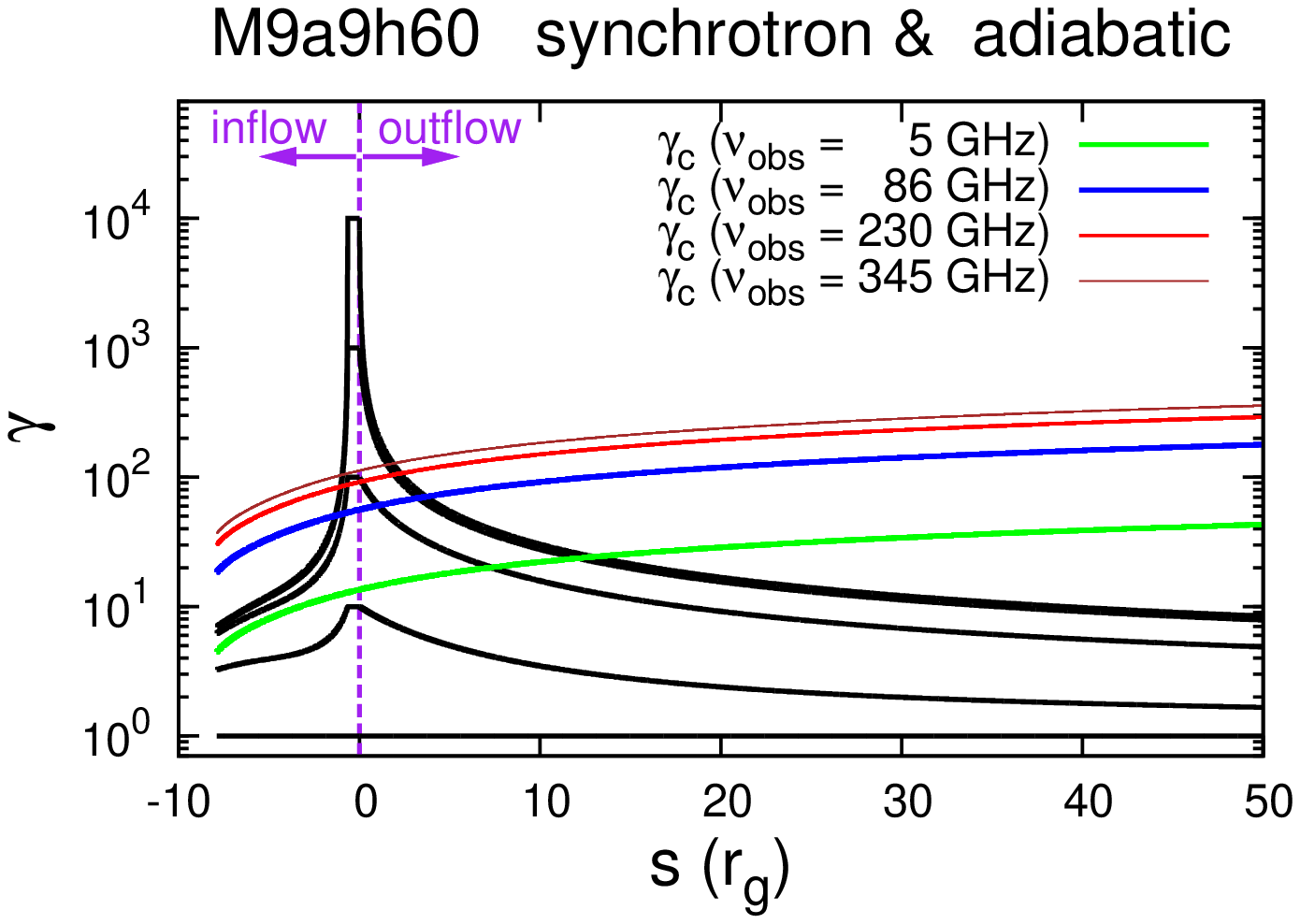}
\caption{Spatial variations of the non-thermal electron energy along a mid-latitude field line for model M9a9h60. 
Selected characteristic curves (black) from equation (\ref{eq:dg_ds}) represent injected electrons with initial energies of $\gamma_{0}=1,~10,~10^{2},~10^{3}$ and $10^{4}$ at the stagnation surface ($s=0$, as indicated by the vertical dashed line), and their spatial variation as they stream along the background GRMHD inflow region ($s<0$) and outflow region ($s>0$). 
Shown separately are the characteristic curves due to synchrotron processes only (top), adiabatic processes only (middle), and both processes (bottom). 
The flattened profile left of the stagnation surface (\S\,\ref{sec:flow_dyn}) is due to the use of $v_{\rm flr}$ (see text) and does not qualitatively change these results.
The characteristic electron energies $\gamma_{c}$ for producing emission at observational frequencies of $\nu_{\rm obs}= 5$~GHz (green), $86$~GHz (blue), $230$~GHz (red), and $345$~GHz (brown) are also overlaid (see \S\,\ref{sec:imp}).}
\label{fig:m9a9h60_plot}
\end{centering}
\end{figure}

\begin{figure} 
\begin{centering}
\includegraphics[width=0.7\columnwidth, angle=-90]{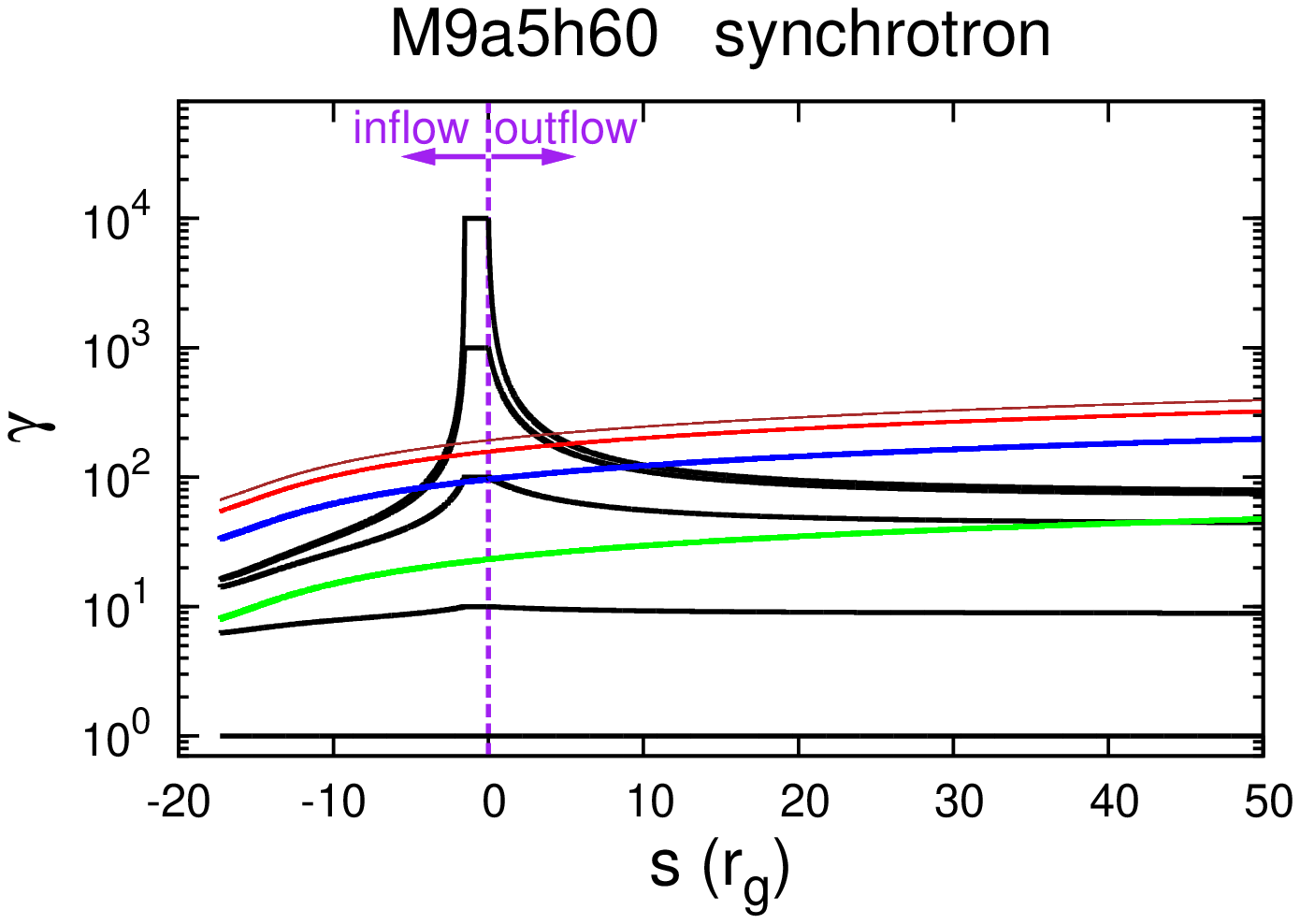}\\
\includegraphics[width=0.7\columnwidth, angle=-90]{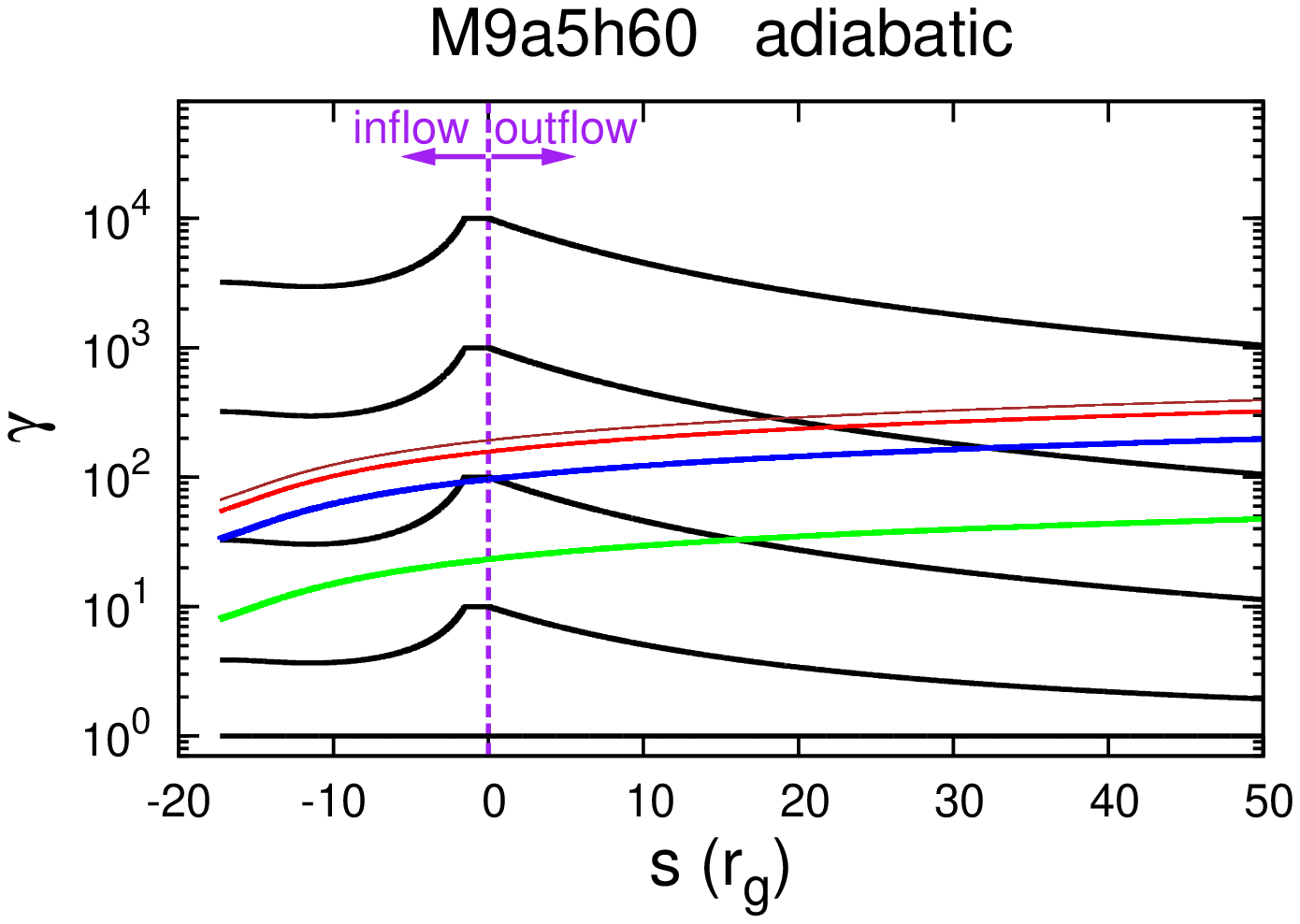}\\
\includegraphics[width=0.7\columnwidth, angle=-90]{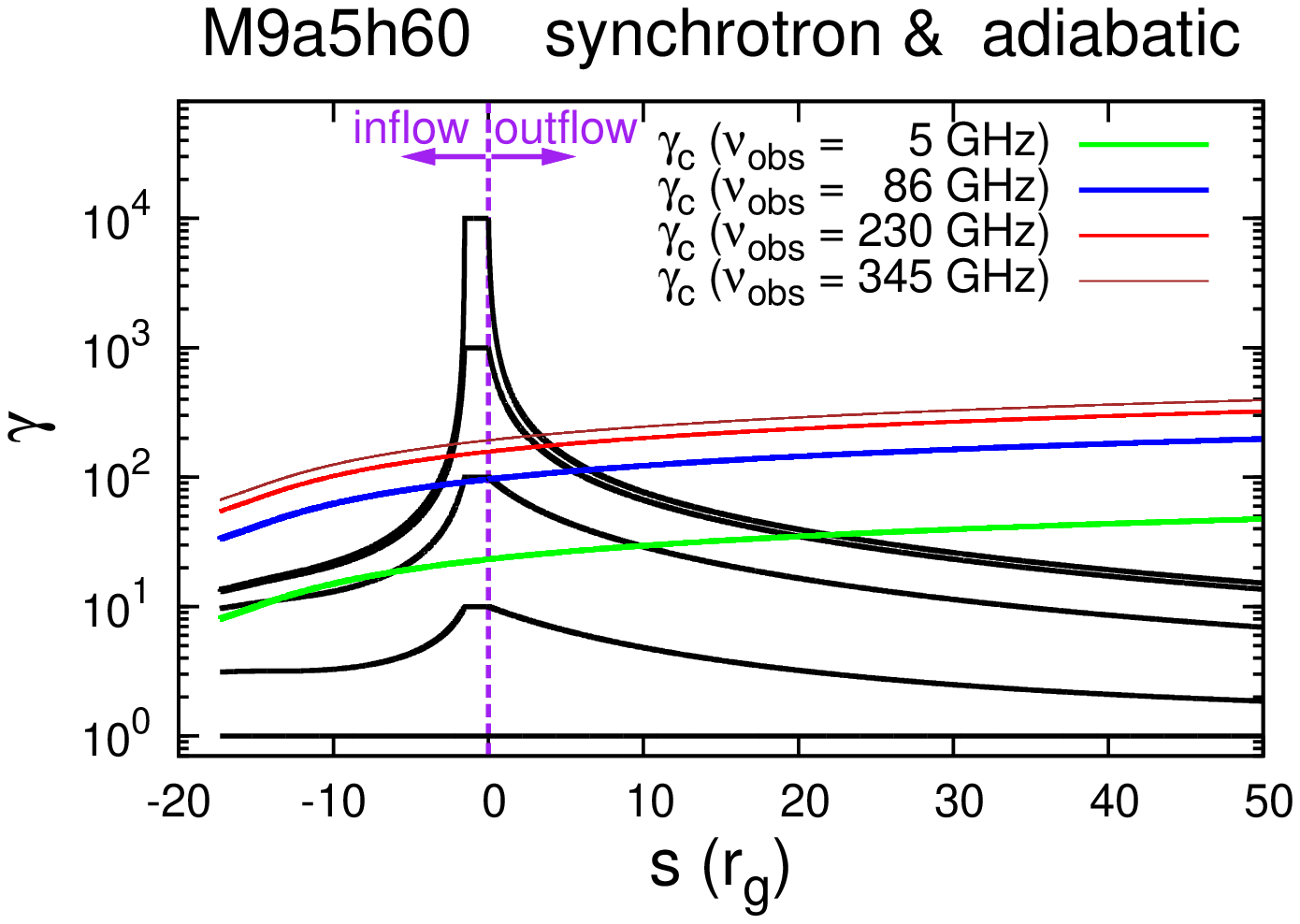}
\caption{Spatial variations of the non-thermal electron energy along a mid-latitude field line for model M9a5h60. 
See Figure~\ref{fig:m9a9h60_plot} for description and comparison.}
\label{fig:m9a5h60_plot}
\end{centering}
\end{figure}

\begin{figure} 
\begin{centering}
\includegraphics[width=0.7\columnwidth, angle=-90]{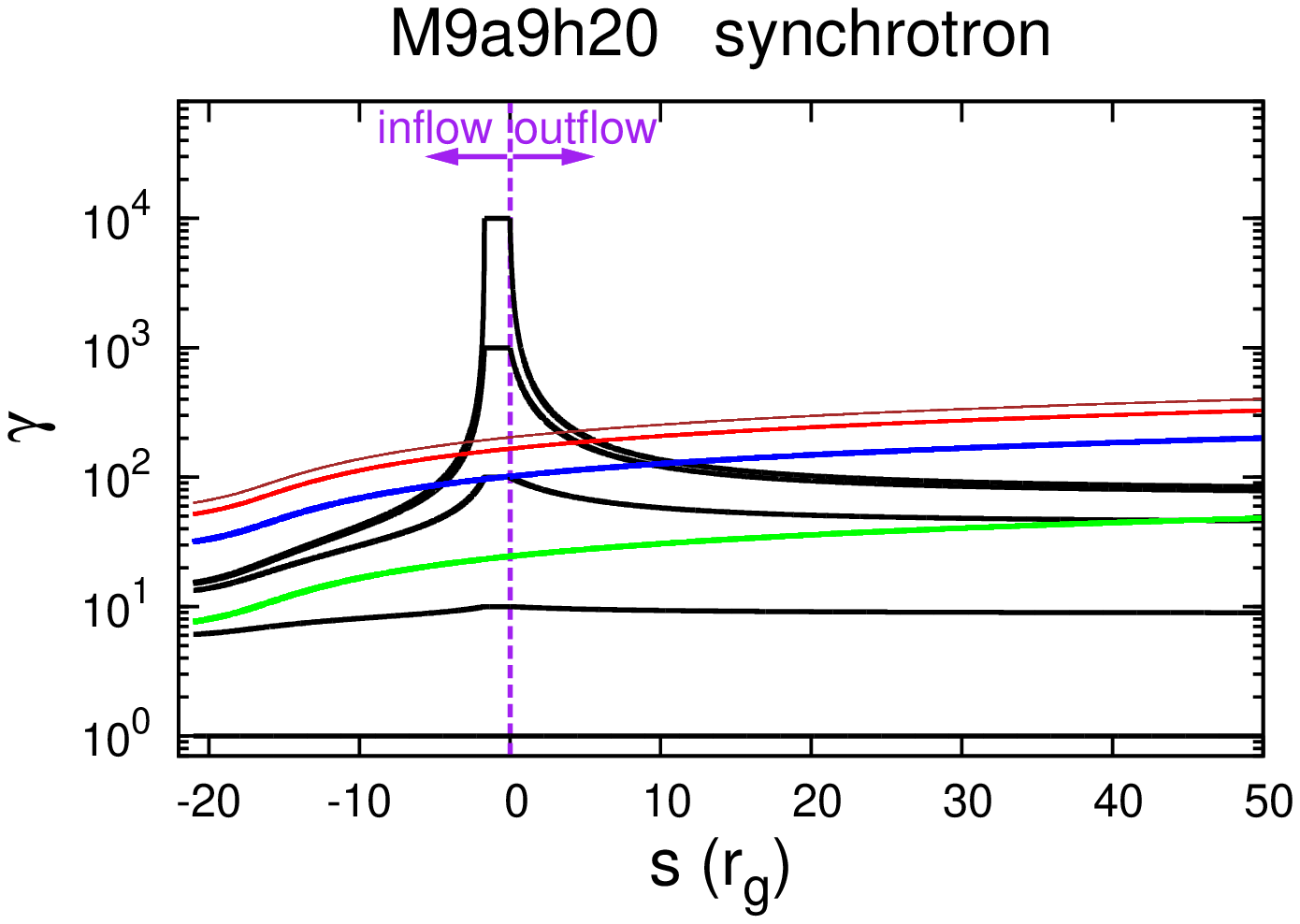}\\
\includegraphics[width=0.7\columnwidth, angle=-90]{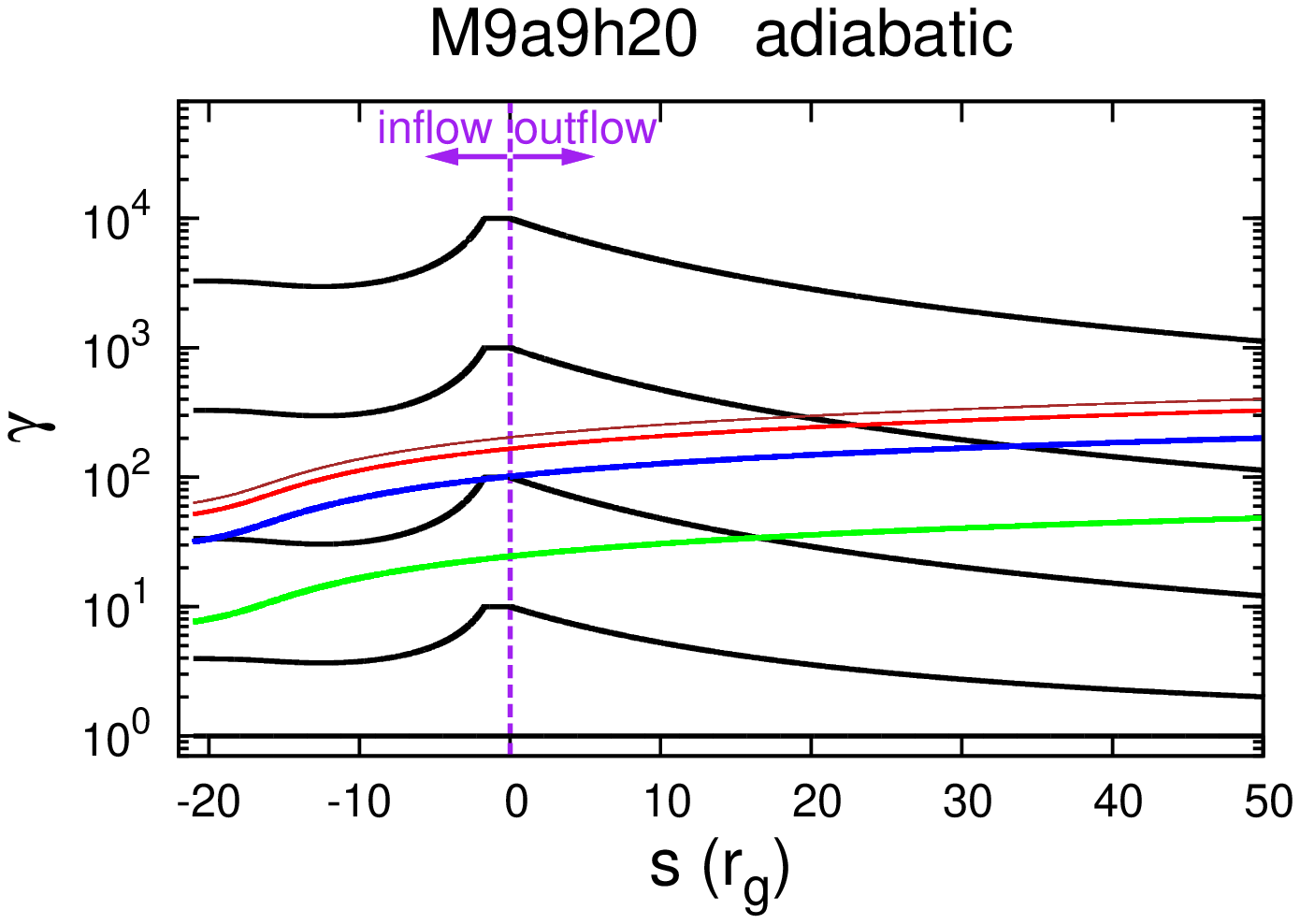}\\
\includegraphics[width=0.7\columnwidth, angle=-90]{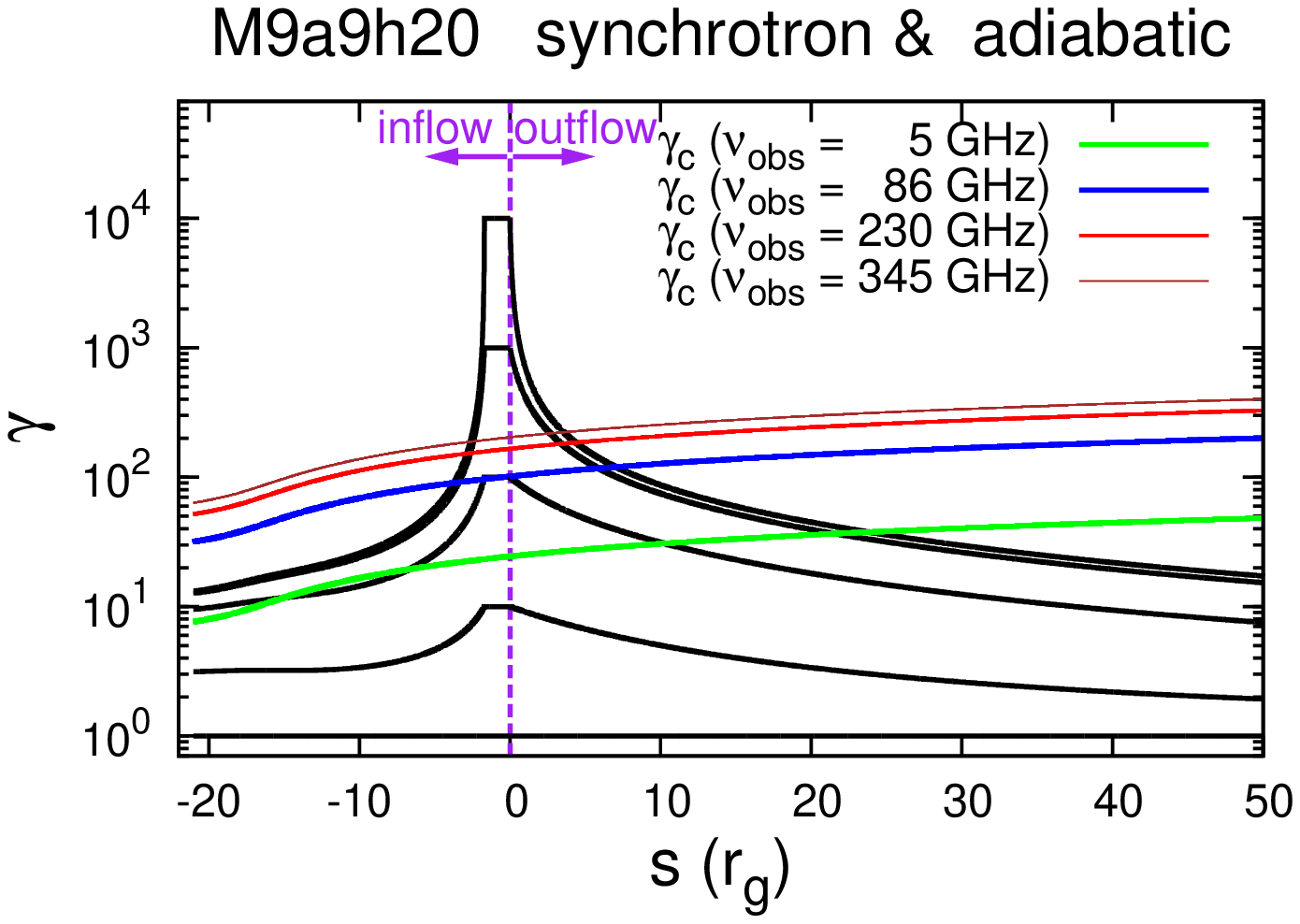}
\caption{Spatial variations of the non-thermal electron energy along a high-latitude field line for model M9a9h20. 
See Figure~\ref{fig:m9a9h60_plot} for description and comparison.}
\label{fig:m9a9h20_plot}
\end{centering}
\end{figure}

\begin{figure} 
\begin{centering}
\includegraphics[width=0.7\columnwidth, angle=-90]{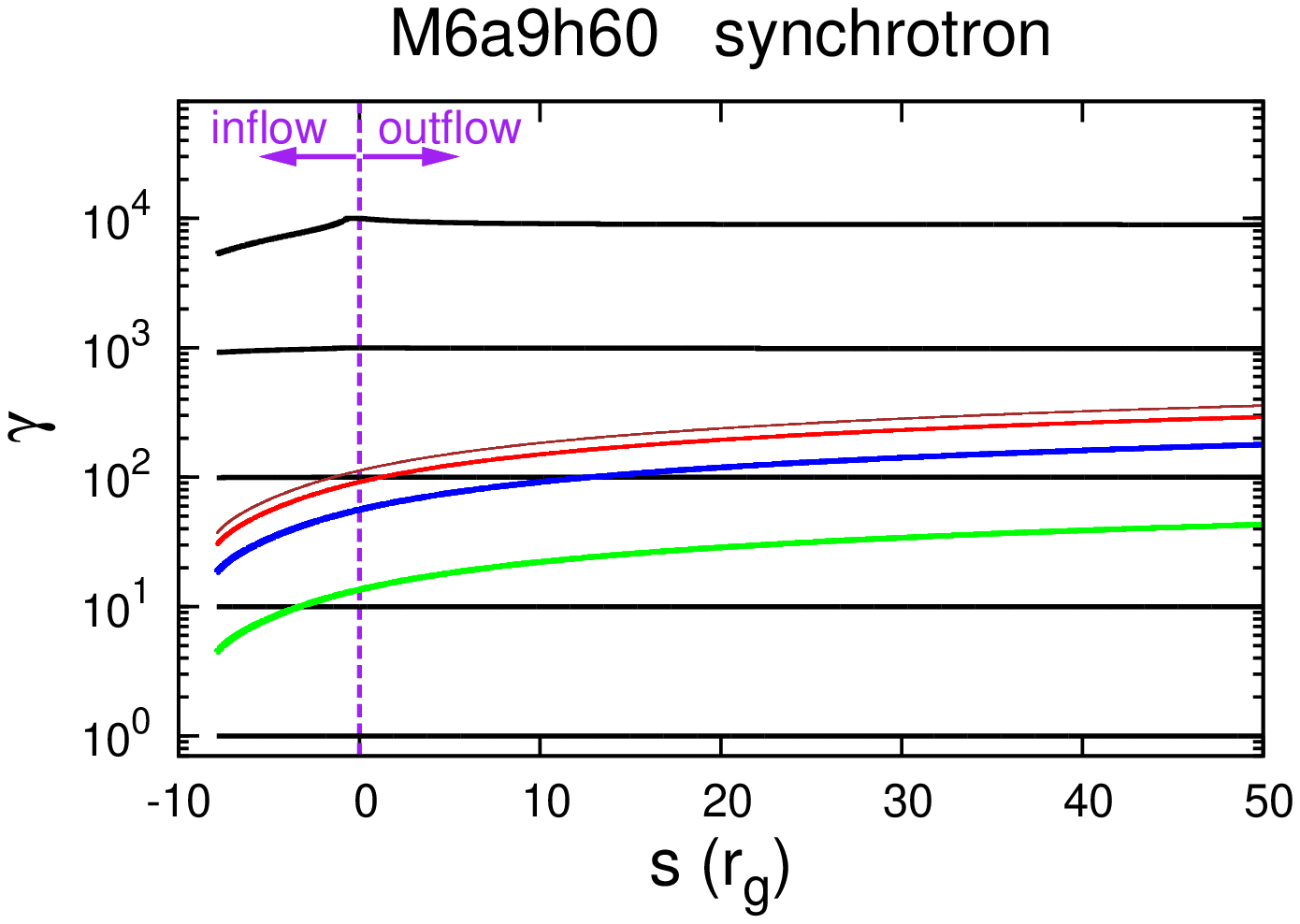}\\
\includegraphics[width=0.7\columnwidth, angle=-90]{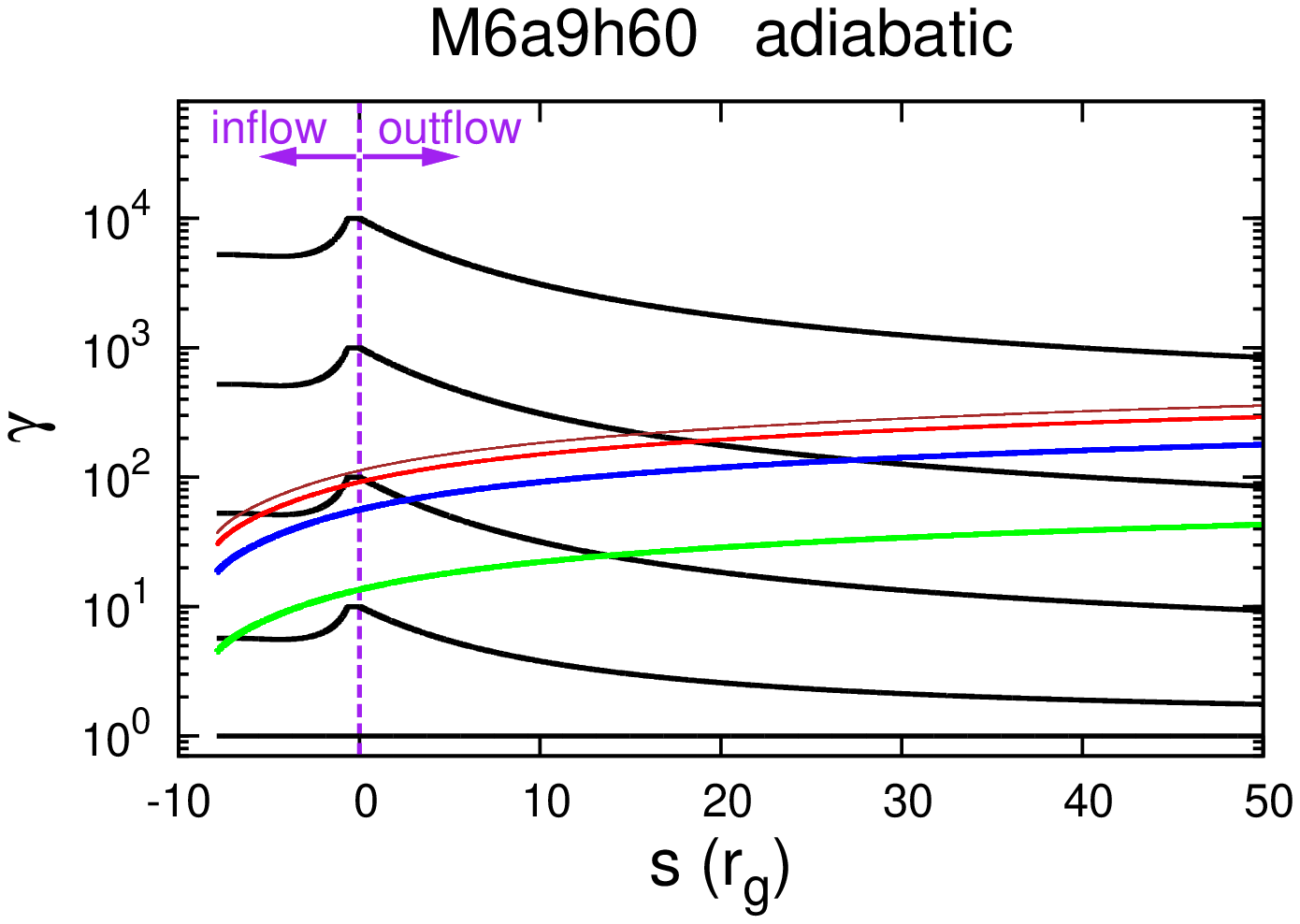}\\
\includegraphics[width=0.7\columnwidth, angle=-90]{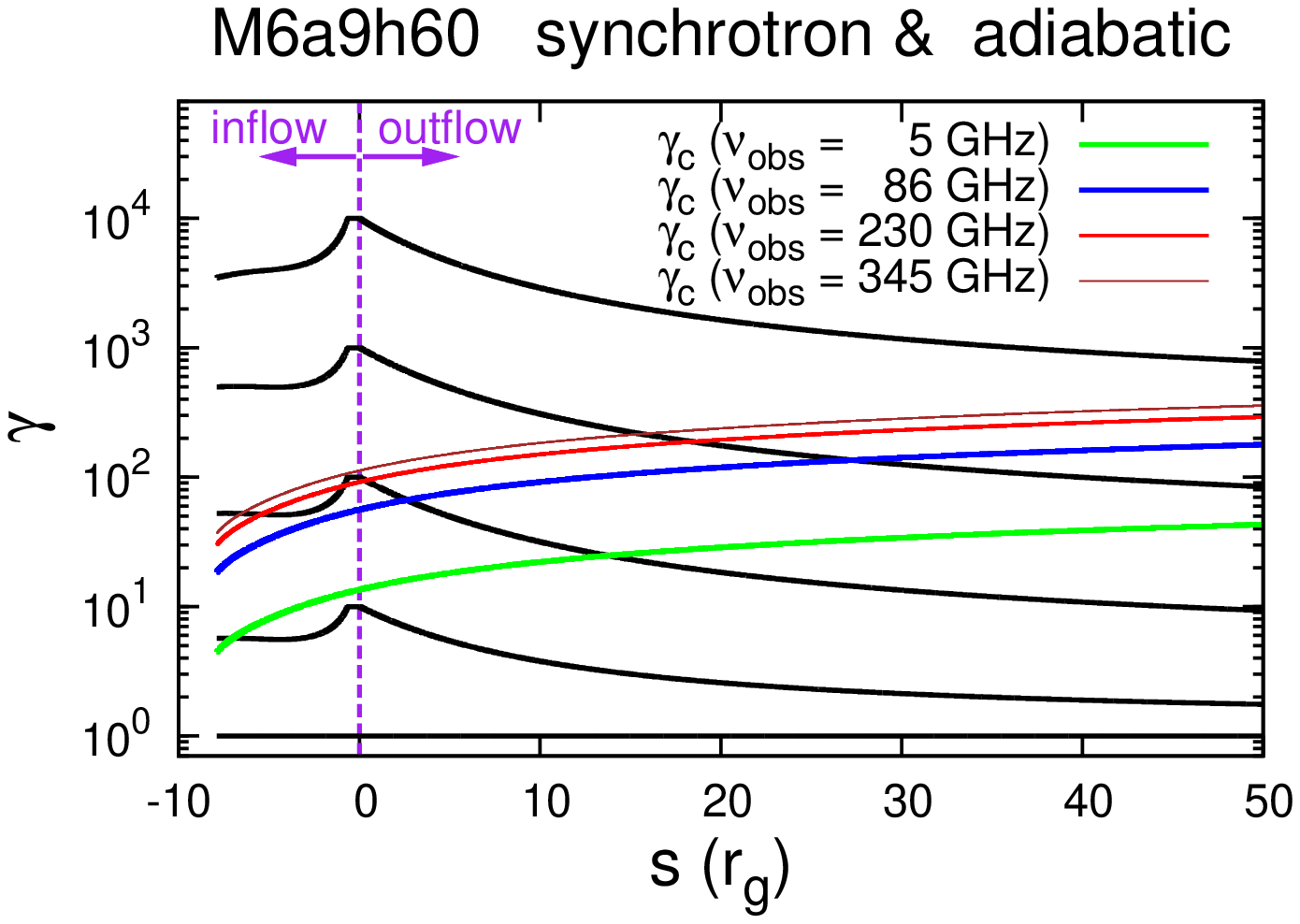}
\caption{Spatial variations of the non-thermal electron energy along a mid-latitude field line for model M6a9h60. 
See Figure~\ref{fig:m9a9h60_plot} for description and comparison.}\label{fig:m6a9h60_plot}
\end{centering}
\end{figure}

\subsection{Model Parameters}\label{sec:par}
We consider jet models with $p=1$ and $\Omega_{\rm F}(\Psi)=0.5 \ \! \Omega_{\rm BH}$, where $\Omega_{\rm BH}$ is the angular velocity of the black hole. The normalisation for the magnetic field strength is chosen to give a field strength $\sim 10$ G (with a corresponding $B_{0}=50$ G), in order to fit the inferred value for M87 \citep{dex12}. 
The model parameters are summarized in Table \ref{tab:model} for different black hole masses and spin parameters, and the calculation is performed along a representative field line $\Psi$ that attaches onto the event horizon at $\theta_{\rm h}$. We present in Figure~\ref{fig:srb} the poloidal distance $s$ and magnetic field strength for different models. The GRMHD flow solutions used for each model are presented in Appendix~\ref{sec:app}. Other choices of $p$ and  $\Omega_{\rm F}$ give a qualitatively similar flow structure.

In the fiducial model, M9a9h60, the assumed parameter values are $M_{\rm BH}=6\times10^{9}\;\!M_{\odot}$ \citep[e.g.,][]{geb11} and a rapid black hole spin of $a=0.998$. 
The solution is computed along a mid-level field line which attaches onto the event horizon at $\theta_{\rm h}=60^{\circ}$.
In contrast to the fiducial settings, a modest black hole spin of $a=0.5$ is chosen in model M9a5h60, and 
a different value of $\theta_{\rm h}=20^{\circ}$ (but with $a=0.998$) is chosen in model M9a9h20.

In addition, although the jet model is independent of the black hole mass, the radiative cooling depends on the physical length scale and is therefore a function of the black hole mass (\S\,\ref{sec:energy_dot}). 
To provide insight into the mass dependence, we consider in model M6a9h60 a smaller black hole mass which is similar to the case of Sgr A$^{*}$, i.e.~$M_{\rm BH}=4\times10^{6}\;\!M_{\odot}$, while keeping the magnetic field the same as in all other cases\footnote{ Note that in general the magnetic field strength depends on the accretion rate and the black hole mass.}.
All calculations include both synchrotron and adiabatic process (i.e.~$\dot \gamma =  \dot \gamma_{\rm syn}  + \dot \gamma_{\rm adi}$)  
since we are concerned with mm/sub-mm VLBI observational frequencies\footnote{
The electron-synchrotron absorption cross-section is  \[
\sigma_{\rm syn}(\nu,B,\sin\theta;\gamma) = 
    \frac{8\pi^2 (3\sin \theta)^{2/3}\; \Gamma(5/3)}{3\sqrt{3}} 
   \left(\frac{e}{B}\right)  \left(\frac{\nu_{\rm L}}{\gamma \nu}\right)^{5/3} \;   \nonumber 
\]
\cite[see e.g.][]{ghi13}, where $\gamma$ is the Lorentz factor of the non-thermal electrons, 
  which gives the optical depth  
\[
  \tau_{\rm syn}\approx 1.09 \times 10^8 \left(\frac{n_{\rm e}}{10^5\;\!{\rm cm}^{-3}} \right) 
  \left(\frac{R}{10^{12}{\;\!\rm cm}} \right) 
     \left(\frac{10\;\!{\rm G}}{B}\right)  \left(\frac{\nu_{\rm L}}{\gamma\;\! \nu}\right)^{5/3} \ ,  
\]
 where $n_{\rm e}$ is the non-thermal electron number density and $R$ is the size of the emission region.
For a magnetic field $B = 10\;\!{\rm G}$, the Larmor frequency  $\nu_{\rm L}$ $(= 28\;\!{\rm MHz}$) 
  is much lower than the observational frequency $\nu$ ($\sim 100\;\!{\rm GHz}$ or higher). 
With $R = 10^{12}\;\!{\rm cm}$, $B=10\;\!{\rm G}$  and $\nu=100\;\!{\rm GHz}$,  
\[
\tau_{\rm syn}\approx 0.06 \;\!  \left(\frac{n_{\rm e}}{10^5{\rm cm}^{-3}} \right)  
 \left(\frac{10^{2}}{\gamma}\right)^{5/3} 
\]
For magnetic aligned flows, $B \sim 10\;\!{\rm G}$ implies $n_{\rm e} \ll 
  \rho/m_{\rm p} \ll10^5\;\! {\rm cm}^{-3}$, where $m_{\rm p}$ is proton mass. 
Hence, synchrotron self-absorption is unimportant in the jets considered in this work.}.
\begin{table}
\caption{Model Parameters (see also \S\,\ref{sec:par}). }
\label{tab:model}
\begin{centering}
\begin{tabular}{lccc} 
\hline
\hline
            & $M_{\rm BH}\;(M_{\odot})$ & $a$ & $\theta_{\rm h}$ \\
\hline
M9a9h60 (fiducial)    &  $6\times10^{9}$ & 0.998 & $60^{\circ}$ \\
M9a5h60		     &  $6\times10^{9}$ & 0.5 & $60^{\circ}$ \\
M9a9h20		     &  $6\times10^{9}$ & 0.998 & $20^{\circ}$ \\
M6a9h60     		   & $4\times10^{6}$ & 0.998 & $60^{\circ}$ \\
\hline
\end{tabular} \\
\end{centering}
\vspace*{2mm}

\end{table}

\subsection{Particle properties of GRMHD jets}\label{sec:particle_results}

\subsubsection{Energy Loss of Non-thermal Electrons}

The energy loss of non-thermal electrons can be described by the characteristic curves ${\rm d}\gamma/{\rm d}s$ in equation (\ref{eq:dg_ds}), which are independent of the choice of $\alpha$ (\S\,\ref{sec:trans_F}).
In Figure~\ref{fig:m9a9h60_plot} we present the spatial variation of non-thermal electrons of M9a9h60, provided that 
non-thermal electrons are generated at the stagnation surface (vertical dashed line).
A family of selected characteristic curves of electron energies $\gamma_{0}=1, 10, 10^{2},10^{3}$ and $10^{4}$ at the stagnation surface are shown (where the subscript ``0" denotes the value at the stagnation surface).
A floor value $v_{\rm flr}=0.03$ is chosen (\S\,\ref{sec:flow_dyn}) so that no numerical stiffness occurs even for the largest values of $\gamma_{0}$ considered in this study. 
This choice of $v_{\rm flr}$ does not change the results qualitatively. Imposing a floor results in a flattened profile near the stagnation surface. 

For both inflow and outflow regions, the electron energy drops as the electrons stream along the flow and away from the stagnation surface.
The synchrotron process, which is $\propto (\gamma^{2}-1)$, results in a more rapid energy decrease for higher electron energies $\gamma$ (top panel). 
As the magnetic field strength decreases along the stream line, the effect of synchrotron cooling become less efficient, and the profile of the characteristic curve flattens at larger distances in the outflow region.
Adiabatic processes on the other hand, which are $\propto (\gamma-1)$, are insensitive to the electron energy (middle panel).
The adiabatic process could result in a heating or cooling process, according to the profile of $\rho$. 
   It is interesting to note that $\rho$ is roughly inversely proportional to $u^{r}$ (Section \S\,\ref{sec:rho}), and $u^{r}$ has in general an increasing profile away from the stagnation surface (see Figure~\S~\ref{fig:app_model} and Figure~7 of \citet{pu12} for examples). As a result, the profile of $\rho$ is always decreasing away from the stagnation surface, and hence adiabatic cooling takes place in both inflow and outflow regions (see equation (\ref{eq:cooling_adi}) and Figure~\S~\ref{fig:app_model}).  
For the combination of these two processes (bottom panel), profiles of $\gamma$ in the outflow region decay more rapidly compared to when only a single process is considered. 
For all cases, electrons with $\gamma=1$ do not gain or lose energy, as expected. In the following we compare the results of other models with those of the fiducial model (M9a9h60).

Figure~\ref{fig:m9a5h60_plot} shows the result of model M9a5h60.
Due to the slower black hole spin ($a=0.5$), the stagnation surface is located further away from the black hole (see also Figure~\ref{fig:srb}).
The location of the stagnation surface further from the black hole affects the relative importance of  synchrotron and adiabatic cooling for electrons with different initial energies.
For synchrotron cooling, the electron energy decays less rapidly from the stagnation surface because the magnetic field is weaker, given that the normalizations for the magnetic fields are the same in both models. 
Due to the energy dependence of synchrotron cooling ($\propto \gamma^{2}-1$) and adiabatic cooling ($\propto \gamma-1$), electrons with higher (or lower) energy more easily lose energy via synchrotron cooling (or adiabatic cooling), for both inflow and outflow regions.

For identical black hole spin parameters, the location of the stagnation surface is further away from the black hole for field lines at higher latitudes (see also Figure~\ref{fig:srb}). 
Figure~\ref{fig:m9a9h20_plot} shows the result for model M9a9h20 ($\theta_{\rm h} = 20^{\circ}$), the case where a field line close to the polar region is considered. 
Similarly, the further-removed location of the stagnation surface causes electrons to lose energy less rapidly, and the electrons with lower energy lose their energy via adiabatic cooling more easily due to the weaker synchrotron cooling.

The spatial variation of the electron energy for M6a9h60 ($M_{\rm BH} = 4\times 10^{6}~M_{\odot}$), the smaller black hole mass case, 
  is shown in Figure~\ref{fig:m6a9h60_plot}. 
The dynamical time scale of the flow in this case is smaller than the previous cases where the black hole is 1000 times more massive. 
The top panel clearly shows that synchrotron loss is insignificant in this case. 
This can be understood by that the synchrotron loss time scale is much longer than the dynamical time scale of the flow.     
Comparing the middle and the bottom panels 
  confirms that the cooling is dominated by adiabatic  processes, 
  of which the adiabatic cooling time scale is governed by the dynamical time scale of the background GRMHD flow 
considered here.

\subsubsection{Spatial variation of the distribution of non-thermal electrons}

The non-thermal electron distribution $n_{\rm nth}(\gamma, s)$ is related to the thermal electron distribution $n_{\rm th}(s)$ determined by the GRMHD flow and the distribution function $\mathcal{G}(\gamma)|_{s}$ 
(equation (\ref{eq:g})).
According to the profile of $\mathcal{F}(\gamma, s)$, the distribution function $\mathcal{G}(\gamma)|_{s}$ can be computed at a given slice of constant $s$ (equation (\ref{eq:g})).  
In the following, we present the spatial evolution of $\mathcal{G}(\gamma)|_{s}$ for different injected electron energy distributions (equation (\ref{eq:inj})) as $\gamma_{0}^{-\alpha}$, with $\alpha=1.2$, $2$, and $2.2$.
In each case, we assume $1\leqslant\gamma_{0}\leqslant1000$, and focus on the result when both synchrotron cooling and adiabatic cooling are included.
Interestingly, the spatial evolution has different characteristics depending on whether synchrotron or adiabatic processes dominate.
Here we select M9a9h60 and M6a9h60 as representative cases for these two situations.

\begin{figure*} 
\begin{centering}
\includegraphics[width=0.9\columnwidth, angle=0]{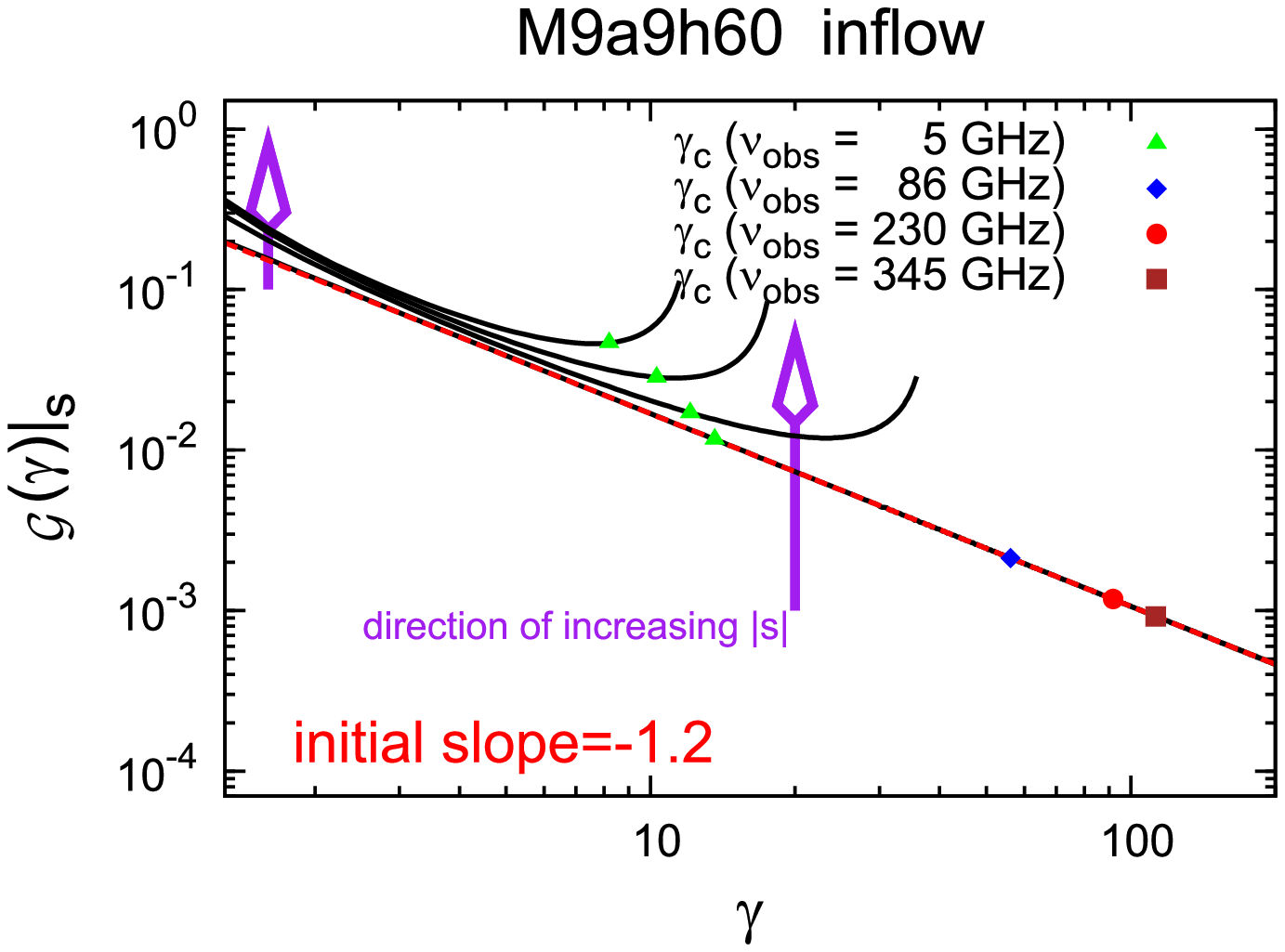}
\includegraphics[width=0.9\columnwidth, angle=0]{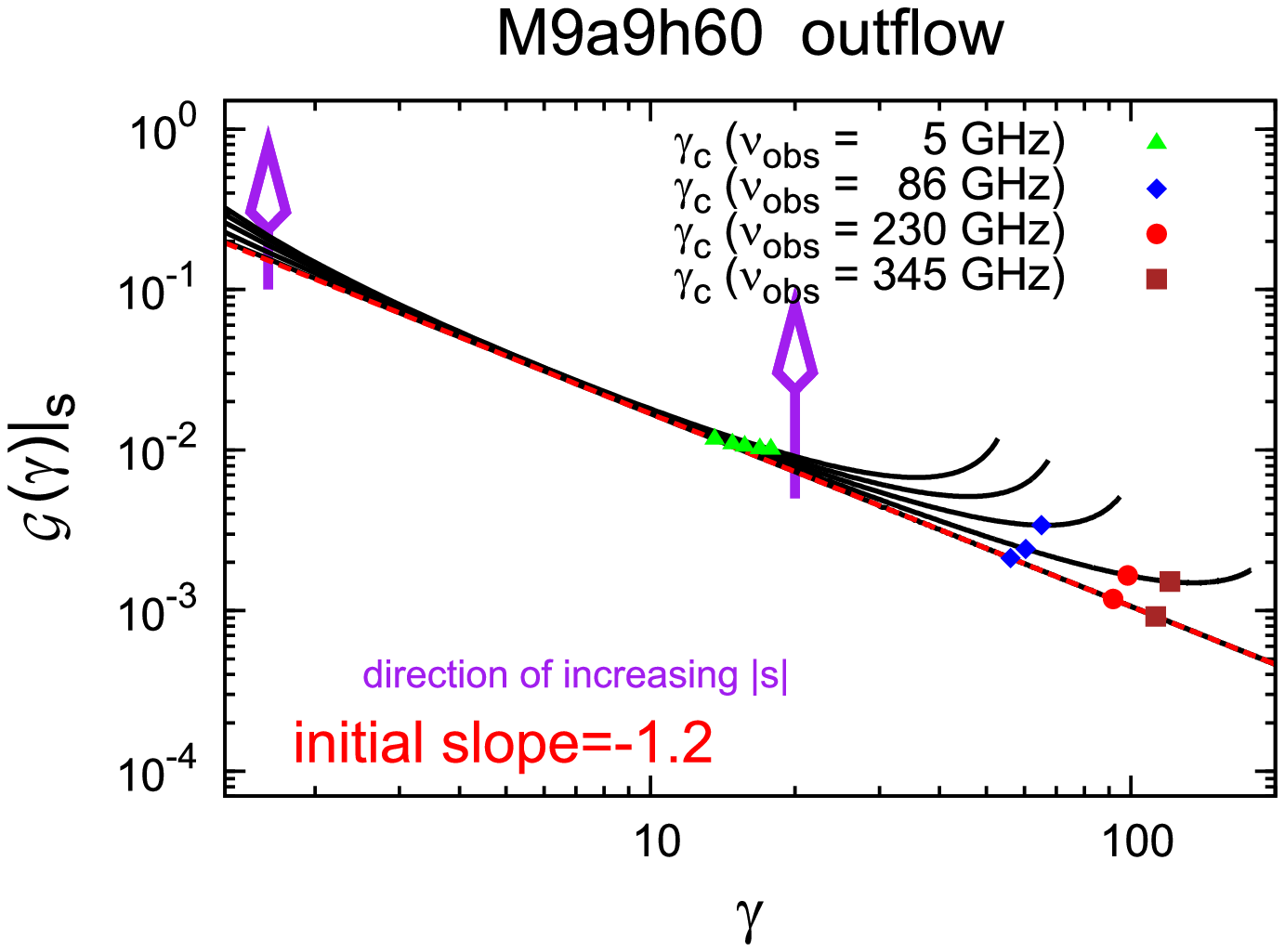}\\
\includegraphics[width=0.9\columnwidth, angle=0]{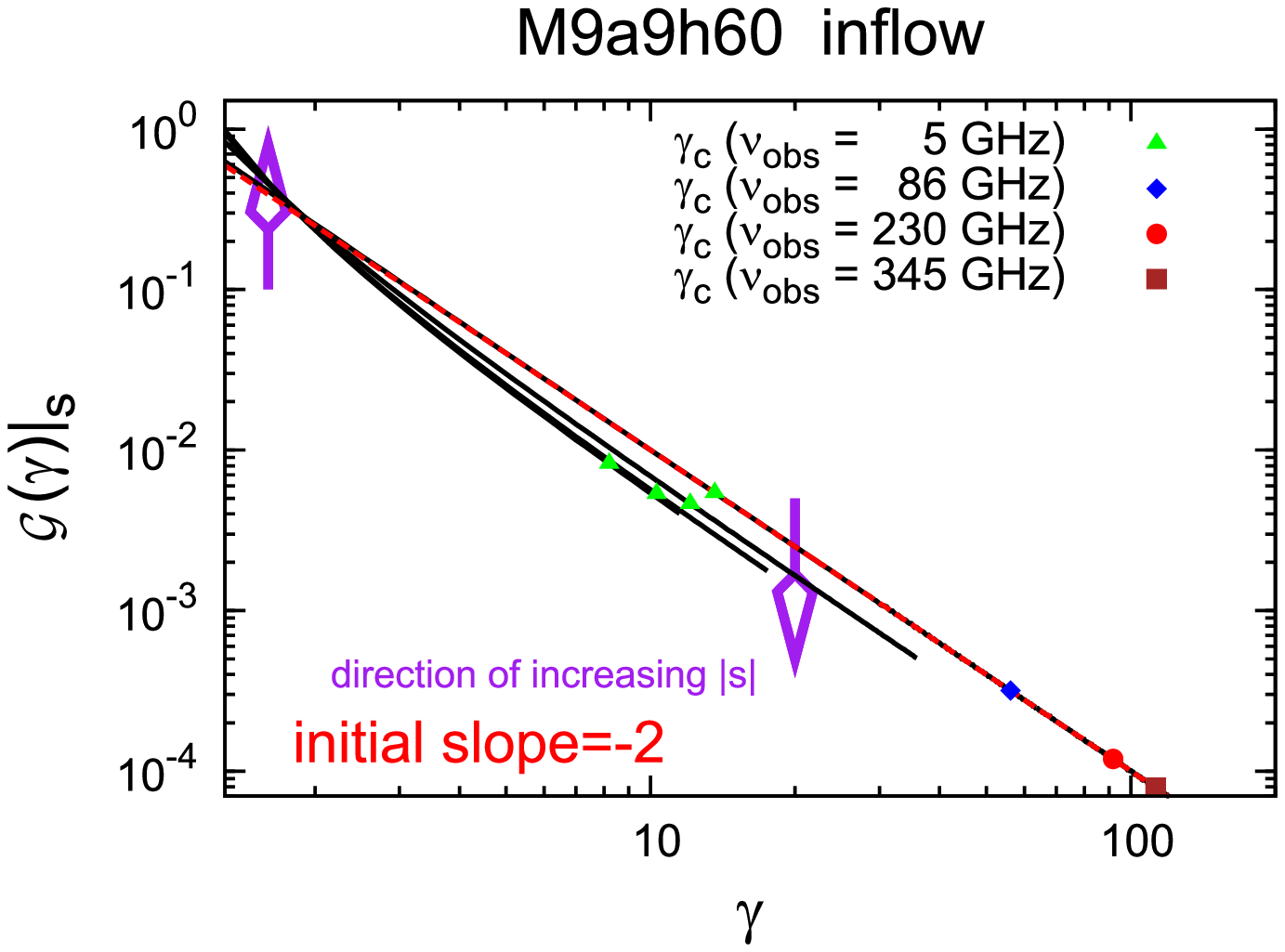}
\includegraphics[width=0.9\columnwidth, angle=0]{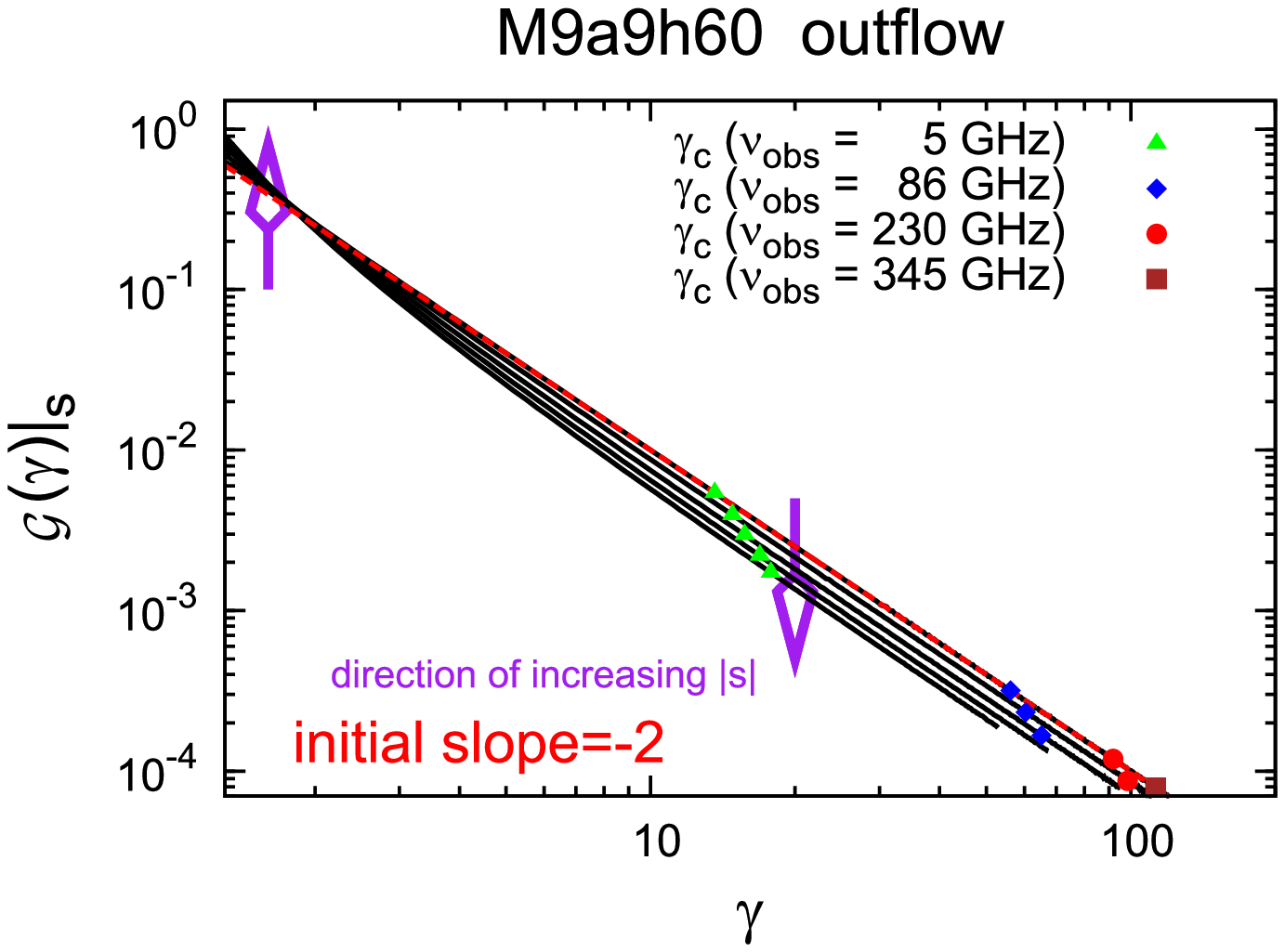}\\
\includegraphics[width=0.9\columnwidth, angle=0]{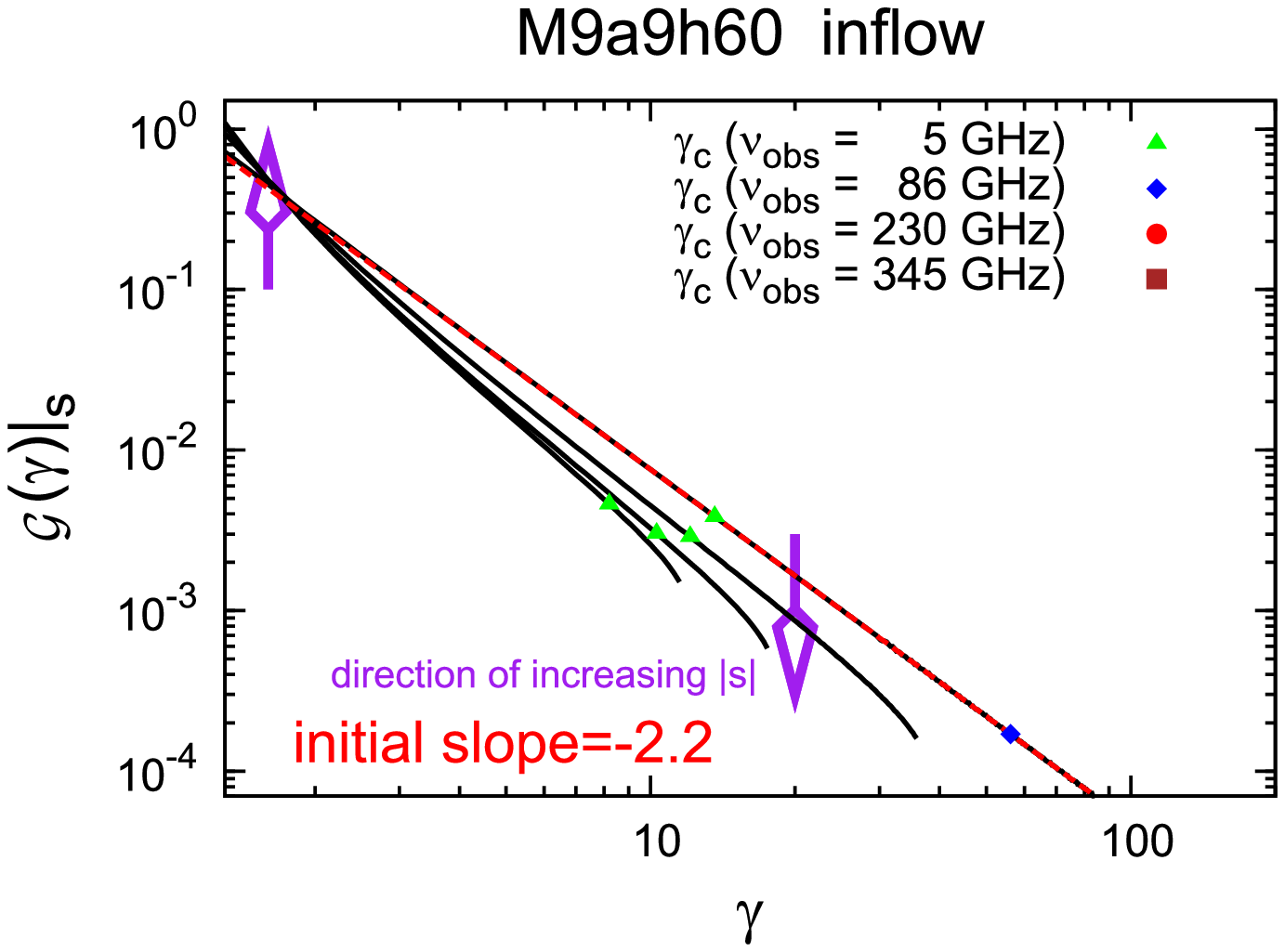}
\includegraphics[width=0.9\columnwidth, angle=0]{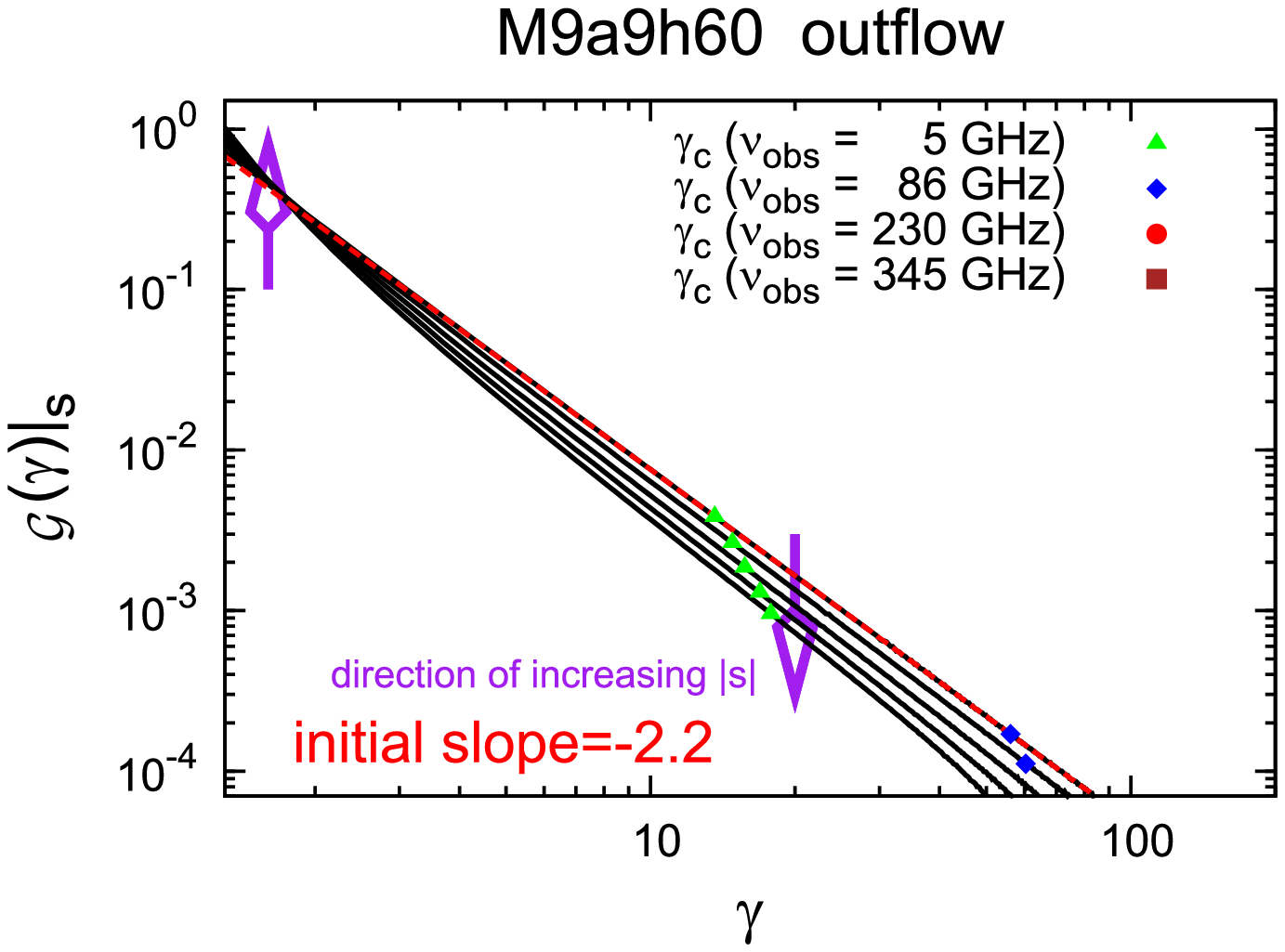}\\
\caption{Spatial variation of  the distribution function $\mathcal{G}(\gamma)$ of model M9a9h60 at different locations along the inflow (left panel) and outflow (right panel). 
The profiles are of $\mathcal{G}(\gamma)|_{s}$, with slices chosen to have a fixed interval of $r$ along the line. 
For inflow, $r= r_{0}, 4, 3$ and $2~r_{\rm g}$ are chosen, where $r_{0}=r|_{s_{0}}\simeq5.2~r_{\rm g}$ is the $r$ coordinate at the stagnation surface $s_{0}$. 
For outflow, $r= r_{0}, 6, 7, 8$ and $9~r_{\rm g}$ are chosen.
At the stagnation surface, the prescribed initial distribution follows $\mathcal{G}(\gamma)|_{s_{0}}\propto\gamma_{0}^{-\alpha}$ and $1\leqslant\gamma_{0}\leqslant10^{3}$, with $\alpha=1.2$ (top panels), $2.0$ (middle panels) and $2.2$ (bottom panels). 
The initial distribution at $r_{0}$ overlaps with the red dashed line, and the subsequent tread of evolution is indicated by the arrows. At each slice of constant $r$, electron energies $\gamma$ that are equal to the corresponding $\gamma_{\rm c}$ for observational frequencies of $\nu_{\rm obs}= 5$~GHz (green), $86$~GHz (blue), $230$~GHz (red) and $340$~GHz (brown) are indicated as colored symbols described in the panel legend.}
\label{fig:m9a9_g}
\end{centering}
\end{figure*}

The spatial variation of  $\mathcal{G}(\gamma)|_{s}$ at different locations of the inflow and outflow region of M9a9h60 are shown in Figure~\ref{fig:m9a9_g}.
Starting from an initial distribution $\gamma_{0}^{-\alpha}$ given at $s_{0}$ (red line), profiles of slices at the locations of decreasing $s$ (downstream from the stagnation surface) no longer follow a power-law profile. 
The variations of each slice with decreasing $s$ are indicated by the arrows in each plot. 
For both the inflow (left panel) and the outflow (right panel), the maximum electron energy $\gamma_{\rm max}$ at each slice is determined by the characteristic curve with $\gamma_{0}=1000$ as shown in the bottom panel of Figure~\ref{fig:m9a9h60_plot}, and is therefore independent of the choice of $\alpha$. 
For all cases, due to both synchrotron cooling and adiabatic cooling, the fraction of electrons of energy close to minimum energy ($\gamma\to1$) increases.
At the high energy end, in both the inflow and outflow regions $\mathcal{G}(\gamma)\vert_{s}$ has a profile with a positive (or negative) slope when $\alpha<2$ (or $\alpha>2$) at high energies, as derived in equation (\ref{eq:app_adi}). When $\alpha=2$, the initial slope $-\alpha=-2$ is preserved.
Note also the conservation of the total area of $\mathcal{G}({\gamma})$ at each slice as required by equation (\ref{eq:g_nor}).
Therefore, a sharp decrease in $\gamma_{\rm max}$ and a flatter slope results in profiles of $\mathcal{G}(\gamma)\vert_{s} > \mathcal{G}(\gamma)\vert_{s_{0}}$ (top panel).

\begin{figure*} 
\begin{centering}
\includegraphics[width=0.9\columnwidth, angle=0]{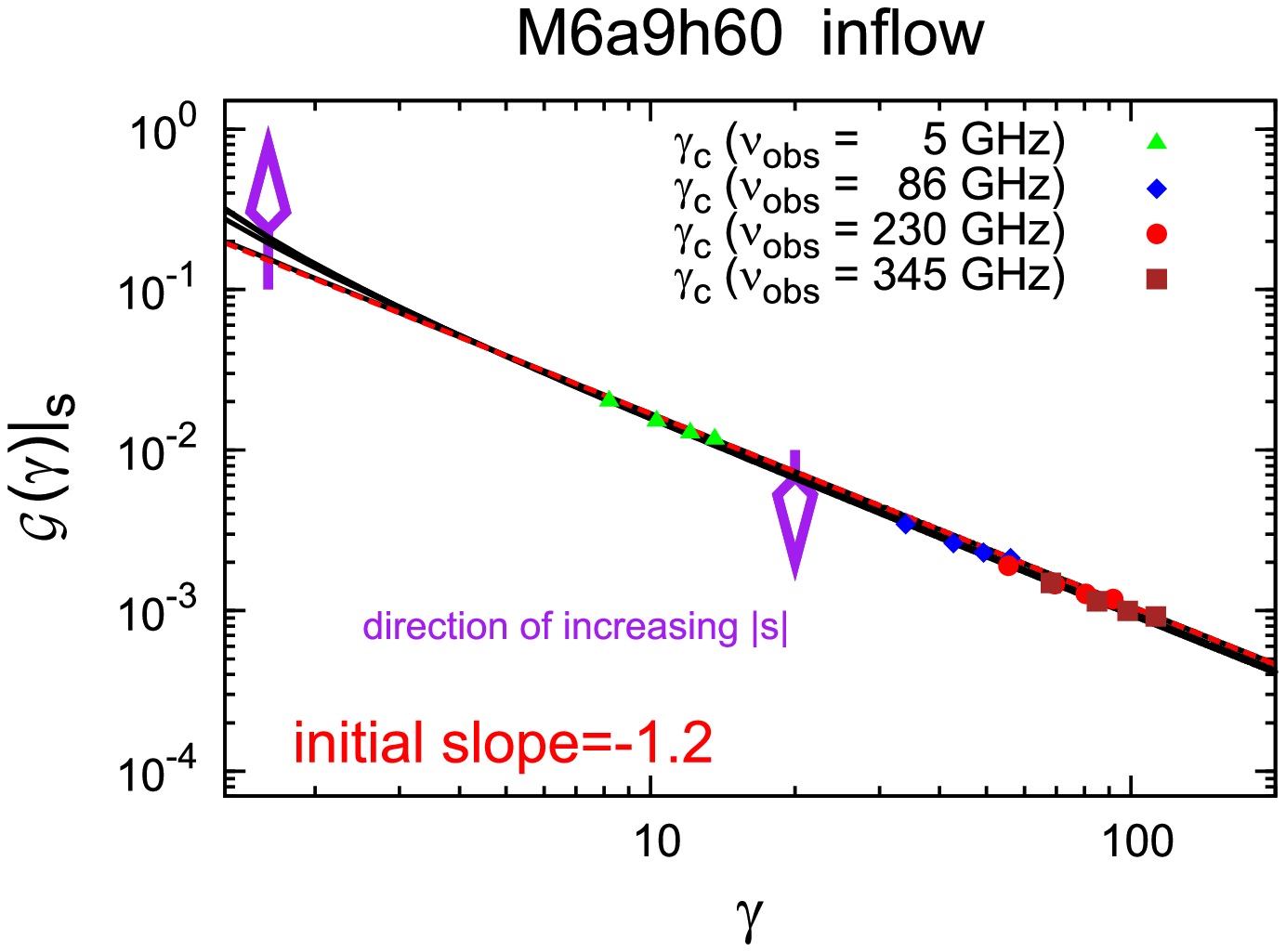}
\includegraphics[width=0.9\columnwidth, angle=0]{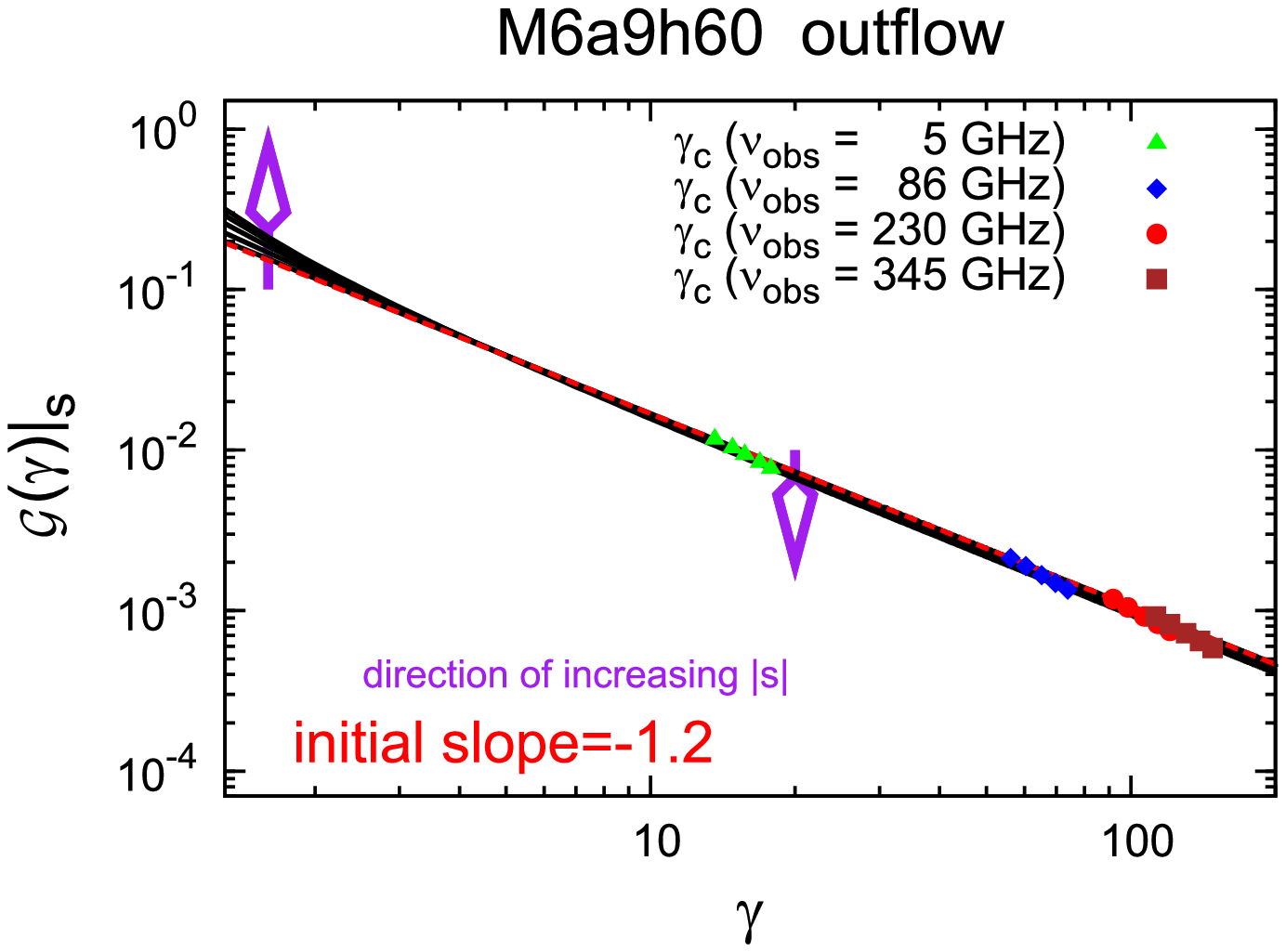}\\
\includegraphics[width=0.9\columnwidth, angle=0]{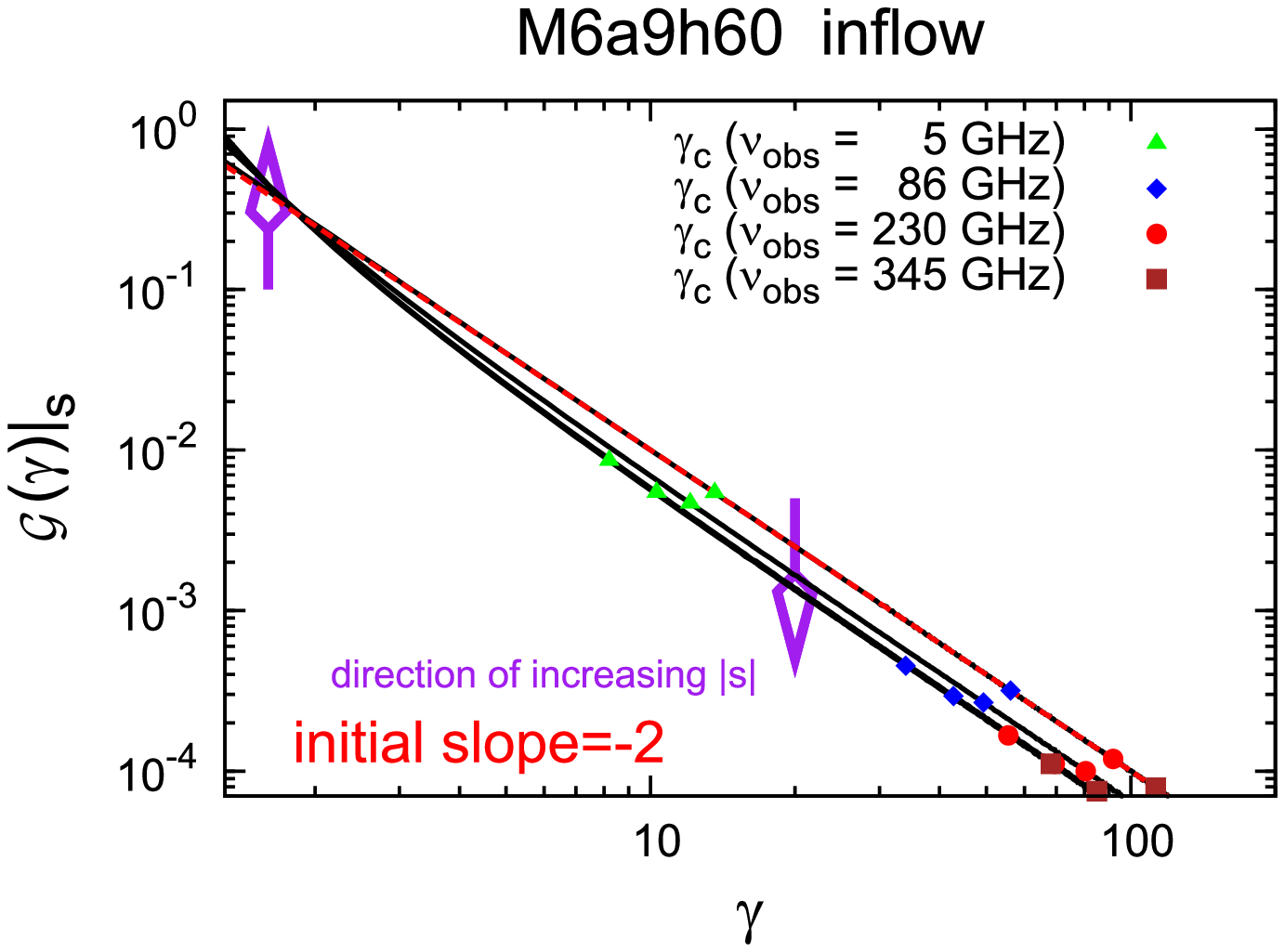}
\includegraphics[width=0.9\columnwidth, angle=0]{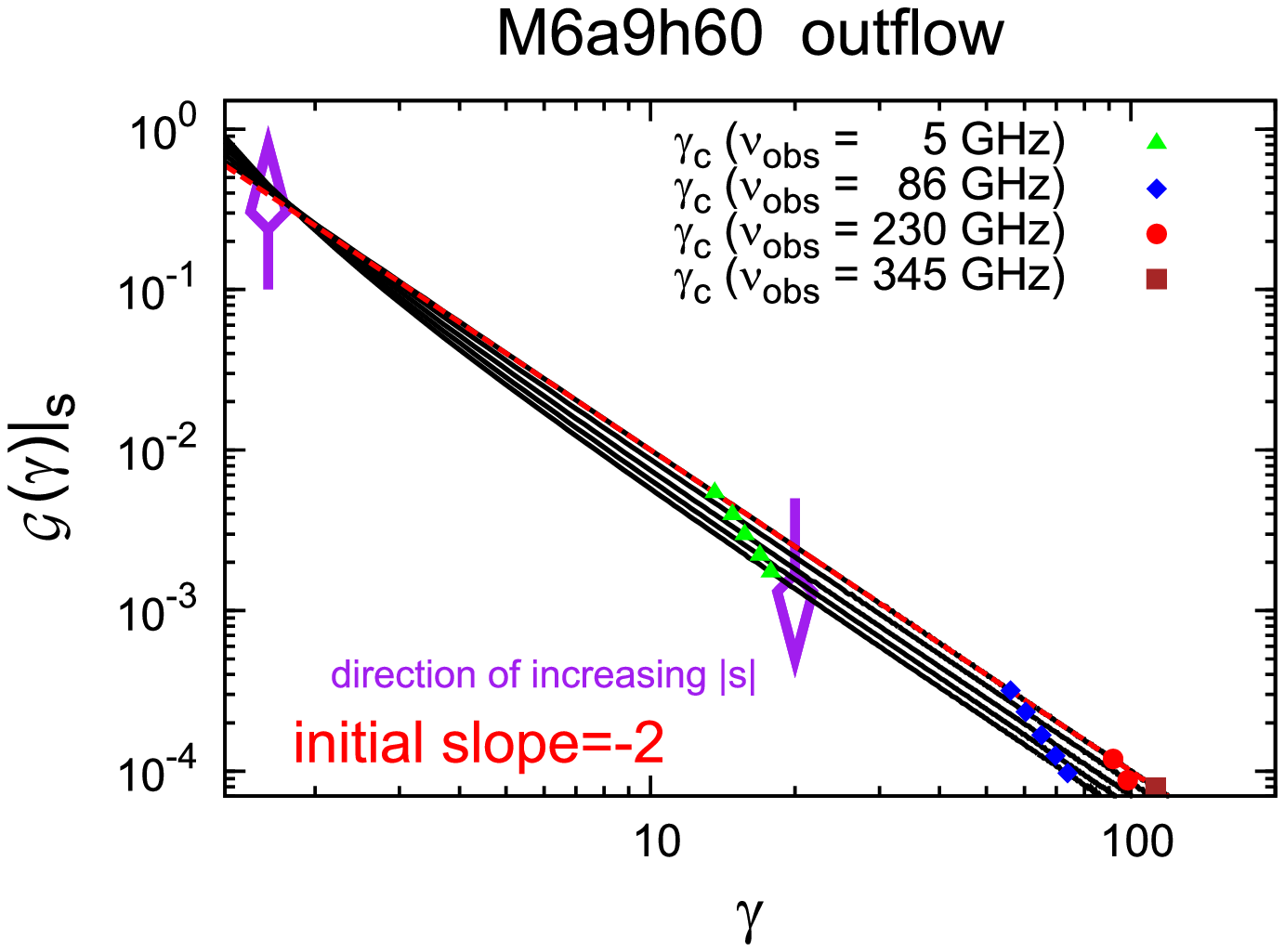}\\
\includegraphics[width=0.9\columnwidth, angle=0]{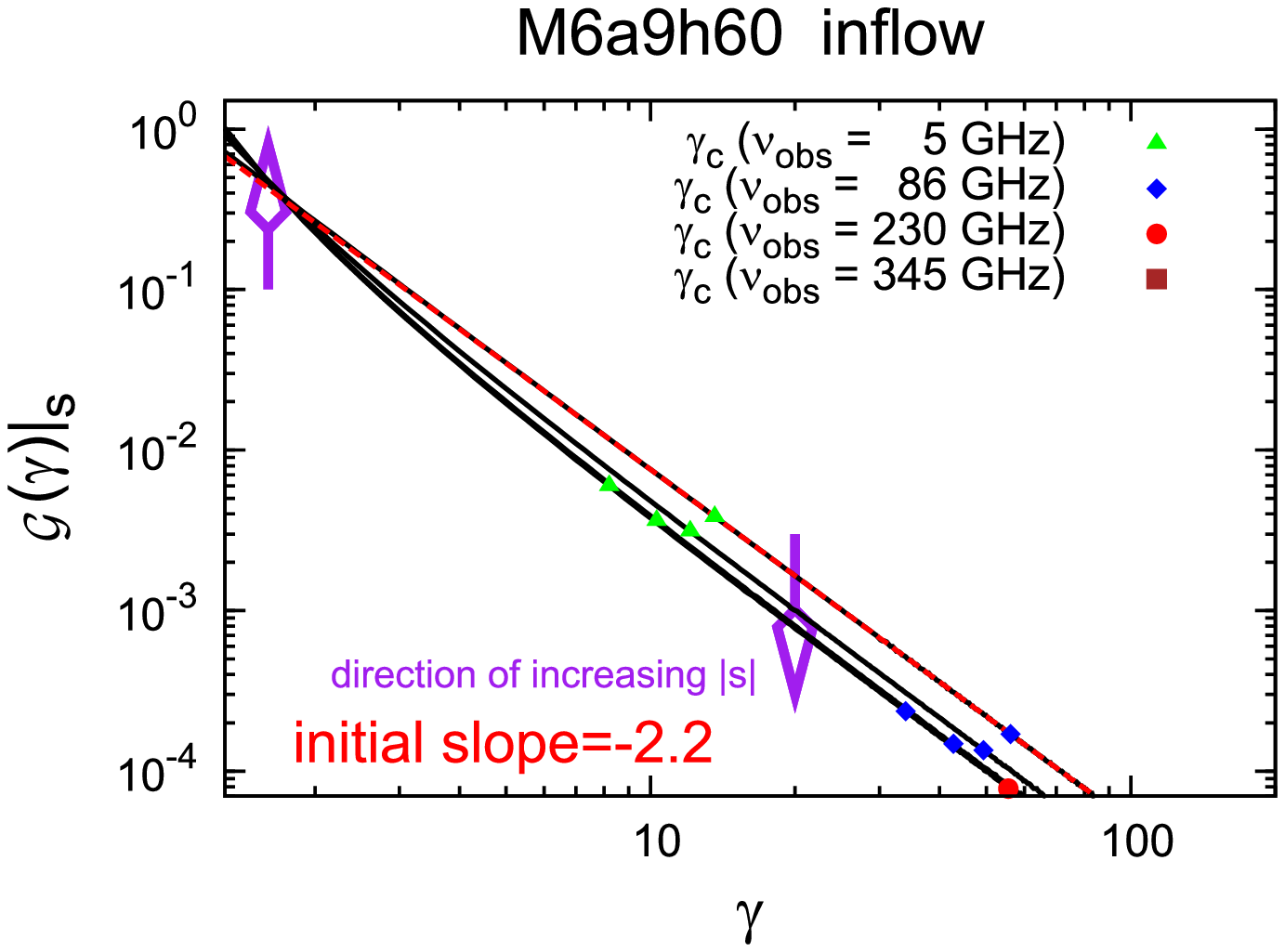}
\includegraphics[width=0.9\columnwidth, angle=0]{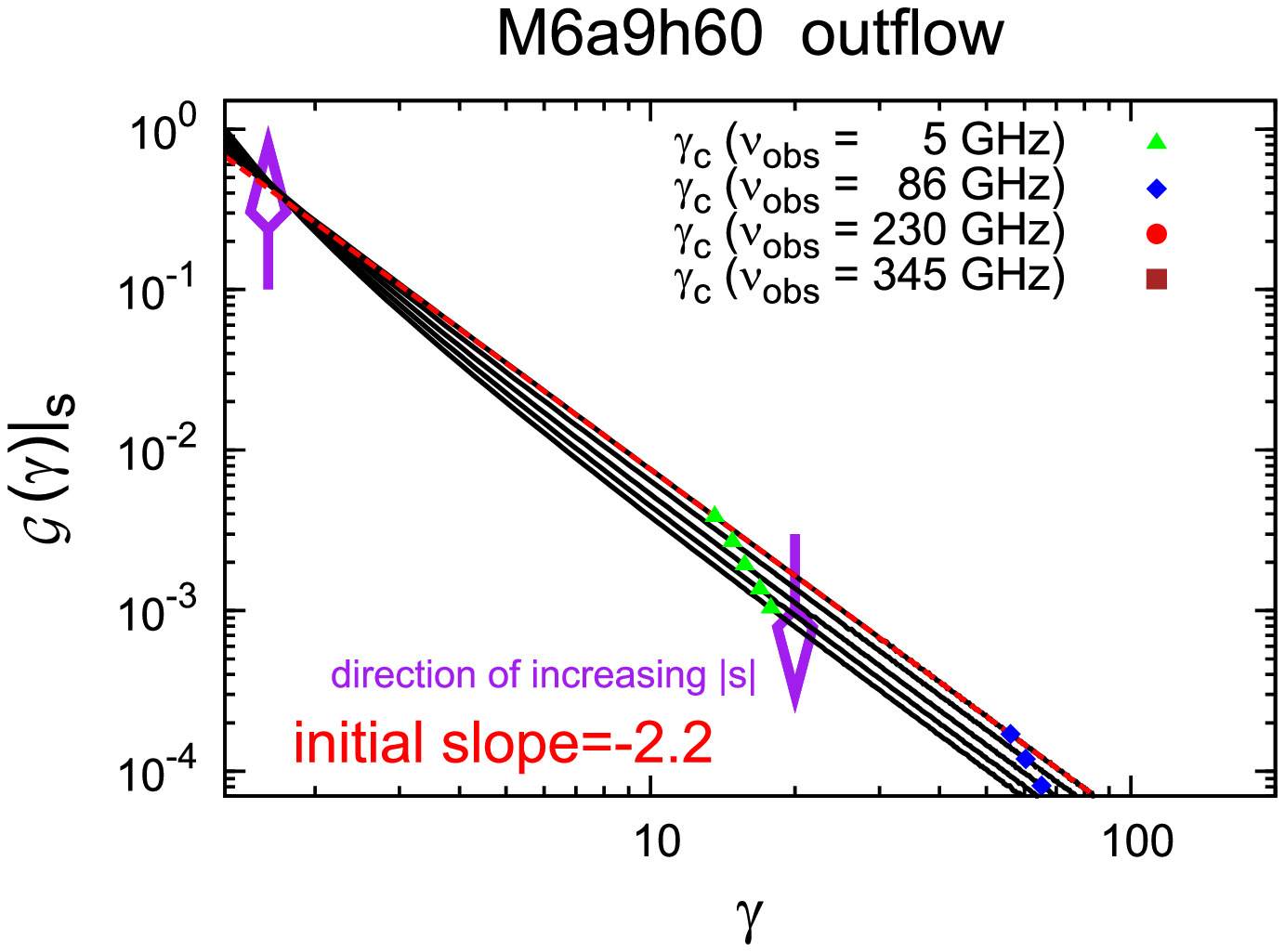}\\
\caption{Spatial variation of the distribution function $\mathcal{G}(\gamma)$ of model M5a9h60 at different locations along the inflow (left panel) and outflow (right panel). See the caption of Figure~\ref{fig:m9a9_g} for further details.}
\label{fig:m6a9_g}
\end{centering}
\end{figure*}

The distribution function of M6a9h60 shown in Figure~\ref{fig:m6a9_g} provides an example of the case where adiabatic processes dominate ($\propto \gamma-1$). 
Similar to the previous discussion, the profile of $\gamma_{0}=1000$ in Figure~\ref{fig:m6a9h60_plot} provides a reference for the maximum value of $\gamma$ on the same slices. 
Due to cooling processes, the fraction of electrons with $\gamma \to1$ increases. 
There are two striking differences when compared to the synchrotron process-dominated case.
Firstly, for all choices of $\alpha$, ${\rm d}(\ln \mathcal{G})/{\rm d}(\ln\gamma)$ is always negative at the high energy end. The slope $-\alpha$ is preserved at different spatial locations when adiabatic processes dominate (equation (\ref{eq:app_adi})).
Secondly, the maximum electron energy does not change as rapidly as in the case of synchrotron processes ($\propto \gamma^{2}-1$).

\subsection{Implications}\label{sec:imp}

For non-thermal electrons,
most synchrotron emission may be attributed to electrons with a characteristic energy 
$\gamma_{\rm c}^{2}\sim\nu_{\rm obs}/\nu_{\rm c}$ \citep[e.g.][]{ryb86, leu11}, 
where $\nu_{\rm obs}$ is the observed frequency and $\nu_{\rm c}\equiv eB/\left(2\pi\;\! m_{\rm e} c\right)$. 
After some computation, we obtain
\begin{equation}\label{eq:gamma_c}
\gamma_{\rm c}\sim59.8\sqrt{\left(\frac{\nu_{\rm obs}}{100\;\! {\rm GHz}}\right)\left(\frac{10\;{\rm G}}{B}\right)}\;\;\;\;.
\end{equation} 
For our frequencies of interest, $\nu_{\rm obs}=5, 86$, $230$, and $345$ GHz, the corresponding profiles of $\gamma_{\rm c}\left(\nu_{\rm obs}\right)$ along the field line for each model overlap in Figures~\ref{fig:m9a9h60_plot}--\ref{fig:m6a9h60_plot}.
With the magnetic field strength decreasing further away from the black hole, $\gamma_{\rm c}$ is monotonically increasing along the field line.
The nature of the decreasing profile of characteristic curves and the increasing profile of $\gamma_{\rm c}$ indicates a finite non-thermal synchrotron emission region within which 
\begin{equation}\label{eq:emission_region}
\gamma_{\max}(s;\Psi) > \gamma_{\rm c}(s;\Psi) \,,
\end{equation}
is satisfied. 
It is clear that the size of the emitting region therefore depends on the observational frequency $\gamma_{\rm c}(\nu_{\rm obs})$ and the variation of electron energy along a given field line $\gamma_{\max}(s;\Psi)$. 
As shown in Figures~\ref{fig:m9a9h60_plot}--\ref{fig:m6a9h60_plot}, the emission region size become larger when the observational frequency decreases, with the stagnation surface being further away, or the flow being dominated by adiabatic processes.
As the observational frequency increases, the emission region size decreases and gradually coincides with the birth place of non-thermal electrons. 
That is to say, the location of the stagnation surface become increasingly apparent as the observational frequency increases. 
For a given field configuration $\Psi$ , the location of the stagnation surface is related to the black hole spin parameter and the angular velocity of the field $\Omega_{\rm F}$ (\S\,\ref{sec:ss}). 
As a result, it may be possible to constrain the spin of the black hole using the information concerning the location of the stagnation surface, as provided by VLBI observations at multiple frequencies.
Observations of non-thermal synchrotron emission of the M87 jet beyond pc scales \citep[e.g.,][]{per99,per01,had11,hom16,mer16} indicate that subsequent re-heating (injecting energy into electrons) far from the emission region discussed here is necessary.

The shape of the emitting region can be inferred as a collection of non-thermal electron distributions on different field lines.
Comparing the results of M9a9h20 ($\theta_{\rm h}=20^{\circ}$) and those of M9a9h60 ($\theta_{\rm h}=60^{\circ}$),
for higher latitude field lines, the stagnation surface is located further away from the black hole and the emission region size appears larger (relatively) along the field line.
It is expected that the emitting region is more extended towards the jet axial region.

In Figures~\ref{fig:m9a9_g} and \ref{fig:m6a9_g} we indicate the electrons with $\gamma$ equal to the corresponding $\gamma_{\rm c}$ for different observational frequencies at each slice of fixed spatial location. 
If $\gamma_{\rm c}$ exceeds $\gamma_{\rm max}$,
the ensemble of electrons at a specific slice can no longer contribute emission at any of the given observational frequencies,
and no indicated  $\gamma_{\rm c}$ appears on that slice.
It is expected that the emission close to the stagnation surface would be exceptionally luminous due to the larger electron number density there.
For the outflow, such a condition will eventually be satisfied due to both synchrotron cooling and adiabatic cooling (left panels of Figures~\ref{fig:m9a9_g} and \ref{fig:m6a9_g}).

For the inflow, 
whether or not the electrons can contribute to the synchrotron emission (i.e.~if $\gamma_{\max}\left(s;\Psi\right)  >  \gamma_{\rm c}\left(s;\Psi\right)$ is satisfied) all the way down to the event horizon depends on how quickly the non-thermal electrons lose their energy.
If synchrotron cooling is efficient (which is the case for a larger black hole mass), the inflow component is only partially observable (as shown by the absence of $\gamma_{\rm c}$ on the slices in the left panels of Figure~\ref{fig:m9a9_g}; see also Figure~\ref{fig:m9a9h60_plot}). 
If adiabatic cooling is more important (which is the case for a smaller black hole mass), the whole inflow component may be observable (as shown by the presence of $\gamma_{\rm c}$ on the slices in the left panels of Figure~\ref{fig:m6a9_g}; see also Figure~\ref{fig:m6a9h60_plot}) since $\gamma_{\max}$ decays at a slower rate ($\propto \gamma-1$) compared to the synchrotron cooling ($\propto \gamma^{2}-1$).

Although the location of the stagnation surface is time-dependent, as shown in GRMHD simulations \citep[][]{bro15}, 
a stationarity assumption may be adopted for M87 because of its long dynamical time scale ($GM_{\rm BH}/c^{3}\sim 8$ hrs) compared to typical VLBI observations. Future VLBI observations will provide an excellent opportunity to constrain the injection site of non-thermal electrons and test the predictions of different jet origin models.
For further comparison, general-relativistic radiative transfer should be included to take into account the energy shift and aberrations due to the fluid motion and the black hole's strong gravitational field \citep[e.g.,][]{fue04,sch04,fue07,dex09,vin11,you12,cha13,pu16}. 

\section{Summary and Conclusion}\label{sec:implication}

We determine the energy spectral evolution of non-thermal electrons in a GRMHD black hole-powered jet 
  and calculate the radio synchrotron radiation from these electrons in the jet inflow and outflow regions  
  near the event horizon.  
Energetic electrons are injected at the stagnation surface, a unique feature predicted by the GRMHD model, 
  and these electrons (\S\ref{sec:model_nth}) are subsequently carried by the GRMHD flow 
  along the magnetic field lines threading the black hole event horizon (\S\ref{sec:model}). 
The energy spectra variation of the non-thermal electrons along the flows 
  is regulated by synchrotron radiative losses and adiabatic processes.

 More specifically, we assume there is no further injection of non-thermal electrons after they are injected from the stagnation surface with a power-law energy spectrum (equation (\ref{eq:inj})), and the scattering and drift along magnetic field lines are unimportant. Without enough time to reach thermalized or equal-partition state, the non-thermal electrons distribution $n_{\rm nth} (\gamma, s)$ is described by equation (\ref{eq:g1}). At each location $s$, the spatial variation of non-thermal electrons are traced by using the normalized energy spectral distribution function $\mathcal{G}(\gamma)|_{s}$ via the presented conservative formula (\S \ref{sec:trans_F}). The spatial variation of non-thermal electrons energy, $d\gamma/ds$, can therefore be described by the characteristic curves, equation (\ref{eq:dg_ds}). We summarize our findings in the following five paragraphs.   

$\mathbf{1.}$ In the {\em outflow} region, 
   the energy of the electrons drops due to radiative losses (via synchrotron radiation, $\dot{\gamma}\propto \gamma^{2}-1$) and mechanical cooling 
   (via adiabatic expansion of the jet fluid, $\dot{\gamma}\propto \gamma-1$).  
Further away from the stagnation surface along the magnetic field lines, 
  the energy of the non-thermal electrons drops significantly, 
  so much so that their contribution to the observable synchrotron radiation become insignificant. 
For imaging observations at a fixed frequency band,  
   the emitting region, if it can be resolved, would appear finite in size. 
The size of the emitting region, however, varies with observational frequency: 
  the smaller the size of the emitting region, the higher the corresponding observational frequency.  
Moreover, adiabatic expansion is the dominant cooling process 
   for the jet outflow far away from the stagnation surface.

$\mathbf{2.}$ In the {\em inflow} region, 
 similar to the outflow region, the energy of electrons drops due to both synchrotron cooling and adiabatic cooling.
If radiative cooling dominates, 
  the electrons can lose energy rapidly and 
  become negligible in their contribution to the observable emission (at a given observational frequency) 
  before entering reaching the event horizon.
If adiabatic cooling dominates, electrons lose energy less rapidly.
As a result, electrons can remain energetic enough to 
contribute to the observable emission when they approach the black hole event horizon, 
  rendering the entire inflow region observable (until the dimming effect due to gravitational red-shift dominates).

$\mathbf{3.}$ Our calculations have shown that the relative importance of radiative losses and adiabatic losses  
   is dependent on the dynamical time scale of the flow and hence the black-hole mass of the system, provided that a similar magnetic field strength is applied (see equation (\ref{eq:cooling_ratio})). If so, 
synchrotron losses are expected to be the dominant cooling process for systems with a more massive black hole, 
  whereas adiabatic losses would be important for the systems with a smaller mass black hole.  
As such, M87-like and Sgr A*-like systems could exhibit different observational properties, 
  even when their jets are driven by the same GRMHD process.  

$\mathbf{4.}$ In studies of emission from relativistic jets in the compact core region, the energy spectra of non-thermal electron populations are often assumed to have a power-law profile with a maximum energy $\gamma_{\rm max }$.     
 We have shown that, when the energy loss is dominated by synchrotron cooling instead of adiabatic cooling,
  the non-thermal electrons, which had an initial power-law energy distribution $\gamma_{0}^{-\alpha}$ when they were injected,  
  cannot maintain a power-law spectrum. 
  Furthermore, $\gamma_{\rm max}$ drops rapidly and the spatial variation of the electron energy density, in terms of $n_{\rm nth}(\gamma, s)$, depends on the value of the initial spectral index $\alpha$.
 On the other hand, if the energy loss is dominated by adiabatic cooling, $\gamma_{\rm max}$ decays less rapidly and the initial power-law energy distribution can be preserved at the higher energy end.

$\mathbf{5.}$ The location of the stagnation surface is further away from the black hole towards to the jet central axis and
  the radio emission (as observed at a specific frequency) will appear more extended accordingly. 
As the observational frequency increases, 
  the emitting region will eventually coincide with the stagnation surface, where the non-thermal electrons are freshly injected.
The high concentration of fresh energetic electrons near the stagnation surface 
   implies that regions adjacent to the stagnation surface should be luminous and more easily observable.

Our calculations have shown strong radio emission from the regions near the stagnation surface.
Predicted observed images of this stagnation surface will be explored with general relativistic radiative transfer calculations as a future work.
Detection of such bright emission features in VLBI imaging observations 
  will unambiguously confirm the GRMHD nature of black hole-powered jets.  
This is achievable with upcoming mm/sub-mm VLBI observations.

\acknowledgements
H.-Y. P., K. A., and M. N. are grateful to the GLT team in ASIAA for their support and encouragement. H.-Y. P. and K. A. are supported by the Ministry of Science and Technology (MOST) of Taiwan, under the grant MOST 103-2112-M-001-038-MY2.
Z. Y. is supported by an Alexander von Humboldt Fellowship.
Z. Y. and Y. M. acknowledge support from the ERC synergy
grant ``BlackHoleCam: Imaging the Event Horizon of Black Holes" (Grant No.~$610058$).
This research has made use of NASA's Astrophysics Data System.

\appendix
\section{Solutions of Background GRMHD Flows}\label{sec:app}
\begin{figure*} 
\begin{centering}
\includegraphics[width=0.3\columnwidth, angle=-90]{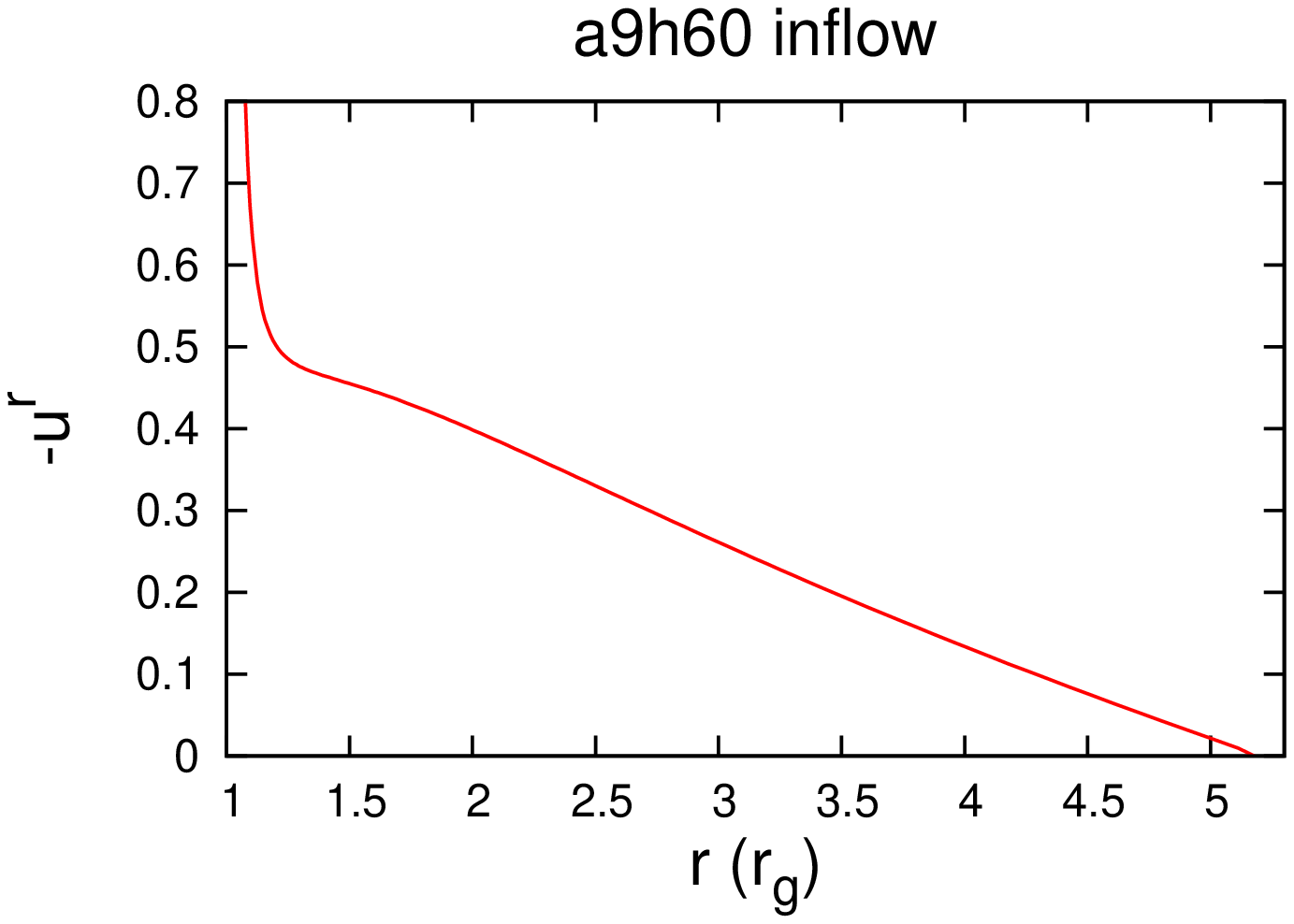}
\includegraphics[width=0.3\columnwidth, angle=-90]{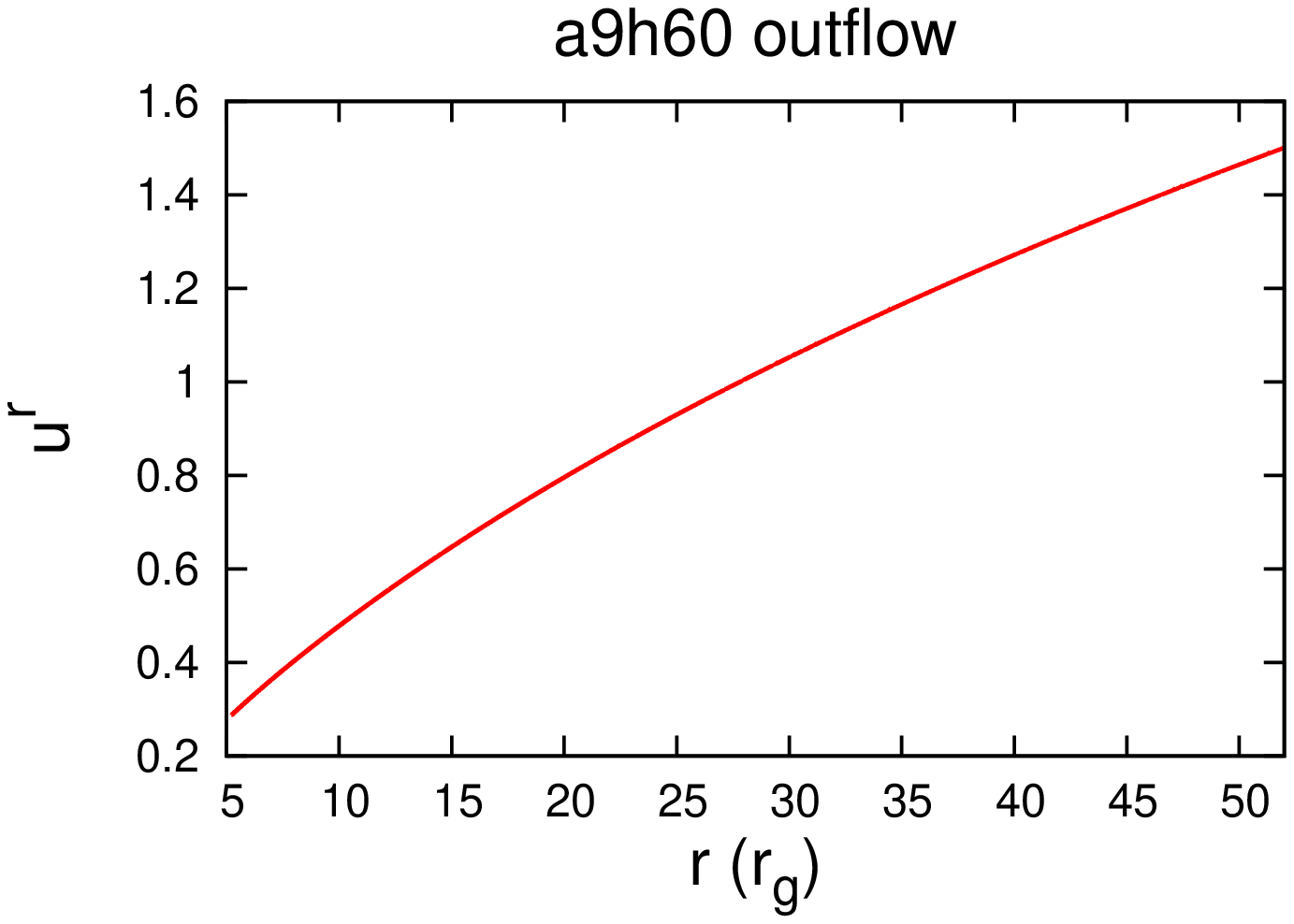}\\
\includegraphics[width=0.3\columnwidth, angle=-90]{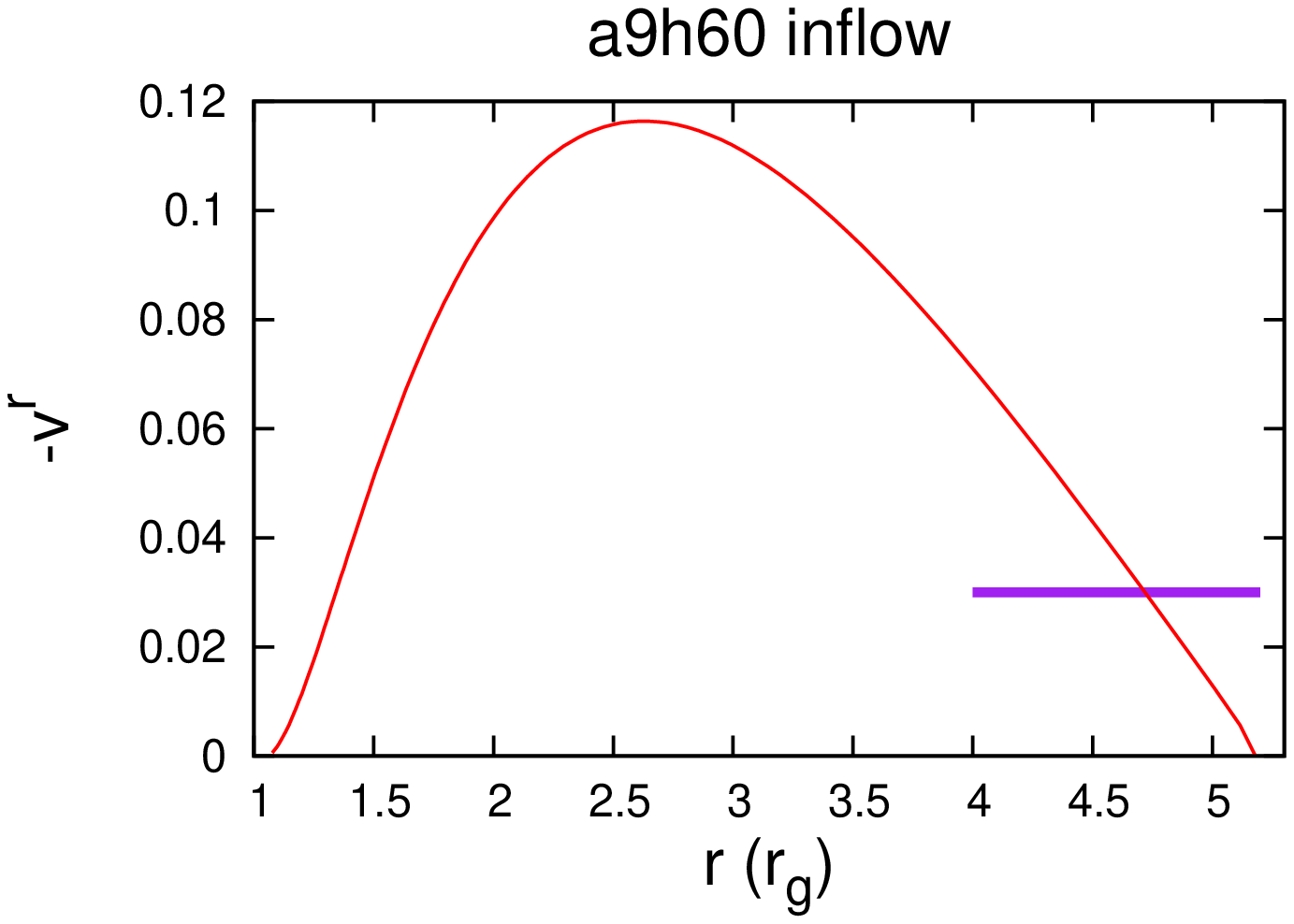}
\includegraphics[width=0.3\columnwidth, angle=-90]{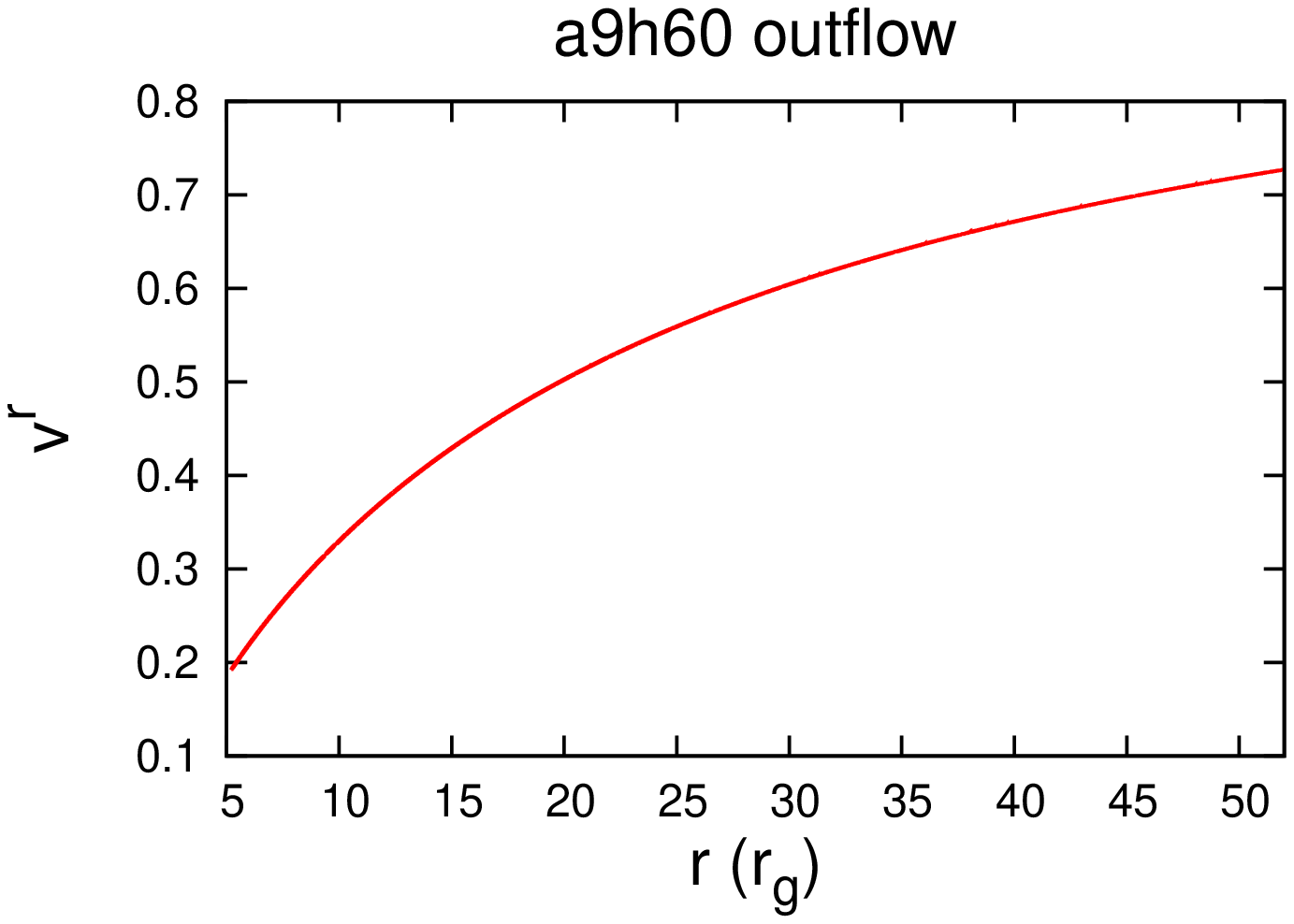}\\
\includegraphics[width=0.3\columnwidth, angle=-90]{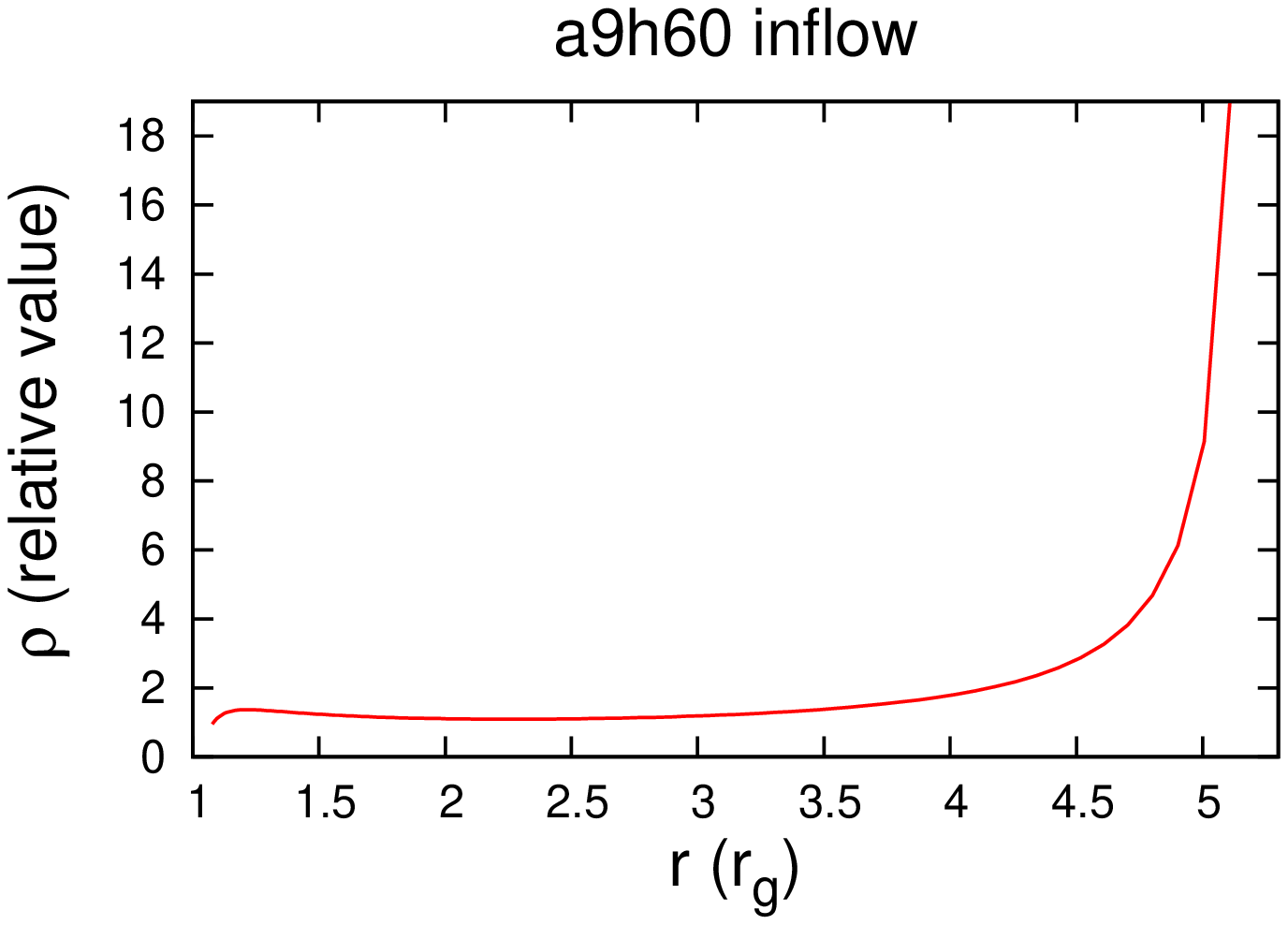}
\includegraphics[width=0.3\columnwidth, angle=-90]{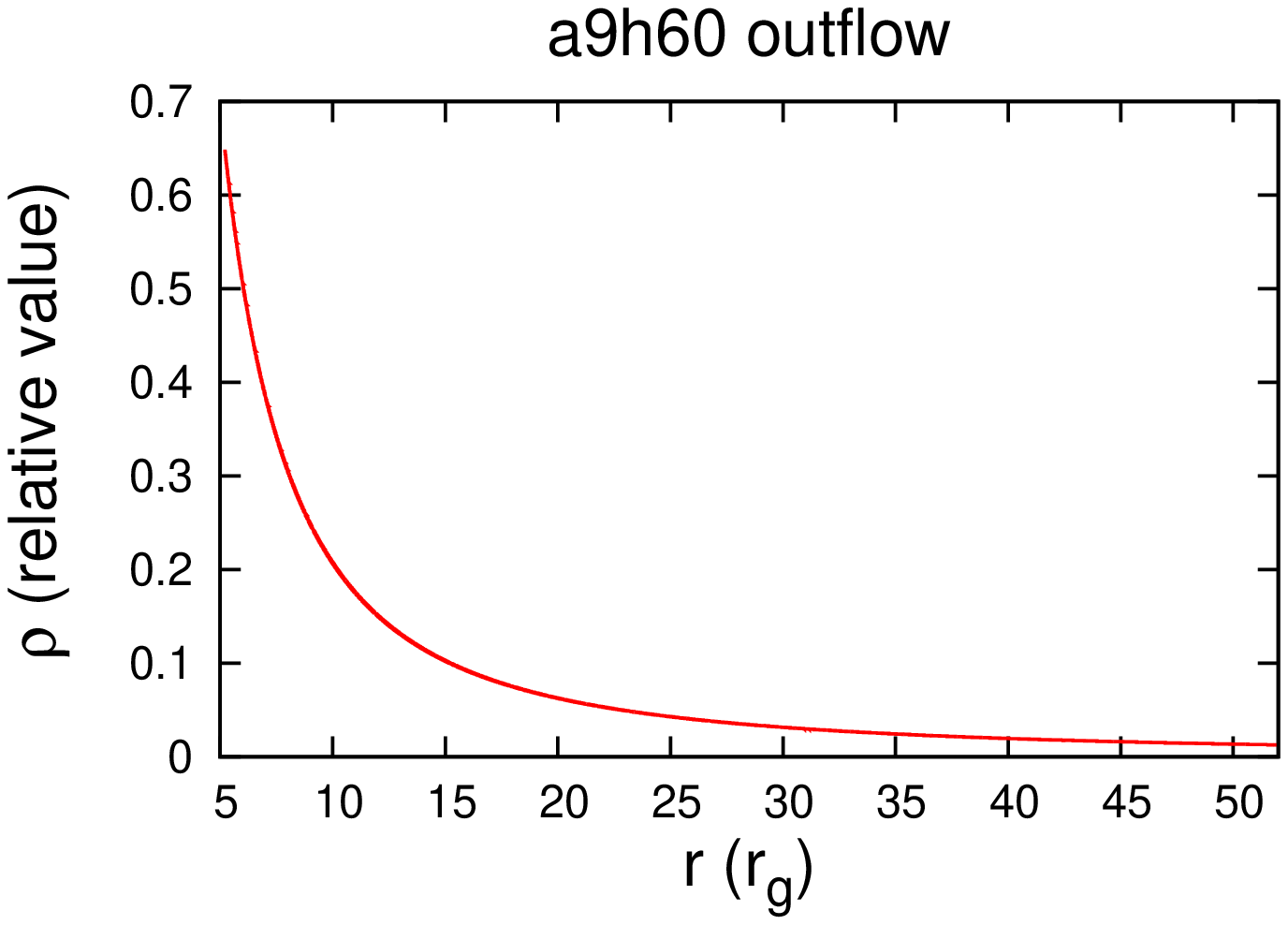}\\
\caption{GRMHD solutions for the inflow region (left panels) and the outflow region (right panels) used for models M9a9h60 and M6a9h60, respectively. The profiles are plotted along the field line, in terms of the Boyer-Lindquist radial coordinate $r$. 
From top to bottom: the radial component of the 4-velocity $u^{r}$, the radial component of the 3-velocity $v^{r}=u^{r}/u^{t}$ (the applied floor value is indicated by the horizontal line), and the number density $\rho$ (relative numerical value).
See \S~\ref{sec:model} and \S~\ref{sec:par} for description and comparison.}\label{fig:app_model}
\end{centering}
\end{figure*}

\begin{figure*} 
\begin{centering}
\includegraphics[width=0.3\columnwidth, angle=-90]{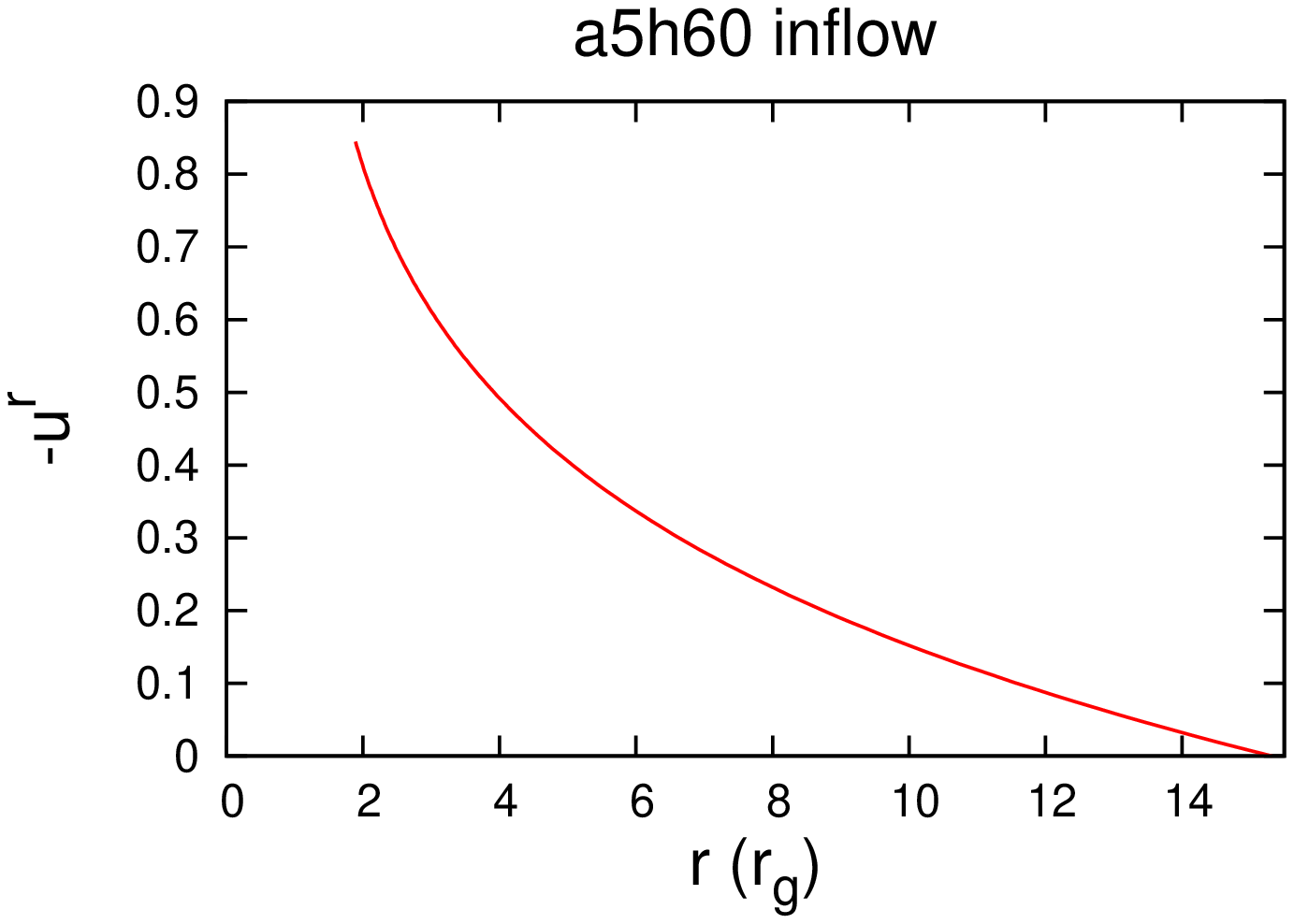}
\includegraphics[width=0.3\columnwidth, angle=-90]{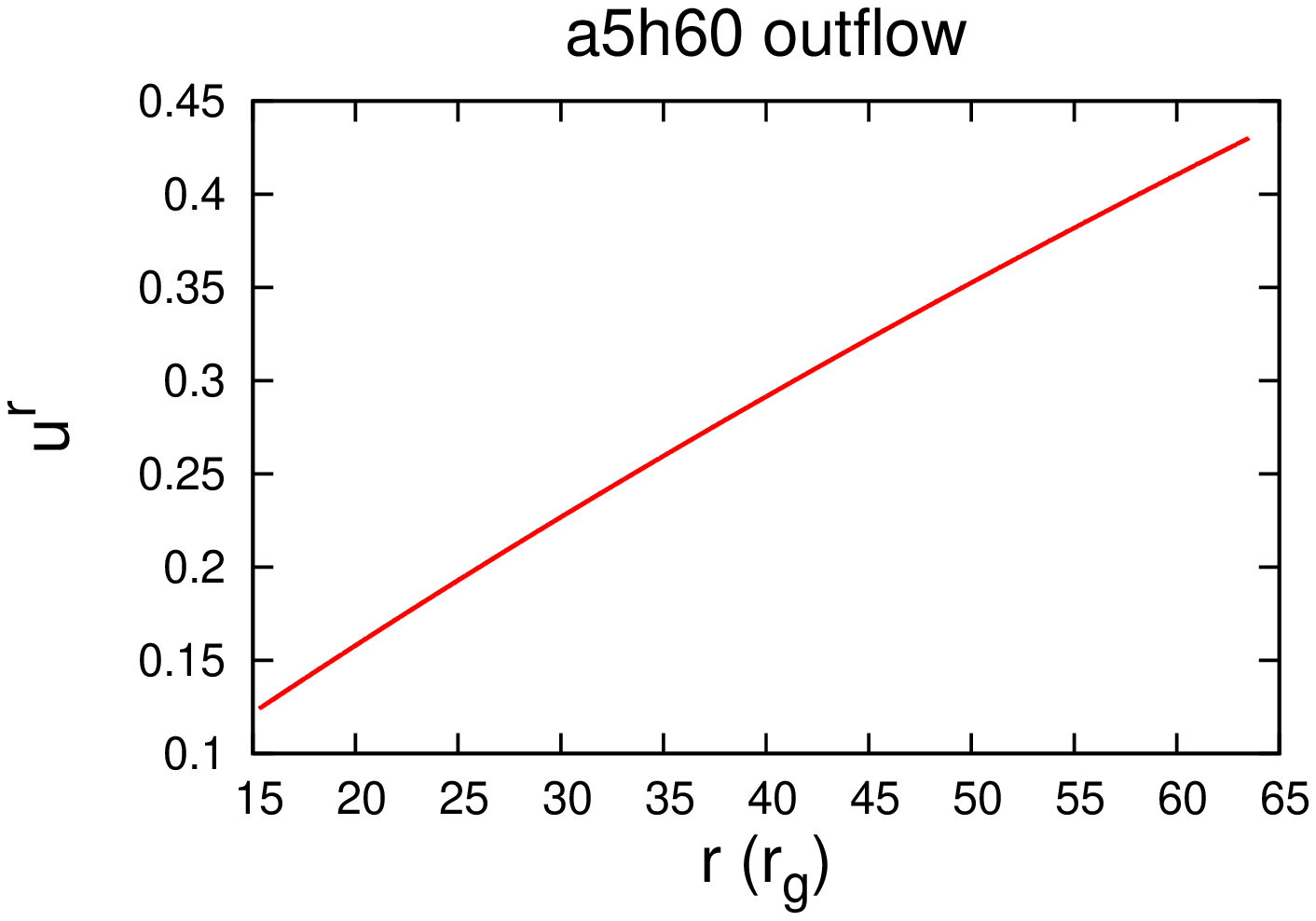}\\
\includegraphics[width=0.3\columnwidth, angle=-90]{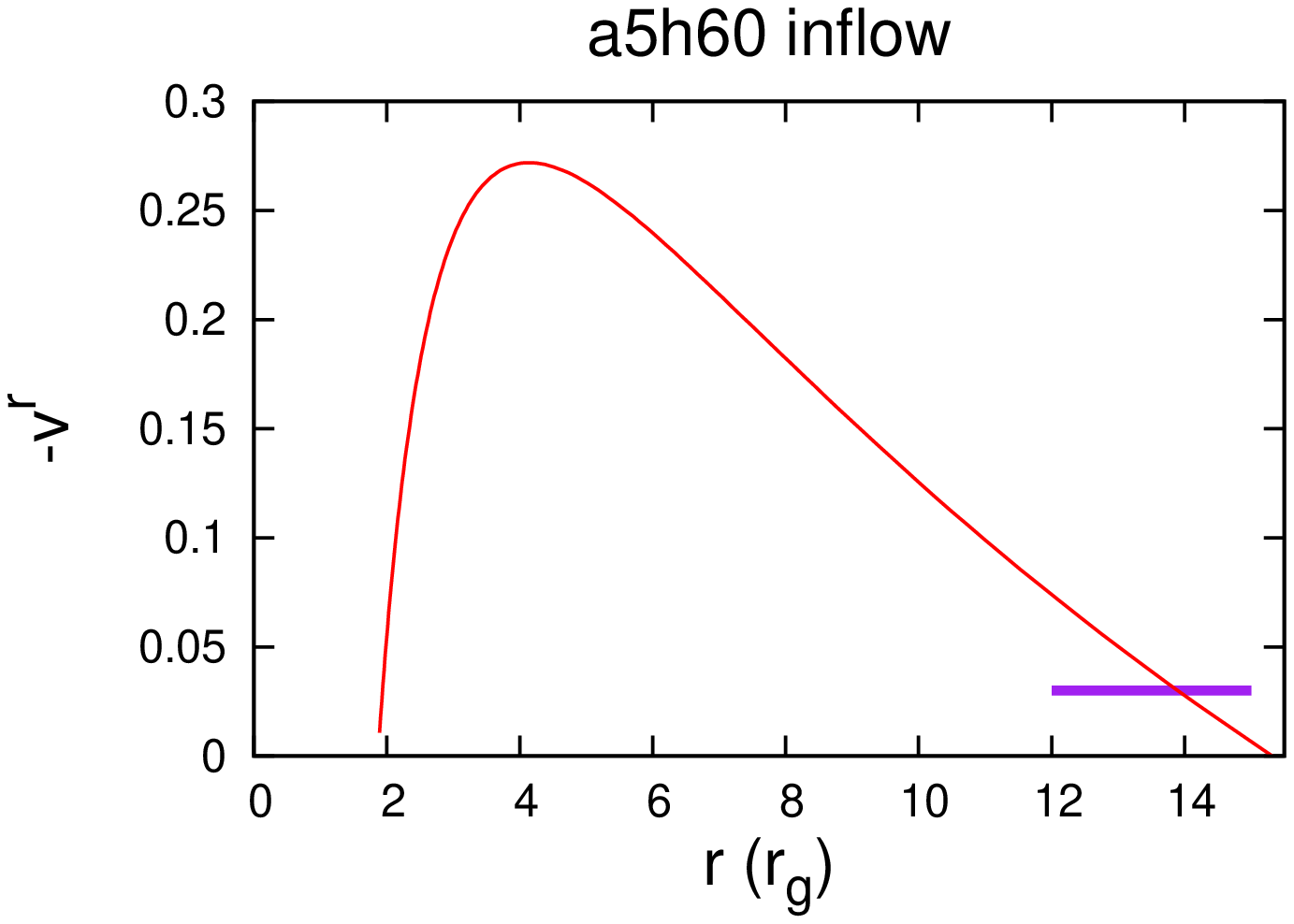}
\includegraphics[width=0.3\columnwidth, angle=-90]{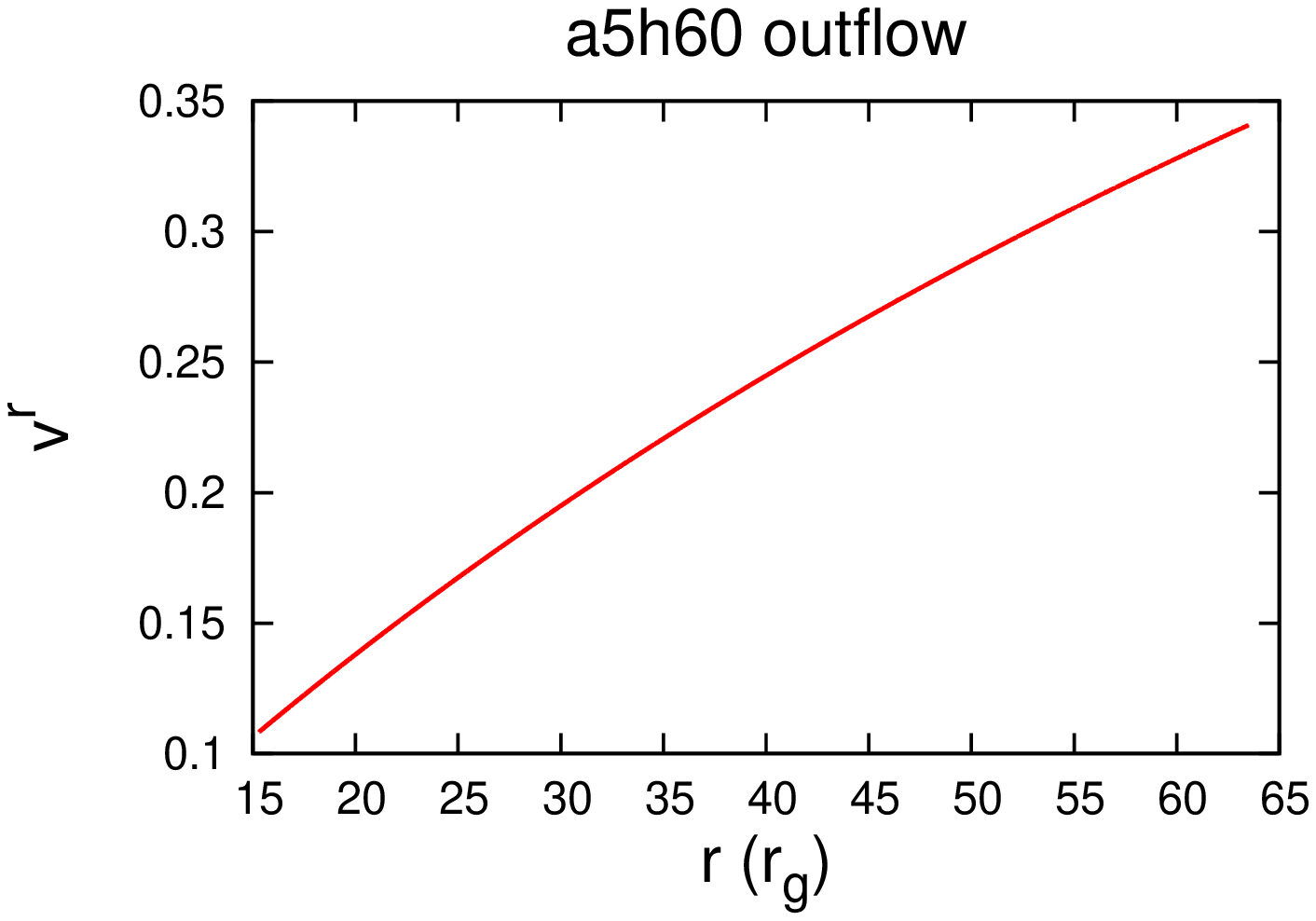}\\
\includegraphics[width=0.3\columnwidth, angle=-90]{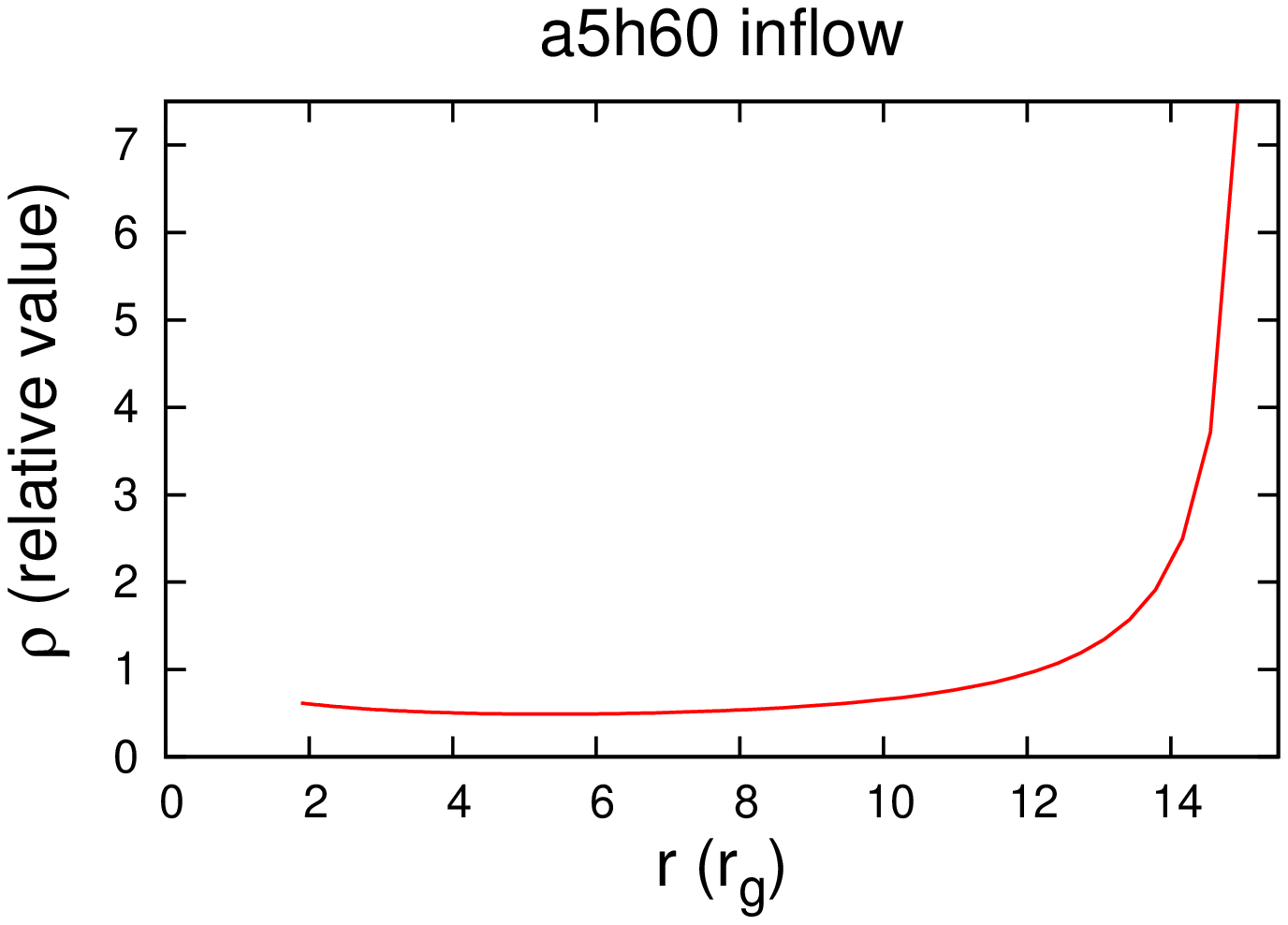}
\includegraphics[width=0.3\columnwidth, angle=-90]{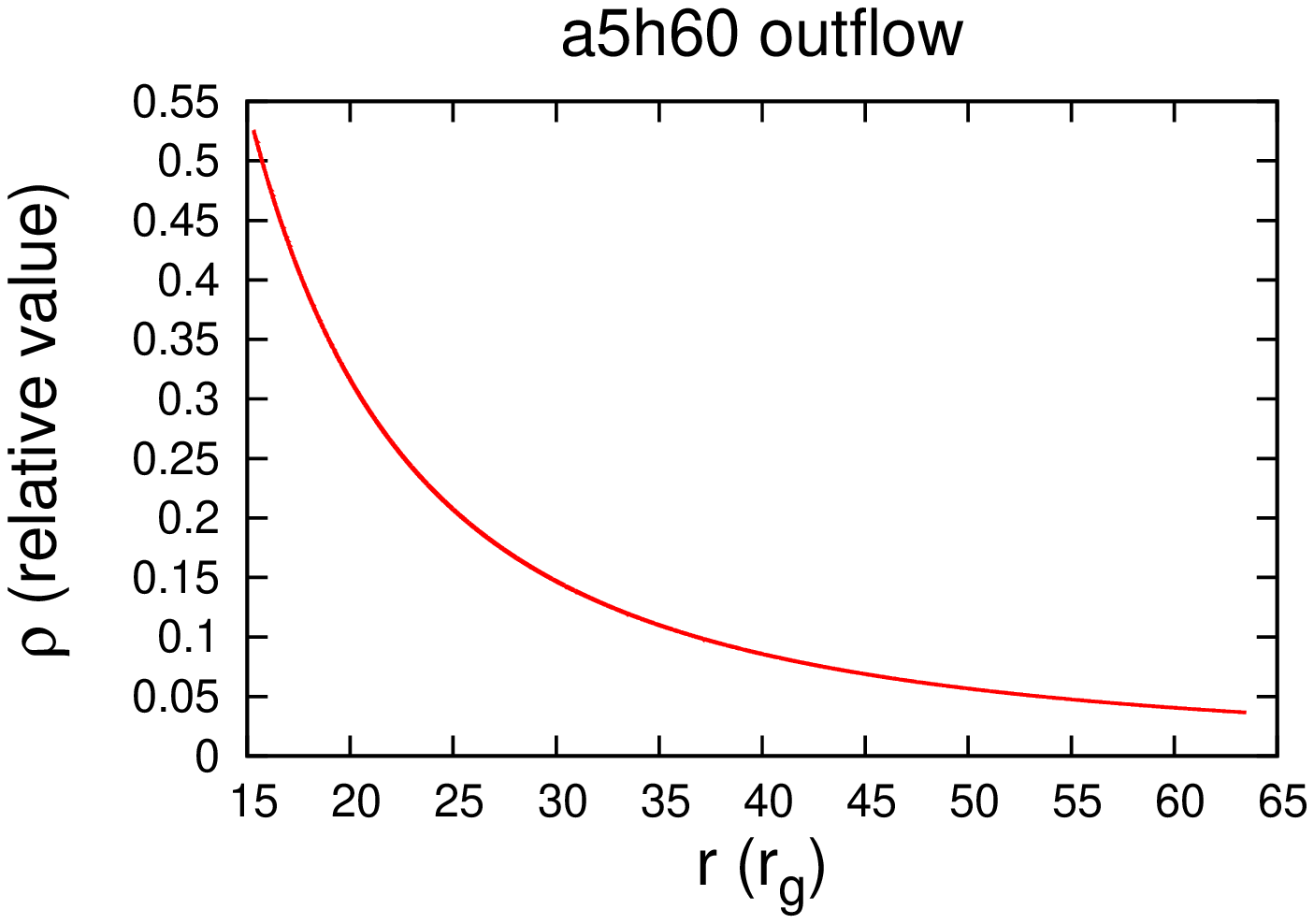}\\
\caption{GRMHD solutions for inflow region (left panel) and outflow region (right panel) used for model M9a5h60. See caption of Figure~\ref{fig:app_model} for description and comparison.} \label{fig:app_model1}
\end{centering}
\end{figure*}

\begin{figure*} 
\begin{centering}
\includegraphics[width=0.3\columnwidth, angle=-90]{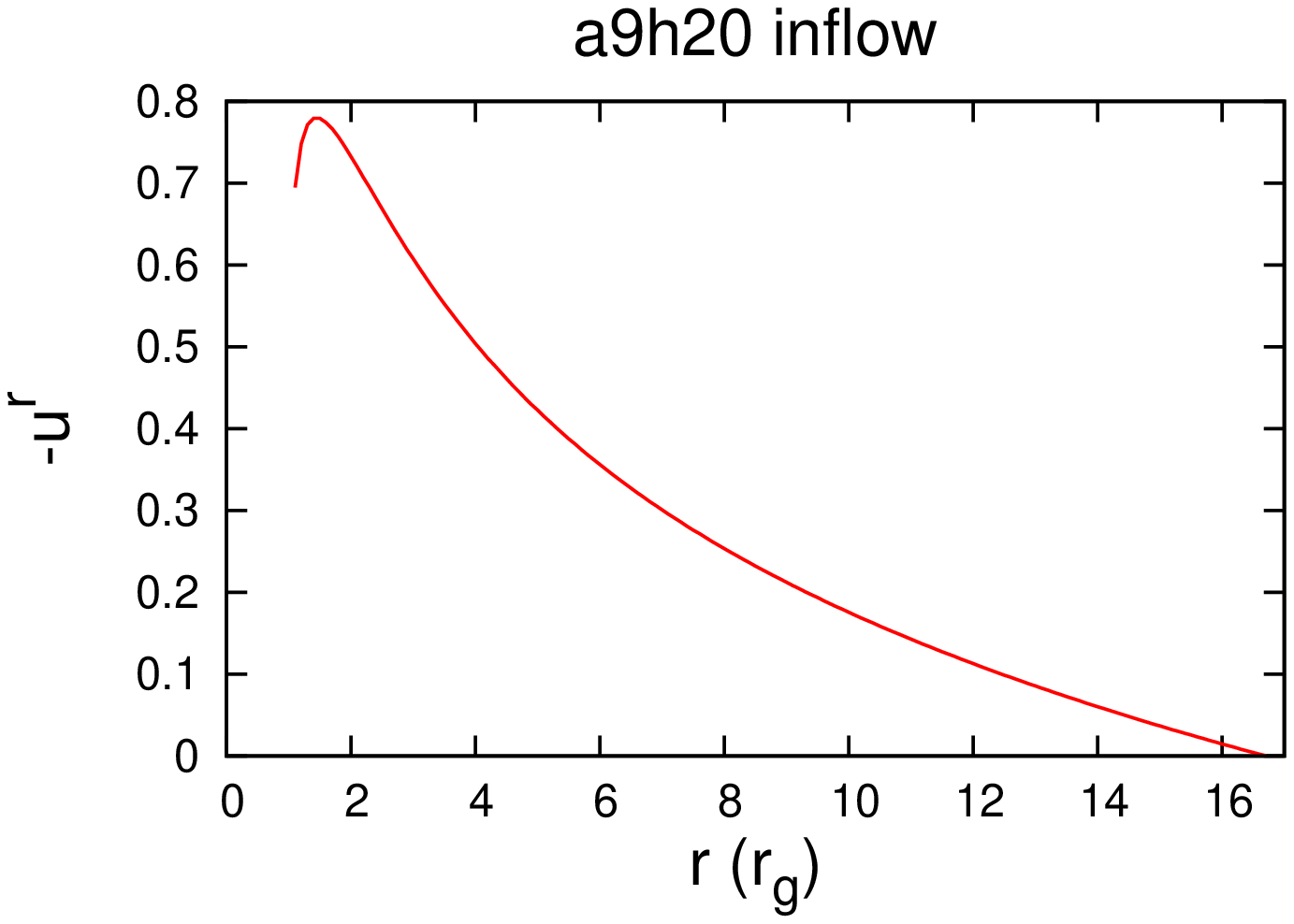}
\includegraphics[width=0.3\columnwidth, angle=-90]{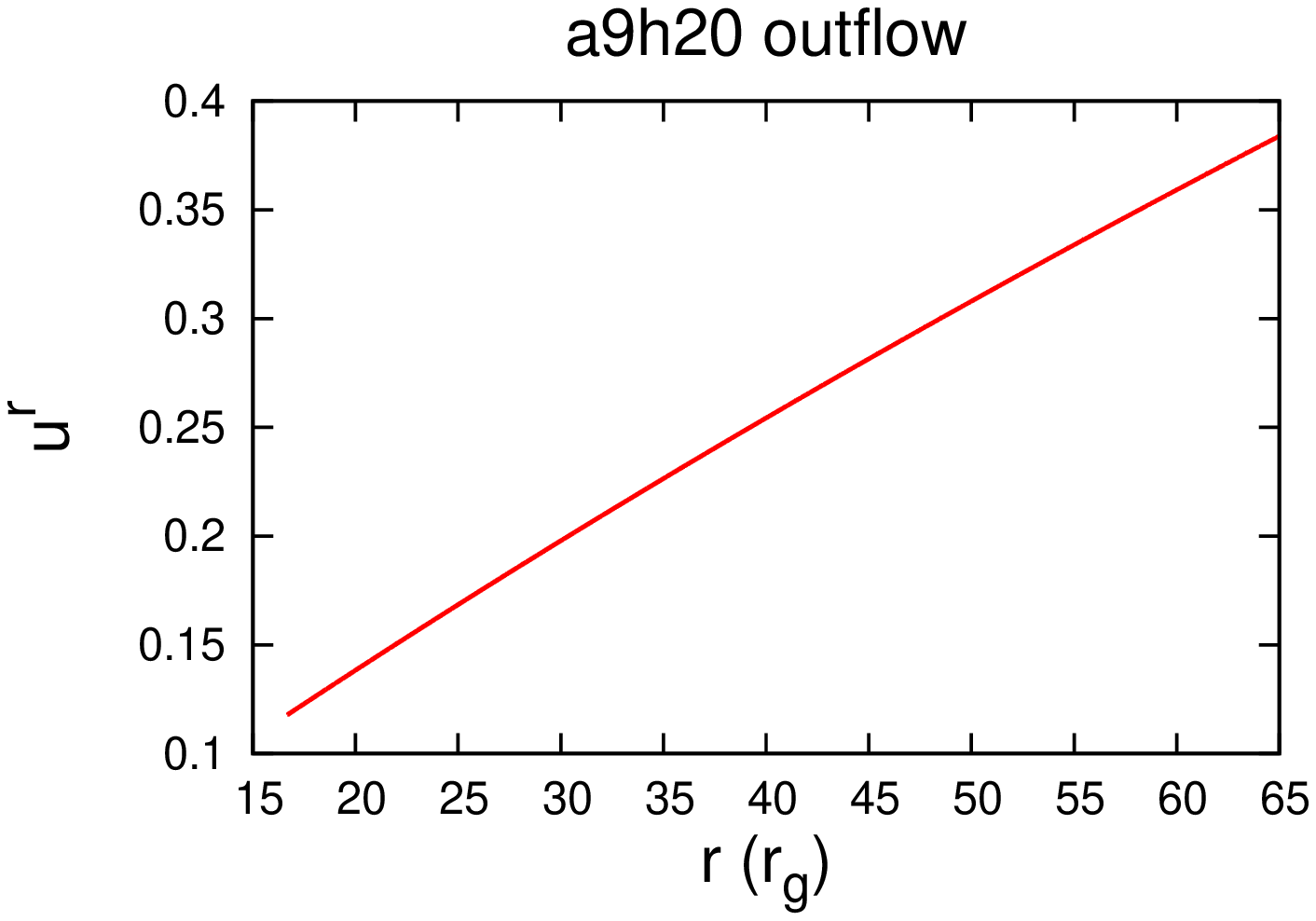}\\
\includegraphics[width=0.3\columnwidth, angle=-90]{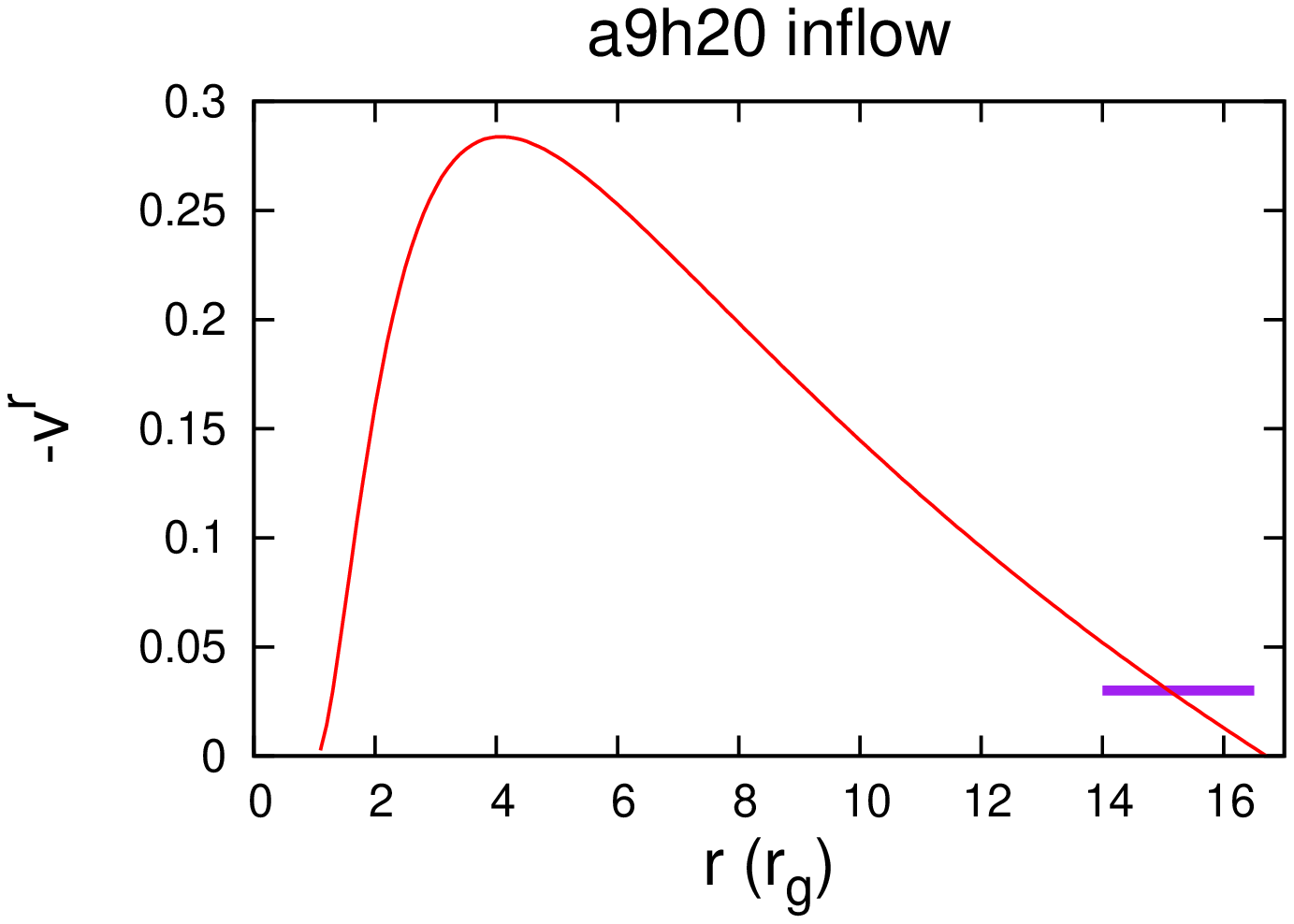}
\includegraphics[width=0.3\columnwidth, angle=-90]{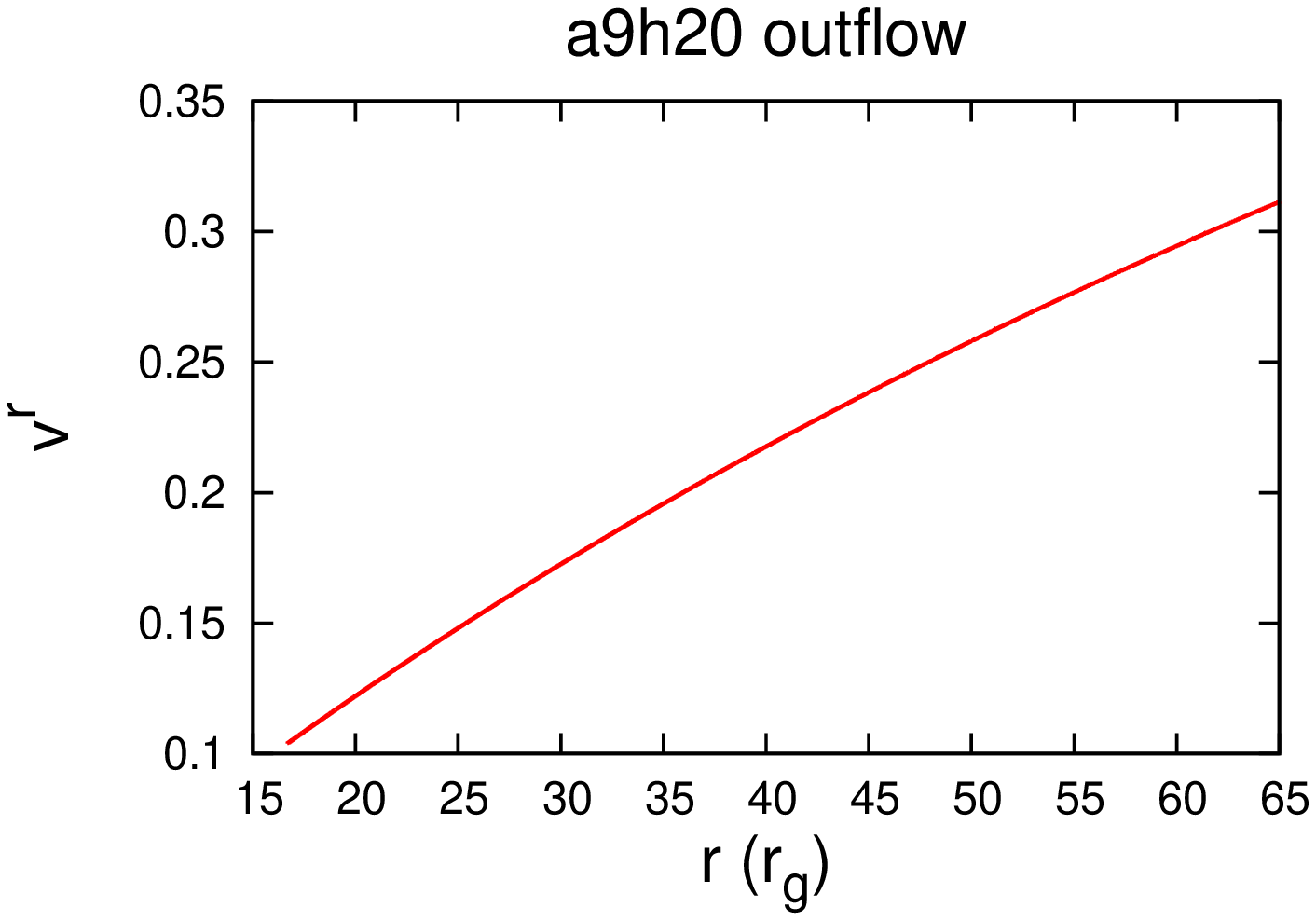}\\
\includegraphics[width=0.3\columnwidth, angle=-90]{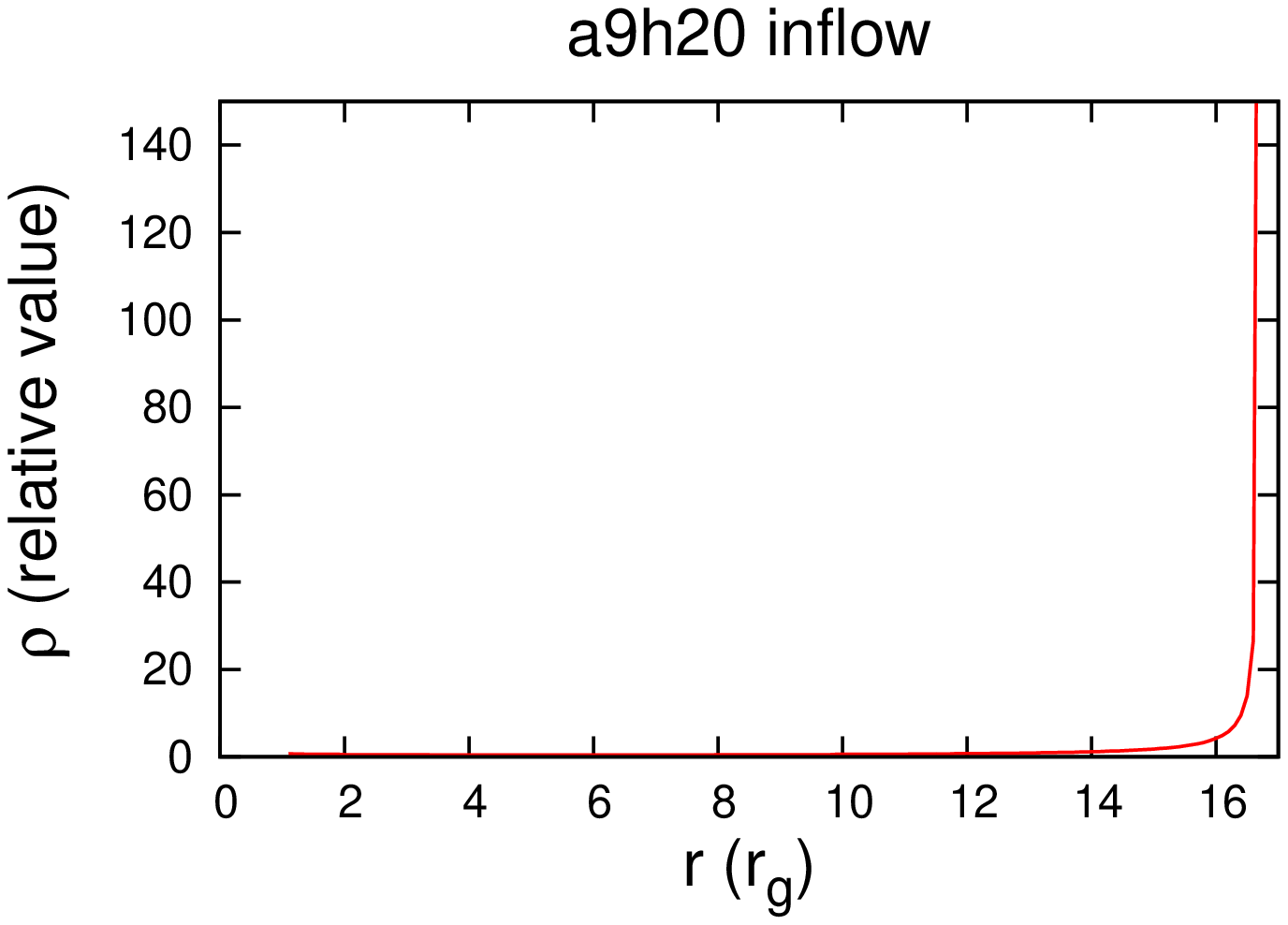}
\includegraphics[width=0.3\columnwidth, angle=-90]{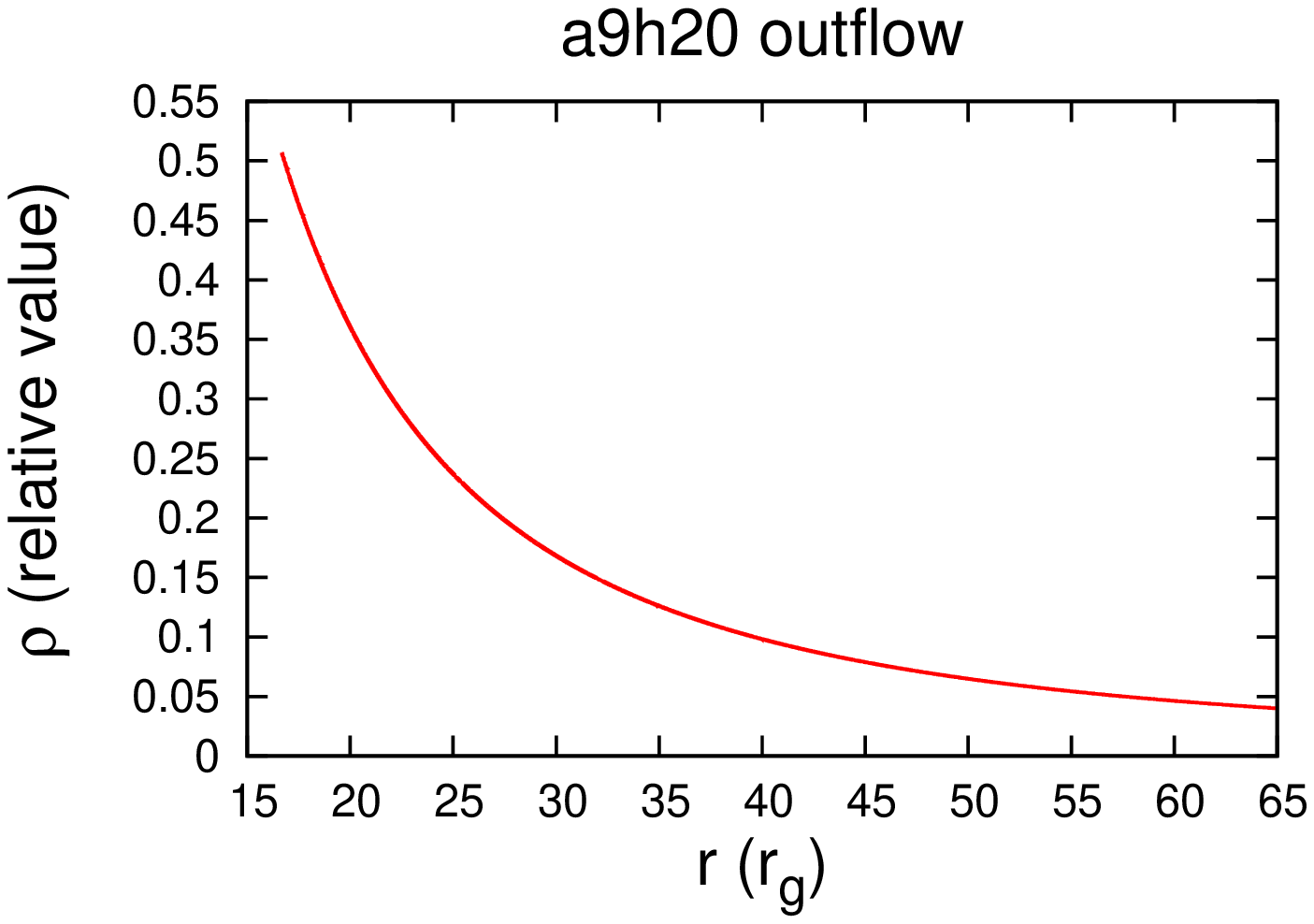}\\
\caption{GRMHD solutions for inflow region (left panel) and outflow region (right panel) used for model M9a9h20. See caption of Figure~\ref{fig:app_model} for description and comparison.} \label{fig:app_model2}
\end{centering}
\end{figure*}

The GRMHD solutions used for the calculation of the electron energy losses for models M9a9h60 and M6a9h60,  model M9a5h60 and model M9a9h20 are presented in Figures~\ref{fig:app_model}, \ref{fig:app_model1} and \ref{fig:app_model2}, respectively.

\end{document}